\documentclass[%
 aip,
 amsmath,amssymb,
 reprint,%
]{revtex4-2}

\usepackage{graphicx}
\usepackage{dcolumn}
\usepackage{bm}
\usepackage{xcolor}
\usepackage{mathrsfs}

\usepackage[utf8]{inputenc}
\usepackage[T1]{fontenc}
\usepackage{mathptmx}
\usepackage{etoolbox}
\usepackage{ulem}

\def\v#1{{\boldsymbol{#1}}}  

\makeatletter
\def\@email#1#2{%
 \endgroup
 \patchcmd{\titleblock@produce}
  {\frontmatter@RRAPformat}
  {\frontmatter@RRAPformat{\produce@RRAP{*#1\href{mailto:#2}{#2}}}\frontmatter@RRAPformat}
  {}{}
}%
\makeatother
\begin{document}

\preprint{AIP/123-QED}

\title[Article]{Viscoelasticity of suspension of red blood cells under oscillatory shear flow}

\author{Naoki Takeishi}
 \altaffiliation[Authors to whom correspondence should be addressed: ]{ntakeishi@kit.ac.jp}
 \affiliation{Faculty of Mechanical Engineering, Kyoto Institute of Technology, Goshokaido-cho, Matsugasaki, Sakyo-ku, Kyoto, 606-8585, Japan}
 \affiliation{Graduate School of Engineering Science, Osaka University, 1-3 Machikaneyama, Toyonaka, Osaka 560-8531, Japan}

\author{Marco Edoardo Rosti}
 \affiliation{Complex Fluids and Flows Unit, Okinawa Institute of Science and Technology Graduate University, 1919-1 Tancha, Onna-son, Okinawa 904-0495, Japan}

\author{Naoto Yokoyama}
 \affiliation{Department of Mechanical Engineering, Tokyo Denki University, 5 Senju-Asahi, Adachi, Tokyo, 120-8551, Japan}


\author{Luca Brandt}
\affiliation{FLOW, Department of Engineering Mechanics, Royal Institute of Technology (KTH), SE 100 44 Stockholm, Sweden}
\affiliation{Department of Energy and Process Engineering, Norwegian University of Science and Technology (NTNU), Trondheim, Norway}
\affiliation{Department of Environmental, Land, and Infrastructure Engineering, Politecnico di Torino, Corso Duca degli Abruzzi 24, 10129, Turin, Italy}

\date{\today}

\begin{abstract}
We present a numerical analysis of the rheology of a suspension of red blood cells (RBCs) for different volume fractions in a wall-bounded, effectively inertialess, small amplitude oscillatory shear (SAOS) flow for a wide range of applied frequencies. The RBCs are modeled as biconcave capsules, whose membrane is an isotropic and hyperelastic material following the Skalak constitutive law. The frequency-dependent viscoelasticity in the bulk suspension is quantified by the complex viscosity, defined by the amplitude of the particle shear stress and the phase difference between the stress and shear. SAOS flow basically impedes the deformation of individual RBCs as well as the magnitude of fluid-membrane interactions, resulting in a lower specific viscosity and first and second normal stress differences than in steady shear flow. Although it is known that the RBC deformation alone is sufficient to give rise to shear-thinning, our results show that the complex viscosity weakly depends on the frequency-modulated deformations or orientations of individual RBCs, but rather depends on combinations of the frequency-dependent amplitude and phase difference. The effect of the viscosity ratio between the cytoplasm and plasma and of the capillary number are also assessed.
\end{abstract}

\maketitle


\section{\label{sec:sec1}INTRODUCTION}

Human blood is a dense suspension consisting of 55\% plasma (which is typically assumed Newtonian fluid), and $45$\% blood cells, with over $98$\% of the cells being red blood cells (RBCs), which are non-spherical deformable particles. The study of the hydrodynamic interactions between individual RBCs and their consequent deformation is of fundamental importance for the bulk rheology of human blood, because these are responsible for the non-Newtonian behavior of the suspension, the well-known shear-thinning behavior~\citep{Chien1967, Chien1970, Cokelet1968, Goldsmith1975}. The RBC deformability highly complicates the suspension rheology, being key to determining the haemorheology. Indeed, the shear-thinning character of the blood disappears when RBCs are hardened~\citep{Chien1967, Chien1970}, with the blood suspension exhibiting an almost Newtonian behavior and higher viscosity compared to a suspension of normal RBCs. Several researchers have attempted to relate the rheological nature of the blood to the dynamics of RBCs, starting from experimental observations of the individual cell behaviors under controlled flows, usually simple steady shear flows, e.g., in Refs.~\onlinecite{ Schmid-Schonbein1969, Fischer1978, Abkarian2007, Lanotte2016}. More recently, numerical simulations have been also adopted to investigate the motion of RBCs at different steady shear rates; in particular, it was found that RBCs experience transitions from rolling/tumbling motions to kayaking (or oscillating/swinging) and swinging, following tank-treading motions as the shear rate increases for a wide range of ratios between the internal and external fluid viscosity ($\lambda = 0.1$--$10$)~\citep{Cordasco2014, Sinha2015, Lanotte2016, Takeishi2019}. Lanotte \textit{et al.}~\cite{Lanotte2016} further showed that tank-treading does occur for low viscosity contrasts ($\lambda \approx 0.25$), but is suppressed at high viscosity contrasts, when rotating multilobar shapes appear instead. The behavior of cells in flow, including vesicles, which are also enveloped by lipid bilayers, from individual dynamics to rheology was outlined in Ref.~\onlinecite{Vlahovska2009}. In our recent numerical analysis, we have related the aforementioned single-cell behavior to the bulk suspension rheology, and successfully demonstrated the shear-thinning character of blood flow examining the cellular-scale dynamics under steady shear~\citep{Takeishi2019}.

In real human blood, RBCs are constantly under mechanical stimulation from the plasma flow due to the heart beat ($\approx$ 1 Hz) and from the vessel walls in various organs. It is also known that RBCs travel in the body as an oxygen carrier, and easily alter their shapes to pass through $3$-to-$4$ $\mu$m capillaries~\citep{Skalak1969}, where the upper limit of shear stress in the human circulatory system has been estimated to be $15$ Pa~\citep{Koutsiaris2013}. However, when they travel through artificial blood pumps, the cells may experience much higher shear stresses, up to $1000$ Pa \citep{Deutsch2006}. 
From a physiological viewpoint, the relationship between the deformation, as a mechanical response to oscillatory loading, and the oxygen transport, as a biological function, is therefore of great interest, and several studies have shed some light on this~\citep{Parthasarathi1999, Wei2016, Zhou2019}. The rheological description of blood under oscillatory flow is thus of fundamental importance not only for a physiological understanding but also in the design of novel artificial blood pumps that minimize mechanical stimuli that may cause the rupture of RBCs, the so-called hemolysis. Using a capsule model consisting of a Newtonian fluid enclosed by a thin elastic membrane, \citet{Matsunaga2015} demonstrated that frequency-dependent deformations of a single spherical capsule become evident for high shear rates and large values of the viscosity contrast between the internal and external fluids. However, it is still unknown whether this knowledge can be used for suspensions of RBCs. The first question in this study is, therefore, whether the cell deformation is reduced or enhanced by varying the oscillatory frequency. Although the recovery of RBCs under oscillatory flow has been investigated in the past via experimental observations~\citep{Nakajima1990, Recktenwald2022, Watanabe2006}, and model analysis~\citep{Noguchi2010, Li2014, Cordasco2016, Zhu2019}, much is still unknown, especially in relation to the bulk suspension rheology and to the dynamical viscoelasticity of suspensions of RBCs under oscillatory shear flow. Hence, our second question is how the viscoelastic character of the suspension of RBCs differs in an oscillatory shear flow with respect to the steady case.

The linear mechanical response of soft materials to weak oscillatory shear strain $\gamma (t)$ is generally frequency dependent and viscoelastic, so that the oscillatory stress $\tau$ can be expressed with the help of a complex shear modulus $G^{\ast}$: $\tau (t) = G^{\ast} \gamma (t)$~\citep{ Squires2010,Mewis2012}. The complex modulus $G^{\ast}$ can be decomposed into its two components, $G^{\ast} = G^{\prime} + i G^{\prime\prime}$, where the real part $G^{\prime}$ is the storage modulus representing the elastic component of the stress, and the imaginary part $G^{\prime\prime}$ is the loss modulus representing the viscous dissipative part~\citep{ Squires2010,Mewis2012}. Traditional (macroscopic) rheometers enable direct mechanical measurements of the frequency dependent $G^{\ast}$. To cite few relevant examples, Mason and Weitz~\cite{Mason1995} extracted the rheological properties from the thermal motion of colloidal probes embedded within the material, whereas Wagner~\cite{Wagner1993} and Lionberger and Russel~\cite{Lionberger1994} calculated $G^{\prime}$ at high frequency in hard-sphere dispersions from theory, and compared the results to the experiments by Shikata and Pearson~\cite{Shikata1994}. Although numerical simulations allow us to quantify the dynamical viscoelasticity by calculating the complex moduli, the analysis of deformable particle suspensions is still a challenge because the hydrodynamic coupling among the particles, their deformation and the solvent motion must all be taken into account~\citep{Wagner1993,Lionberger1994}.

As regards the role of deformability, Frohlich and Sack~\cite{Frohlich1946} were the first to investigate a suspension of Hookean elastic spheres and reported a linearized Oldroyd-type constitutive equation derived to relate the macroscopic extensional stress and the strain rate. Oldroyd~\cite{Oldroyd1953} reported that a suspension of Newtonian droplets in another Newtonian liquid exhibits viscoelastic behavior with an Oldroyd-type constitutive relation. The rheological behavior of such suspensions is similar to that of a suspension of elastic particles in a Newtonian fluid. Based on the coupled solutions for the flow field around the particle and the deformation of the particle, both analytical and numerical approaches have been applied to investigate the dilute suspension rheology, especially under steady shear flow~\citep{Barthes-Biesel1980, Goddard1967, Misbah2006, Roscoe1967}.
Among others, Goddard and Miller\cite{Goddard1967} derived constitutive equations for the rheological behavior of a suspension of slightly deformed viscoelastic spheres in the dilute limit. Misbah~\cite{Misbah2006} derived equations which describe the vesicle orientation in the flow and its shape evolution, and outlined a rheological law for a dilute vesicle suspension. Along with these analytical and numerical studies, recent computer simulation approaches have successfully been used to investigate rheological properties of a dilute and semi-dilute suspension of deformable particles in steady shear flow, e.g., elastic initially spherical particles~\citep{Gao2011, Rosti_Brandt_Mitra2018, Rosti_Brandt2018} and capsule~\citep{Omori2014}.
Numerical analyses of semi-dilute and jammed particle suspensions under oscillatory shear flow have been reported recently, e.g., rigid spherical particles at finite inertia~\citep{Villone2021}, soft particle glasses~\citep{Khabaz2018}, viscoelastic particles~\citep{Kammer2020}, bubbles~\citep{Lin2019}, vesicles~\citep{Farutin2012}, and spherical capsule in dilute condition~\citep{Matsunaga2020}. The recent theoretical work by Armstrong \textit{et al.}~\cite{Armstrong2018} compared viscoelastic moduli obtained with several viscoelastic(-thixotropic) models and laboratory measurements under oscillatory shear flow. \citet{Matsunaga2020} systematically investigated the effect of viscosity ratio on the viscoelastic character of capsule suspension for a wide range of oscillatory shear rate frequencies. By employing continuum modeling, in particular the Oldroyd 8-constant framework, Saengow \textit{et al.}~\cite{Saengow2019} assessed the non-Newtonian character of human blood under uni-directional large-amplitude oscillatory shear (LAOS) flow,  which is generated by superposing LAOS onto a steady shear flow. Clarifying the cellular-scale dynamics under oscillatory flow allows us to build precise continuum models of suspensions~\citep{Oldroyd1958, Anand2013, Alves2021}, and may lead us to novel biomedical applications~\citep{Mutlu2018}, such as phenotype cell screening, and circulating tumor cell isolation in a chip~\citep{Martel2014}. However, this can be achieved only after fully grasping the effect of the particle deformations induced by the oscillations of the bulk suspension.

Therefore, the objective of this study is to clarify the relationship between the behavior of individual RBCs especially under SAOS flows and the viscoelastic character of dense suspension of RBCs. The contribution of the individual deformed RBC to the bulk suspension rheology is quantified by the stresslet tensor~\citep{Batchelor1970}. Here, the RBC is modeled as a biconcave capsule, whose membrane follows the Skalak constitutive law~\citep{Skalak1973}. The internal and external fluid of the RBCs are modeled as Newtonian fluids and solved by the lattice-Boltzmann method (LBM), while the membrane mechanics with a  finite-element method. The fluid and structure are fully coupled by an immersed boundary method~\citep{Peskin2002}. We resort to graphics processing unit (GPU) computing to speed-up the above-mentioned numerical procedure. 
The same tools have been successfully applied to the analysis of the bulk suspension rheology of RBCs under steady shear flow in Takeishi \textit{et al.}~\cite{Takeishi2019}.


\section{PROBLEM STATEMENTS}
\subsection{\label{sec:setup}Flow and cell models}

\begin{figure}
  \centering
  \includegraphics[height=5.0cm]{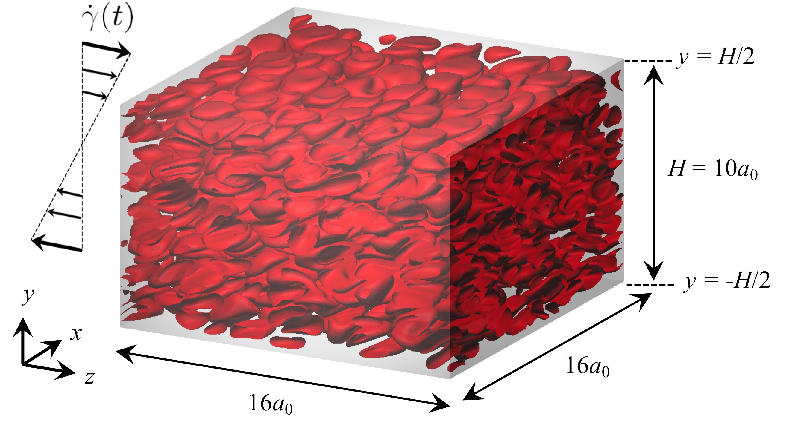}
  \caption{Computational domain: a rectangular box of size 16$a_0$ $\times$ 10$a_0$ $\times$ 16$a_0$ along the span-wise $x$, wall-normal $y$, and stream-wise $z$ directions. Periodic boundary conditions are imposed on the two homogeneous directions ($x$ and $z$ directions). The oscillatory shear flow is generated by moving the top and bottom walls located at $y = \pm H/2$, where $H$ (= 10$a_0$) is the domain height. The snapshot depicts a dense suspension of RBCs ($\phi$ = 0.41) with viscosity ratio $\lambda$ = 5 and capillary number $Ca_0$ = 0.05.}
  \label{fig:domain}
\end{figure}

We consider a cellular flow consisting of plasma and RBCs with radius $a_0$ in a rectangular box of size 16$a_0$ $\times$ 10$a_0$ $\times$ 16$a_0$ along the span-wise $x$, wall-normal $y$, and stream-wise $z$ directions, with a resolution of 16 fluid lattices per radius of RBC. The size of the domain and the numerical resolution have been justified in our previous works~\citep{Takeishi2014, Takeishi2019}. Here, we further verified the effect of the wall-to-wall distance $H$ as well as the cell-depleted peripheral layer (CDPL), see table~\ref{tab:effect_height} and Fig.~\ref{fig:effect_height} in appendix~\S\ref{appA_domain_size}. Periodic boundary conditions are imposed on the two homogeneous directions ($x$ and $z$ directions). Each RBC is modeled as a biconcave capsule, i.e.,~a Newtonian fluid enclosed by a thin elastic membrane, with a major diameter of 8 $\mu$m (= 2$a_0$), and maximum thickness of 2 $\mu$m (= $a_0$/2). The initial shape of the RBC is set to be the classical biconcave shape. A sketch of the computational domain, with the coordinate system used, is shown in Fig.~\ref{fig:domain} with an instantaneous visualisation of one of the dense cases, volume fraction $\phi$ = 0.41 (the so-called discharge hematocrit, $Hct_D$).

The oscillatory shear flow is generated by moving the top and bottom walls located at $y = \pm H/2$ with velocity $U_{wall} = \pm \dot{\gamma}(t)H/2$, where $H$ (= 10$a_0$) is the domain height and $\dot{\gamma}(t)$ the shear rate, defined as
\begin{equation}
	\dot{\gamma}(t) = \dot{\gamma}_0 \exp{(i 2 \pi \mathrm{f} t)}.
\end{equation}
Here, $\dot{\gamma}_0$ is the shear-rate amplitude and $\mathrm{f}$ its frequency. The oscillatory strain $\gamma (t)$ is therefore defined as
\begin{equation}
	\gamma (t)
	= \int \dot{\gamma} (t) dt
	= \frac{\dot{\gamma}_0}{i 2 \pi \mathrm{f}} \exp{(i 2 \pi \mathrm{f} t)}
	= \gamma_0 \exp{(i 2 \pi \mathrm{f} t - \pi/2)},
\end{equation}
where $\gamma_0 (= \dot{\gamma}_0/(2 \pi \mathrm{f}))$ is the strain amplitude. For the following analysis, we define the non-dimensional input frequency as $f_{in} = \mathrm{f}/\dot{\gamma}_0$.

The membrane is modeled as an isotropic and hyperelastic material. The surface deformation gradient tensor $\mathbf{F}_s$ is given by
\begin{equation}
  d\v{x}_m = \mathbf{F}_s \v{\cdot} d\v{X}_m,
  \label{Fs}
\end{equation}
where $\v{X}_m$ and $\v{x}_m$ are the membrane material points of the reference and deformed states, respectively. The local deformation of the membrane can be measured by the Green--Lagrange strain tensor
\begin{equation}
  \mathbf{E} = \frac{1}{2} \left( \mathbf{C} - \mathbf{I}_s \right),
  \label{e}
\end{equation}
where $\mathbf{I}_s$ is the tangential projection operator. The two invariants of the in-plane strain tensor $\mathbf{E}$ can be given by 
\begin{equation}
  I_1 = \lambda_1^2 + \lambda_2^2 - 2, \quad I_2 = \lambda_1^2 \lambda_2^2 - 1 = J_s^2 - 1,
  \label{I1andI2}
\end{equation}
where $\lambda_1$ and $\lambda_2$ are the principal extension ratios. The Jacobian $J_s = \lambda_1\lambda_2$ indicates the ratio of the deformed to the reference surface areas. The elastic stresses in an infinitely thin membrane are replaced by elastic tensions. The Cauchy tension $\mathbf{T}$ can be related to an elastic strain energy per unit area, $w_s \left( I_1, I_2 \right)$:
\begin{equation}
  \mathbf{T} = \frac{1}{J_s} \mathbf{F}_s \cdot \frac{\partial w_s \left( I_1, I_2 \right)}{\partial \mathbf{E}} \cdot \mathbf{F}_s^T,
  \label{T}
\end{equation}
where the strain energy function $w_s$ satisfies the Skalak (SK) constitutive law~\citep{Skalak1973}
\begin{equation}
  \frac{w_s}{G_s} = \frac{1}{4} \left( I_1^2 + 2 I_1 - 2I_2 + C I_2^2\right).
  \label{SK}
\end{equation}
In the expression above, $G_s$ is the surface shear elastic modulus, and $C$ a coefficient representing the area incompressibility. Substituting equation~\eqref{SK} into equation~\eqref{T}, we obtain the principal tensions in the membrane $T_1$ and $T_2$ with $T_1 \geq T_2$:
\begin{equation}
  \frac{T_i}{G_s} = \frac{\lambda_i}{\lambda_j} \left[ \lambda_i^2 - 1 + C \left( \lambda_i^2 \lambda_j^2 - 1 \right) \right],
  \quad \text{for} \ (i , j) = (1, 2) \ \text{or} \ (2, 1).
  \label{eq:Ti}
\end{equation}
In this study, we set 
$C$ = 10$^2$~\citep{Barthes-Biesel2002}.  Bending resistance is also considered~\citep{Li2005}, with a bending modulus $k_b$ = 5.0 $\times$ 10$^{-19}$ J~\citep{Puig-de-Morales-Marinkovic2007}; these values have been shown to successfully reproduce the deformation of RBCs in shear flow and  the thickness of the CDPL~\citep{Takeishi2014}.
Note that the nodal forces on discrete elements, produced by bending energy, preserve the force- and torque-free principle. The reader is referred to~\citet{Pozrikidis2010} for a precise description of bending resistance.

The internal (cytoplasm) and external (plasma) fluids are treated as Newtonian viscous fluids, which obey the incompressible Navier--Stokes equations. Although the large standard deviation of the values of the geometric and mechanical properties of healthy RBCs were presented depending on the use of different techniques, cytoplasmic viscosity of RBCs was summarized as $6.07\pm3.8$ mPa$\cdot$s in~\citet{Tomaiuolo2014}.
The normal plasma viscosity, on the other hand, is $\mu_0$ = 1.1--1.3 cP (1.1--1.3 mPa$\cdot$s) at 37 $^\circ$C~\citep{Harkness1970}. 
Hence, physiologically relevant values of the viscosity ratio lie in the range $\lambda$ (= $\mu_1/\mu_0$) = 1.89--8.23 if the plasma viscosity is set to be $\mu_0$ = 1.2 cP. 
The viscosity ratio $\lambda$ = 5 is adopted for most of the results presented in this study, except when specifically investigating the role of $\lambda$ on the suspension rheology. In this case,
 the range $\lambda$ = 0.1--10 is investigated.

In the inertialess limit, the problem is characterised by the capillary number,
\begin{equation}
  Ca_0 = \frac{\mu_0 \dot{\gamma}_0 a_0}{G_s},
  \label{Ca}
\end{equation}
in addition to the non-dimensional input frequency $f_{in} = \mathrm{f}/\dot{\gamma}_0$ and the viscosity ratio $\lambda$ introduced above. 
The capillary  number quantifies the relative importance of elastic and viscous forces, with highly deformable cells characterised by large values of $Ca_0$.
To limit the computational cost and yet obtain results not affected by inertial effects, we set the Reynolds number $Re = \rho \dot{\gamma}_0 a_0^2/\mu_0$ = 0.2, where $\rho$ is the plasma density. This value well reproduces the capsule dynamics in unbounded shear flows obtained with the boundary integral method in Stokes flow (see also Figures~12 and 13 in~\citet{Takeishi2019}). We further investigate the effect of $Re$ on the Stokes boundary layer in appendix~\S\ref{appA_stokes_problem} and confirmed that the layer remained the same even at lower $Re (= 0.05)$ as shown in Fig.~\ref{fig:stokes_layer}.


\subsection{Numerical method}

The finite-element method (FEM) is used to solve the  weak form of the  equations governing the inertialess membrane dynamics and obtain the load $\v{q}$ acting on the membrane:
\begin{equation}
  \int_S \v{\hat{u}} \v{\cdot} \v{q} dS = \int_S \mathbf{\hat{\epsilon}} : \mathbf{T} dS,
  \label{WeakForm}
\end{equation}
where $\v{\hat{u}}$ and $\mathbf{\hat{\epsilon}} = ( \nabla_s \v{\hat{u}} + \nabla_s \v{\hat{u}}^T )\big/2$ are the virtual displacement and virtual strain, respectively. 
The in-plane elastic tension $\mathbf{T}$ is obtained from the Skalak constitutive law, see equation (\ref{SK}).

The Navier--Stokes equations for the internal and external fluids are discretised by the LBM based on the D3Q19 model~\citep{Chen1998, Dupin2007}. In the  LBM, the macroscopic flow is obtained by collision and streaming of hypothetical particles described by the lattice-Boltzmann-Gross-Krook (LBGK) equation~\citep{Bhatnagar1954}, which is given as
\begin{align}
  &f_i \left( \v{x}_f + \v{c}_i \Delta t, t + \Delta t \right) - f \left( \v{x}_f, t \right) \nonumber \\
  &\qquad = - \frac{1}{\tau} \left[ f_i \left( \v{x}_f, t \right) - f_i^{eq} \left( \v{x}_f, t \right) \right] + F_i \Delta t,
  \label{LB}
\end{align}
where $f_i$ is the particle distribution function for particles with velocity $\v{c}_i$ ($i$ = 0--18) at the fluid node $\v{x}_f$, $\Delta t$ is the time step size, $f_i^{eq}$ is the equilibrium distribution function, $\tau$ is the non-dimensional relaxation time, and $F_i$ is the external force applied from the membrane material points, obtained with the immersed boundary method (IBM)~\citep{Peskin2002}. The membrane force at the membrane node $\boldsymbol{x}_m$ is distributed to the neighboring fluid nodes $\boldsymbol{x}_f$ using a smoothed delta function approximating the Dirac delta function as
\begin{align}
  D \left( \boldsymbol{x} \right) =
  \begin{cases}
  \dfrac{1}{64 \left( \Delta x_f \right)^3} \prod_{k = 1}^3 \Bigl( 1 + \cos \dfrac{\pi x_k}{2 \Delta x_f} \Bigr) & \text{if} \ \left| x_k \right| \leq 2 \Delta x_f, \\
  0 & \text{otherwise},
  \end{cases}
  \label{delta_f}
\end{align}
where $k \in [1,3]$ and $x_1 = x$, $x_2 = y$, $x_3 = z$. Similarly, the velocity at the membrane node $\boldsymbol{v} \left( \boldsymbol{x}_m \right)$ is obtained by interpolating the velocities at the fluid nodes.
The membrane node $\v{x}_m$ is updated by Lagrangian tracking with the no-slip condition, i.e.
\begin{equation}
\frac{d \v{x}_m}{dt} = \v{v} \left( \v{x}_m \right).
\end{equation}
The explicit fourth-order Runge--Kutta method is used for time-integration.

In our coupling of method of LBM and IBM, the hydrodynamic interaction between individual RBCs is solved without modeling a non-hydrodynamic inter-membrane repulsive force in the case of vanishing inertia, as also shown in Figure~13 in~\citet{Takeishi2019}. The volume-of-fluid (VOF) method~\citep{Yokoi2007} and front-tracking method~\citep{Unverdi1992} are employed to update the viscosity in the fluid lattices. A volume constraint is implemented to counteract the accumulation of small errors, leading to changes in the volume of the individual cells~\citep{Freund2007}: in our simulation, the volume error is always maintained lower than 10$^{-3}$\%.  All numerical procedures are fully implemented on GPU to accelerate the numerical simulation.  For further details of the methods we refer to our previous works~\citep{Takeishi2019, Takeishi2022}. The mesh size of the LBM for the fluid solution is set to be 250 nm, and that of the finite elements describing the membrane is also approximately 250 nm (an unstructured mesh with 5,120 elements is used for each cell).  This resolution has been shown to successfully represent single- and multi-cellular dynamics~\citep{Takeishi2014}; we have verified that the results of multi-cellular dynamics are not changing with twice the resolution for both the fluid and membrane~\citep[see also][]{Takeishi2014}. We have further investigated the distance between neighboring membrane nodes. The result of time-averaged distance of neighboring membrane nodes $\langle \Delta_\mathrm{node} \rangle$ for different volume fractions $\phi$ is discussed in appendix~\S\ref{appA_distance}.

\subsection{Analysis of capsule suspensions}
For the following analysis, the behavior of RBCs subjected to oscillatory shear flow is quantified by the length of the semi-major axis $a_{max}$ and orientation angle $\theta$ between the major axis of the deformed RBC and the shear direction. The length of the semi-major axis $a_{max}$ of the deformed RBC is obtained from the eigenvalues of the inertia tensor of an equivalent ellipsoid approximating the deformed RBC~\citep{Ramanujan1998}. The orientation angle $\theta$ of a single deformable spherical capsule converges to $\pi$/4 in shear flow as $Ca \to$ 0~\citep{Barthes-Biesel1980, Barthes-Biesel1985}.

The suspension rheology of RBCs, or the contribution of the suspended RBCs to the bulk viscosity, is quantified by the particle stress $\boldsymbol{\Sigma}^{(p)}$~\citep{Batchelor1970}. Specifically, for a deformable capsule and any viscosity ratio, Pozrikidis~\cite{Pozrikidis1992} analytically derived an expression for the corresponding stresslet, so that the particle contribution to the total stress can be written as:
\begin{align}
  \boldsymbol{\Sigma}^{(p)}
  &= \frac{1}{V} \sum_{i = 1}^N \mathbf{S}_i, \\
  &= \frac{1}{V} \sum_{i = 1}^N \int_{A_i}
  \left[
  \frac{1}{2} \left( \v{r}\v{\hat{q}} + \v{\hat{q}}\v{r} \right) - \mu_0 \left( 1 - \lambda \right) \left( \v{v} \v{n} + \v{n} \v{v} \right)
  \right] dA_i,
  \label{eq:stresslet}
\end{align}
where $V$ is the volume of the domain, $\mathbf{S}_i$ the stresslet of the {\it i}-th RBC (or capsule), $\v{r}$ is the membrane position relative to the centre of the RBC, $\v{\hat{q}}$ the load acting on the membrane including the contribution from the bending rigidity, $\mu_0$ the outer fluid (plasma) viscosity, $\v{v}$ the interfacial velocity of the membrane, and $A_i$ the membrane surface area of the {\it i}-th RBC.  Here, the suspension shear viscosity $\mu_{all} (= \mu_0 + \delta \mu)$ is represented by the viscosity $\mu_0$ of the carrier fluid (plasma) and a perturbation $\delta \mu$. This leads to the introduction of the relative viscosity $\mu_{re}$ and of the specific viscosity $\mu_{sp}$ defined as:
\begin{align}
  \mu_{re} &= \frac{\mu_{all}}{\mu_0} = 1 + \mu_{sp}, \\
  \mu_{sp} &= \frac{\delta \mu}{\mu_0} = \frac{\Sigma^{(p)}_{12}}{\mu_0 \dot{\gamma}_0},
\end{align}
where the subscript $_1$ represents the streamwise direction (i.e., the $z$-direction in this study) and the subscript  $_2$ the wall-normal direction  (i.e., the $y$-direction in this study). Using the diagonal components of the particle stress $\Sigma^{(p)}_{ii}$, the first and second normal stress differences, typically used to quantify the suspension viscoelastic behavior, can be defined as:
\begin{align}
  \frac{N_1}{\mu_0 \dot{\gamma_0}} &= \frac{\Sigma^{(p)}_{11} - \Sigma^{(p)}_{22}}{\mu_0 \dot{\gamma}_0}, \\
  \frac{N_2}{\mu_0 \dot{\gamma_0}} &= \frac{\Sigma^{(p)}_{22} - \Sigma^{(p)}_{33}}{\mu_0 \dot{\gamma}_0}.
\end{align}

In the case of small oscillations, the particle stress $\Sigma_{12}^{(p)} (t)$ is also an oscillatory function, defined by the amplitude $|\Sigma_{12}^{(p)}|^{amp}$ and the phase difference $\delta$ ($-\pi$/2 $\leq \delta \leq$ 0) between $\Sigma_{12}^{(p)}(t)$ and the input shear rate $\dot{\gamma}(t)$:
\begin{align}
	\Sigma_{12}^{(p)} (t)
	= |\Sigma_{12}^{(p)}|^{amp} \exp{\left\{ i (2 \pi \mathrm{f} t + \delta) \right\}}.
  \label{eq:Sigma}
\end{align}
Following classical rheological analyses,  the frequency-dependent   rheology of particle suspensions is expressed in terms of a complex modulus $G^\ast$ and a complex viscosity $\eta^\ast$. Using the fluid strain $\gamma (t)$, $G^{\ast}$ is given by:
\begin{align}
	G^\ast
	&= \frac{\Sigma_{12}^{(p)} (t)}{\gamma (t)}
	= \frac{i |\Sigma_{12}^{(p)}|^{amp}}{\gamma_0} \exp{(i \delta)}
	= G^{\prime} + i G^{\prime\prime},
  \label{eq:G}
\end{align}
where $G^\prime$ is the storage modulus and $G^{\prime\prime}$ is the loss modulus defined as:
\begin{align}
	\left\{
	\begin{array}{l}
	G^{\prime}
	=-\dfrac{2 \pi \mathrm{f} |\Sigma_{12}^{(p)}|^{amp}}{\dot{\gamma}_0} \sin{\delta}, \\
	G^{\prime\prime}
	= \dfrac{2 \pi \mathrm{f} |\Sigma_{12}^{(p)}|^{amp}}{\dot{\gamma}_0} \cos{\delta}.
	\end{array}
	\right.
  \label{eq:G_def}
\end{align}
Using the applied shear $\dot{\gamma} (t)$, instead, one can introduce a complex viscosity $\eta^\ast$, given by:
\begin{align}
	\eta^\ast
	&= \frac{\Sigma_{12}^{(p)} (t)}{\dot{\gamma} (t)}
	= \frac{|\Sigma_{12}^{(p)}|^{amp}}{\dot{\gamma}_0} \exp{(i \delta)}
       = \eta^{\prime} - i \eta^{\prime\prime},
  \label{eq:eta}
\end{align}
where $\eta^\prime$ and $\eta^{\prime\prime}$ are the viscous and elastic component of $\eta^\ast$, respectively~\citep{Thurston1972}, and are defined as:
\begin{align}
	\left\{
	\begin{array}{l}
	\eta^{\prime}
	= \dfrac{|\Sigma_{12}^{(p)}|^{amp}}{\dot{\gamma}_0} \cos{\delta} = \dfrac{G^{\prime\prime}}{2\pi \mathrm{f}}, \\
	\eta^{\prime\prime}
	= -\dfrac{|\Sigma_{12}^{(p)}|^{amp}}{\dot{\gamma}_0} \sin{\delta} = \dfrac{G^\prime}{2\pi \mathrm{f}}.
	\end{array}
	\right.
  \label{eq:eta_def}
\end{align}
In the following analysis, $\eta^\prime$ and $\eta^{\prime\prime}$ are normalized by $\mu_0$, and hence, the moduli are rewritten as:
\begin{align}
	\left\{
	\begin{array}{l}
	\dfrac{\eta^{\prime}}{\mu_0}
	= |\mu_{sp}|^{amp} \cos{\delta}, \\
	\dfrac{\eta^{\prime\prime}}{\mu_0}
	= -|\mu_{sp}|^{amp} \sin{\delta},
	\end{array}
	\right.
  \label{eq:eta_def2}
\end{align}
where $|\mu_{sp}|^{amp} (=|\Sigma_{12}^{(p)}|^{amp}/(\mu_0 \dot{\gamma}_0))$ is the magnitude of the specific viscosity.

The oscillations are examined after the fully-developed regime has been reached. At least after non-dimensional time $\dot{\gamma}_0t \geq 40$, the influence of the initial conditions on the specific viscosity was negligible~\citep{Takeishi2019}. In this study, therefore, the oscillations will be applied after this time, which is redefined as $\dot{\gamma}_0t = 0$ for simplicity. The dominant frequency for the input $\dot{\gamma} (t)$ and output $\mu_{sp} (t)$ are obtained with a discrete Fourier transform (DFT), using the {\it Fastest Fourier Transform in the West} ($FFTW$) library~\citep{Frigo2005}. The frequency peak of the output signal corresponds to the frequency of the applied shear for all cases considered here. The details of the DFT analysis adopted in this study are reported in the appendix~\S\ref{appA_DFT}. To reduce the influence of the initial conditions, the DFT analysis  and the spatial-temporal average denoted as $\langle \cdot \rangle$ start once the amplitude of $\mu_{sp}$ becomes basically constant; as an example, for $f_{in}$ = 0.1, the analyses start from the non-dimensional time $\dot{\gamma}_0 t$ = 800, and continues for 12 periods (see Fig.~\ref{fig:time_hct04ca005lam5}c). For all the cases, one output wave is resolved by at least 100 discrete times, and over 10 wave periods are considered in the DFT analysis. The effect of the number of periods on the calculated values are discussed in the appendix~\S\ref{appA_wave_num}. When presenting the results, we will initially focus on the analysis of suspensions in physiological conditions, with volume fraction $\phi$ = 0.41, $\lambda$ = 5, and $Ca_0$ = 0.05, and later consider variations of the viscosity ratio $\lambda$, volume fraction $\phi$ and capillary number $Ca_0$.

\section{Results}

\subsection{Dense suspension of RBCs under oscillatory flow}
\begin{figure*}
  \centering
  \includegraphics[height=6.0cm]{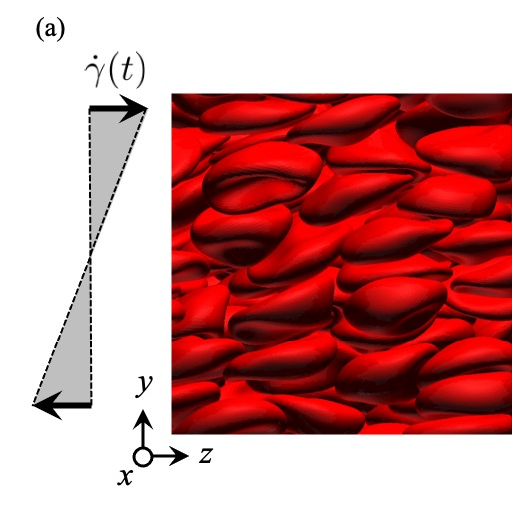}
  \includegraphics[height=6.0cm]{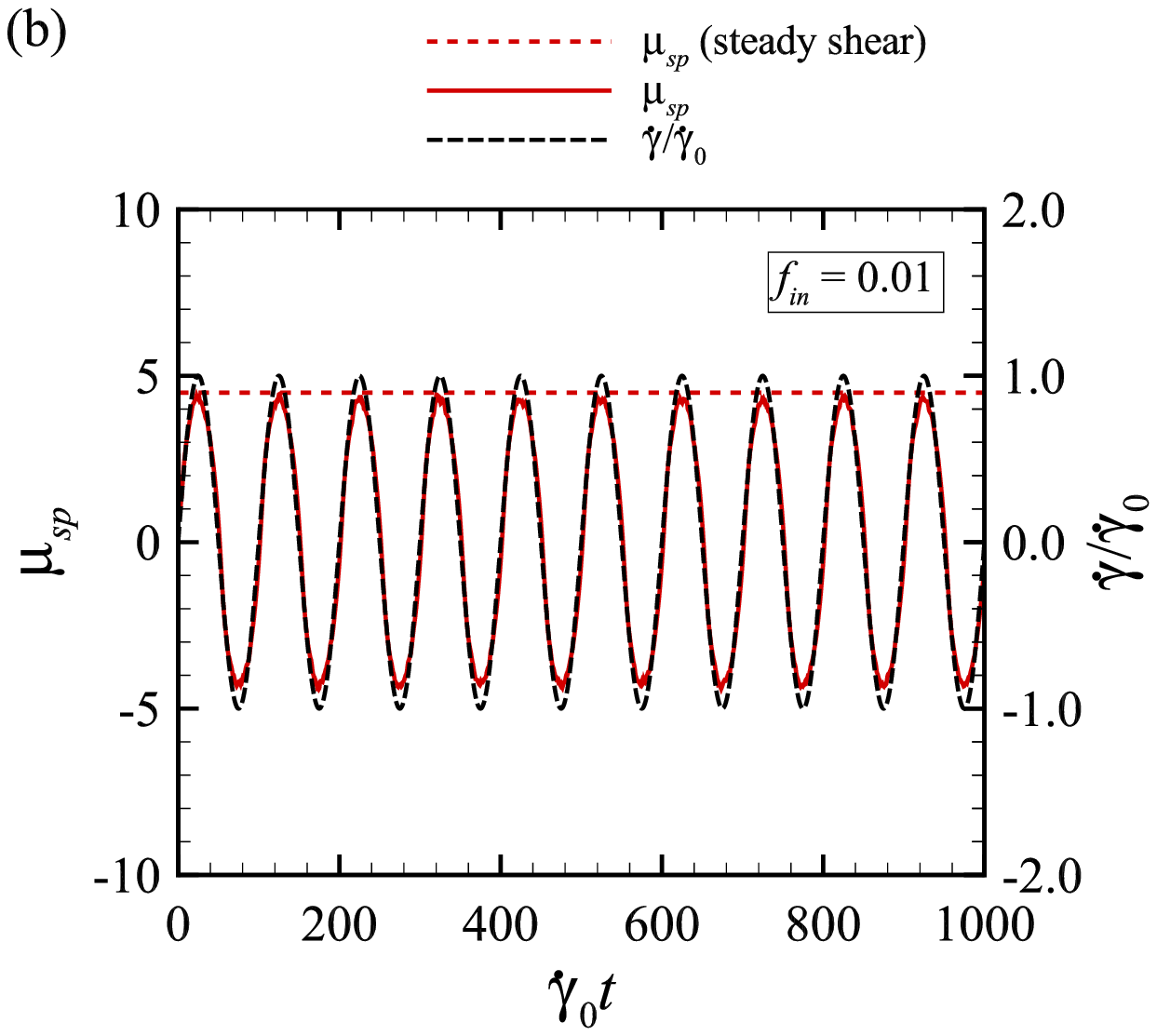}
  \includegraphics[height=5.5cm]{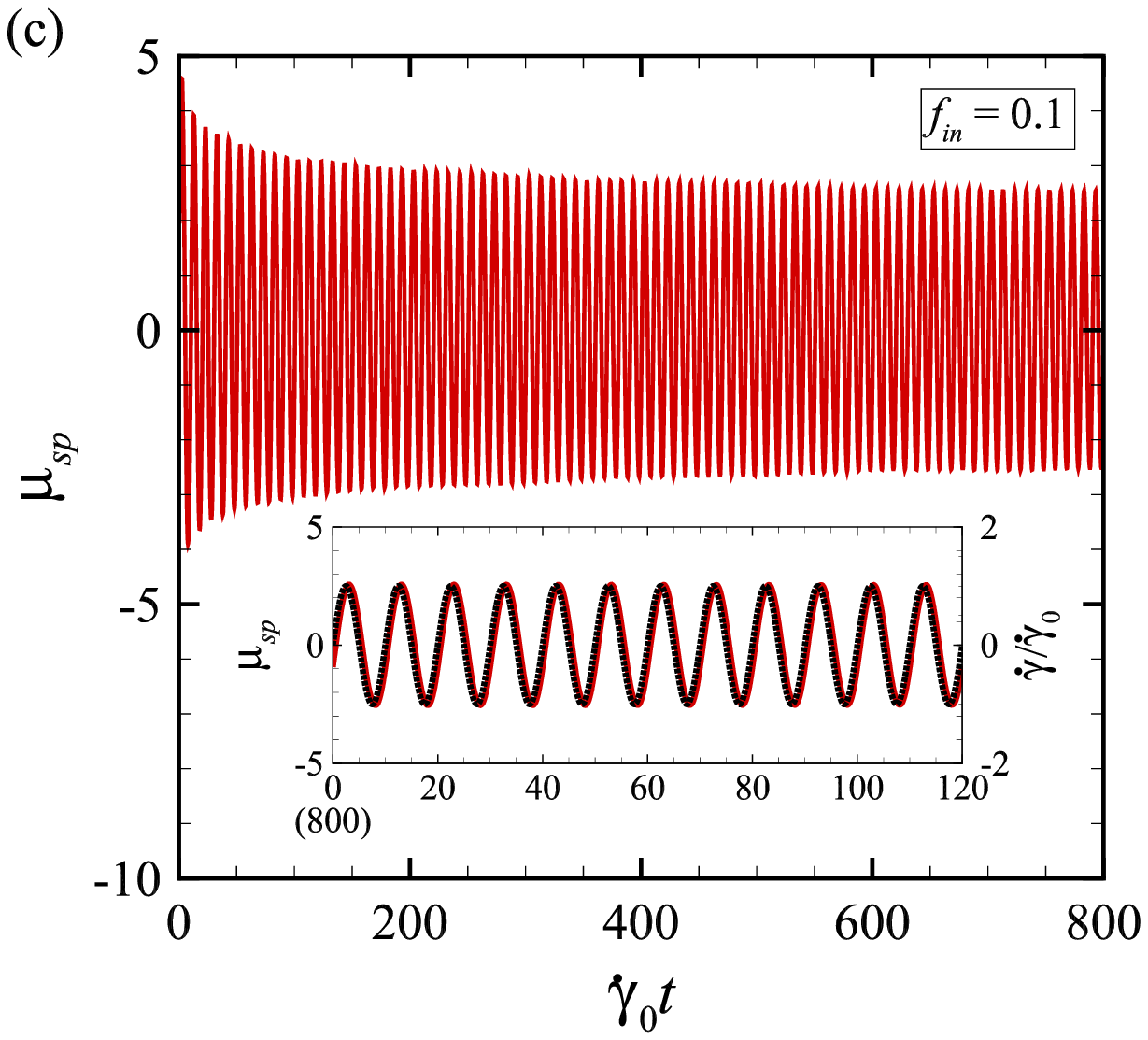}
  \includegraphics[height=5.5cm]{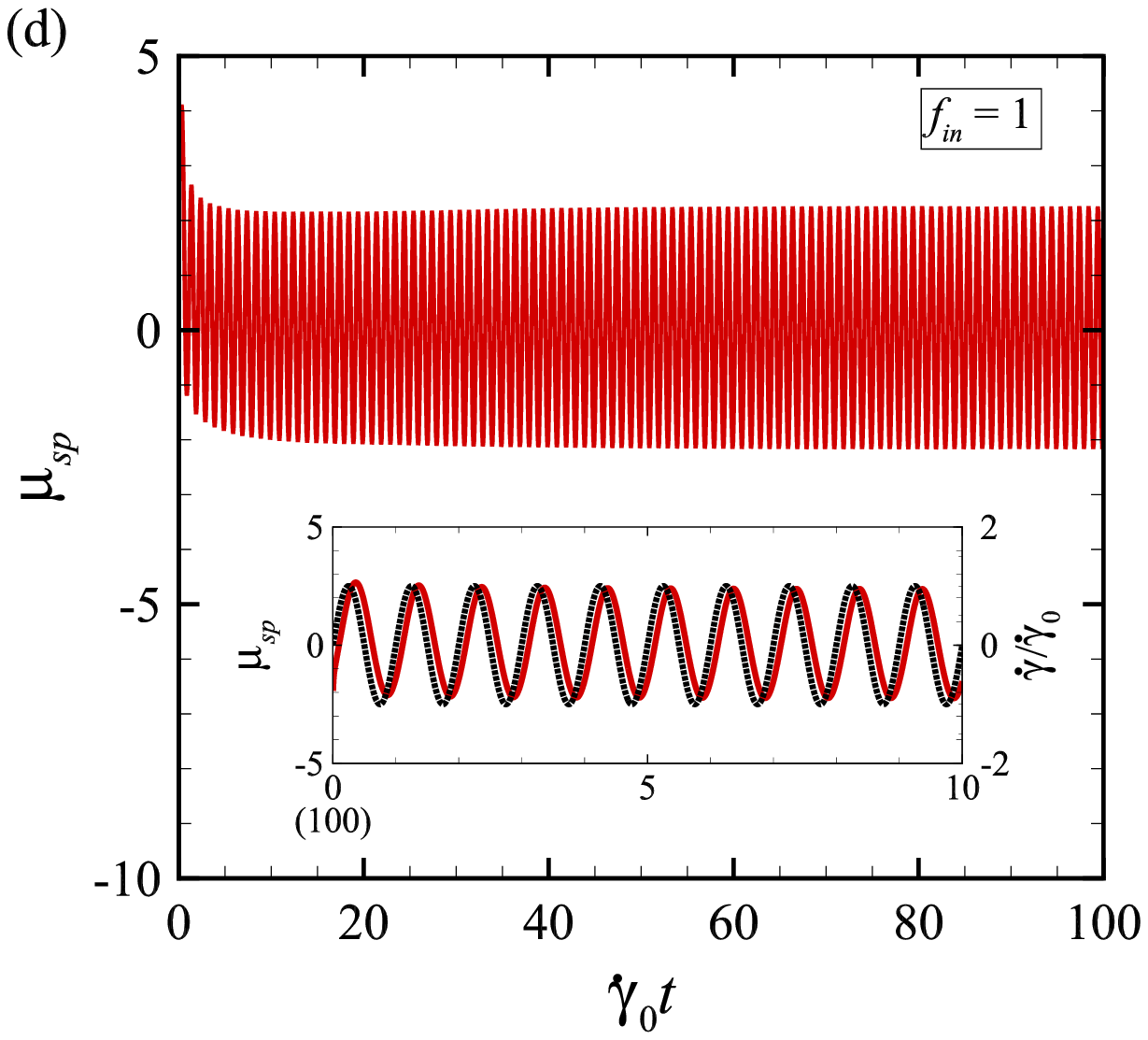}
  \caption{
	  (a) Snapshot of a dense suspension  of RBCs ($\phi$ = 0.41) zooming on the central part of the rectangular computational box for the lowest shear frequency $f_{in}$ = 0.01 under consideration.
	  (b) Time history of the specific viscosity $\mu_{sp}$ and input shear rate $\dot{\gamma}/\dot{\gamma}_0$ under $f_{in}$ = 0.01, where $\dot{\gamma}$ is normalized by the amplitude $\dot{\gamma}_0$.  The result of $\mu_{sp}$ obtained under steady shear flow is also displayed as red dashed line.
	  (c and d) Time history of $\mu_{sp}$ and $\dot{\gamma}$ for shear frequencies $f_{in}$ = (c) 0.1 and (d) 1.
	  The results are obtained with $\phi$ = 0.41, $\lambda$ = 5, and $Ca_0$ = 0.05.
  }
  \label{fig:time_hct04ca005lam5}
\end{figure*}

First, we investigate the oscillatory behavior of dense suspension of RBCs under SAOS flow at  $Ca_0$ = 0.05. A snapshot of the numerical results for the lowest input shear frequency investigated in this study $f_{in}$= 0.01 is shown in Fig.~\ref{fig:time_hct04ca005lam5}(a), while the time history of the specific viscosity $\mu_{sp} (t)$ and shear-rate $\dot{\gamma} (t)$ are displayed in Fig.~\ref{fig:time_hct04ca005lam5}(b), where the value of $\mu_{sp}$ obtained under steady shear flow is also displayed with dashed line for comparison.  As expected, the magnitude of $\mu_{sp}$ at the lowest shear frequency $f_{in}$ is similar to that obtained under steady shear flow, and the output wave, $\mu_{sp}$,  responds to the imposed shear, $\dot{\gamma}$, without significant delay. When the shear frequency $f_{in}$ increases, the resultant amplitude $|\mu_{sp}|^{amp}$ is reduced, as shown in Figs.~\ref{fig:time_hct04ca005lam5}(c) and \ref{fig:time_hct04ca005lam5}(d). Furthermore, there is a clear phase delay between the $\mu_{sp}$ and $\dot{\gamma}$ oscillations, with the phase difference increasing as $f_{in}$ increases. The rate of decrease in the amplitude of the specific viscosity $|\mu_{sp}|^{amp}$ shown in Fig.~\ref{fig:time_hct04ca005lam5}(c) is only 0.00781 over a non-dimensional time of $\dot{\gamma}_0 t$ = 100, starting from $\dot{\gamma}_0 t$ = 800. Thus, the results that we discuss later, e.g., in Fig.~\ref{fig:hct04ca005lam5} can be considered to be fully converged.

\begin{figure}
  \centering
  \includegraphics[height=5.5cm]{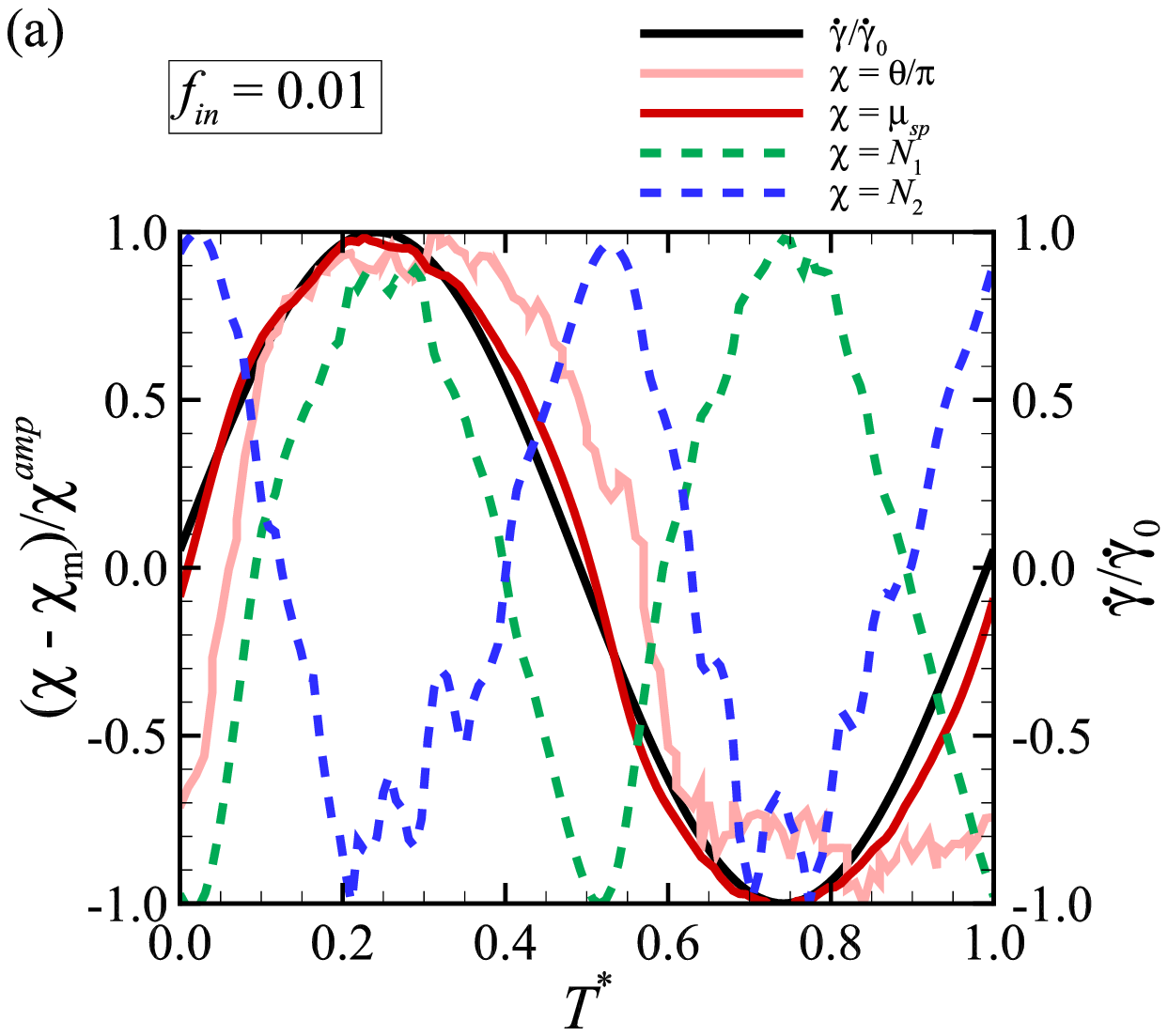} 
  \includegraphics[height=5.5cm]{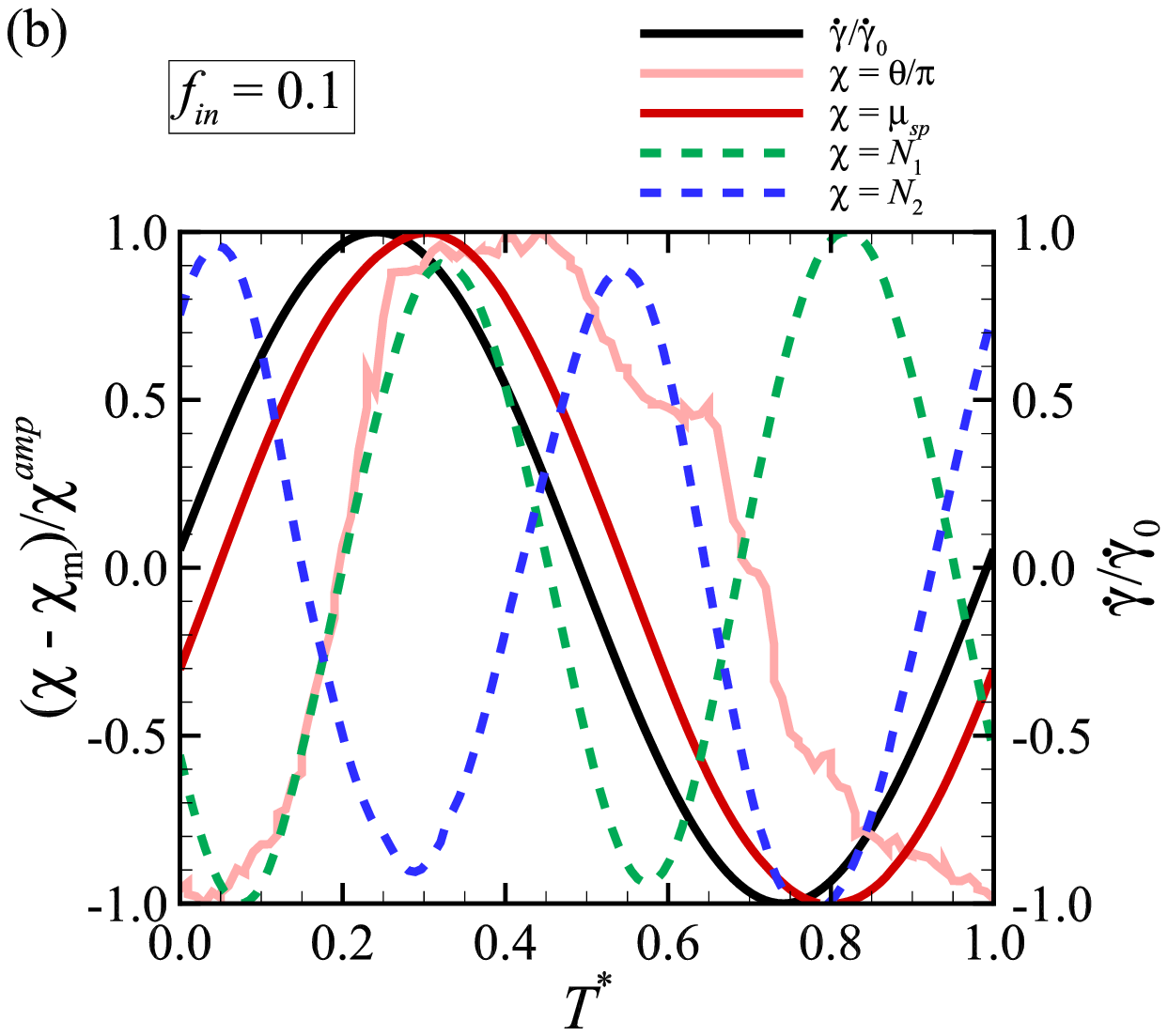}
  \includegraphics[height=5.5cm]{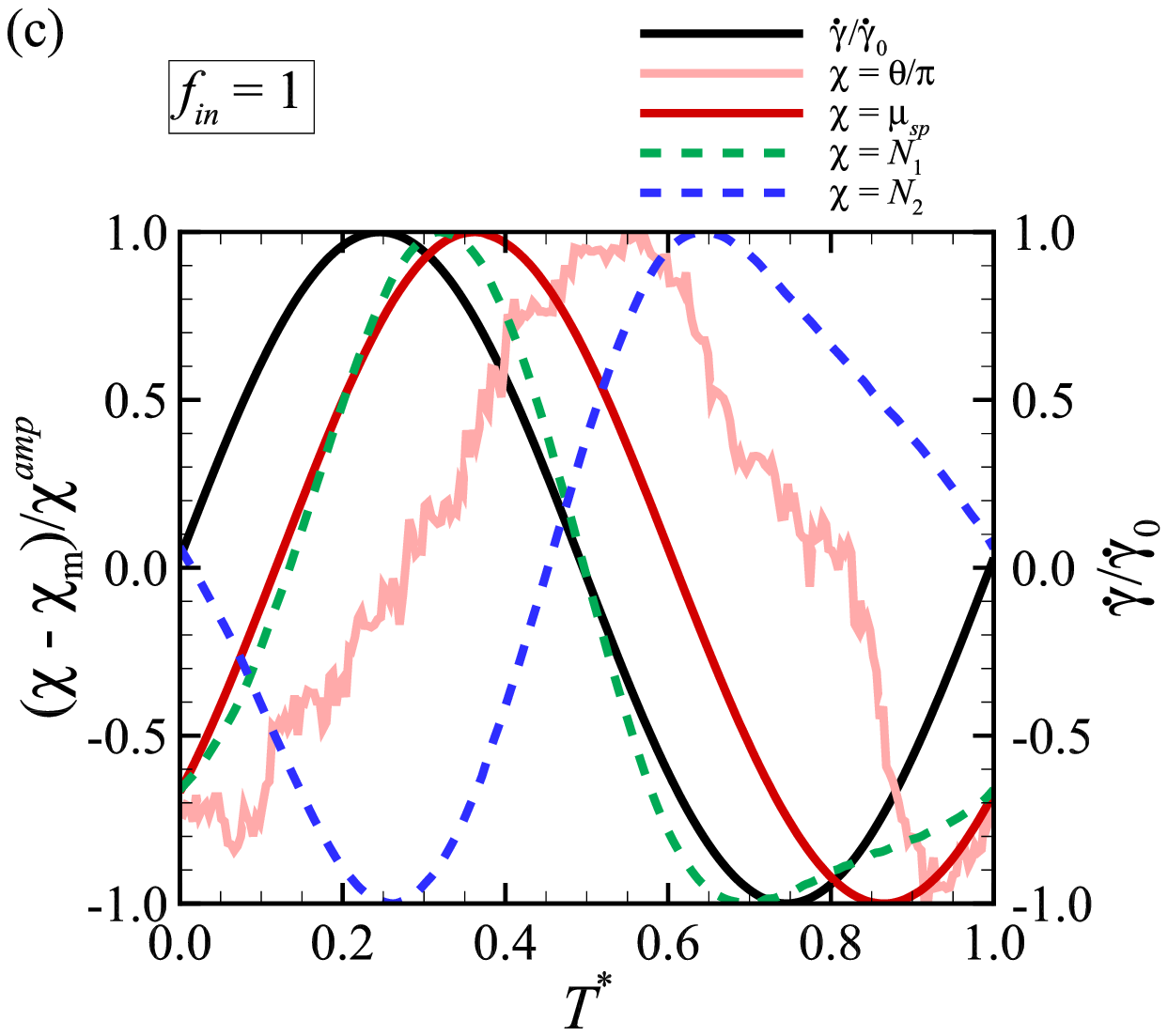}
  \caption{
	  Time history of the orientation angle $\theta$ and rheological parameters [the specific viscosity $\mu_{sp}$ and the normal stress differences $N_i$ ($i$ = 1, 2)], during a specific period $T^\ast$ at statistical steady state after the initial  transient, for different frequencies $f_{in}$ = ($a$) 0.01, ($b$) 0.1 and ($c$) 1, where the period is normalized by the inverse frequency $1/f_{in}$. The values are averaged between individual RBCs, normalized by each amplitude $\chi_{amp}$ and shifted so that each baseline is the mean value $\chi_m$.
	  The results are obtained with $\phi$ = 0.41, $\lambda$ = 5, and $Ca_0$ = 0.05.
  }
  \label{fig:phase_hct04ca005lam5}
\end{figure}

\begin{figure*}
  \centering
  \includegraphics[height=5.5cm]{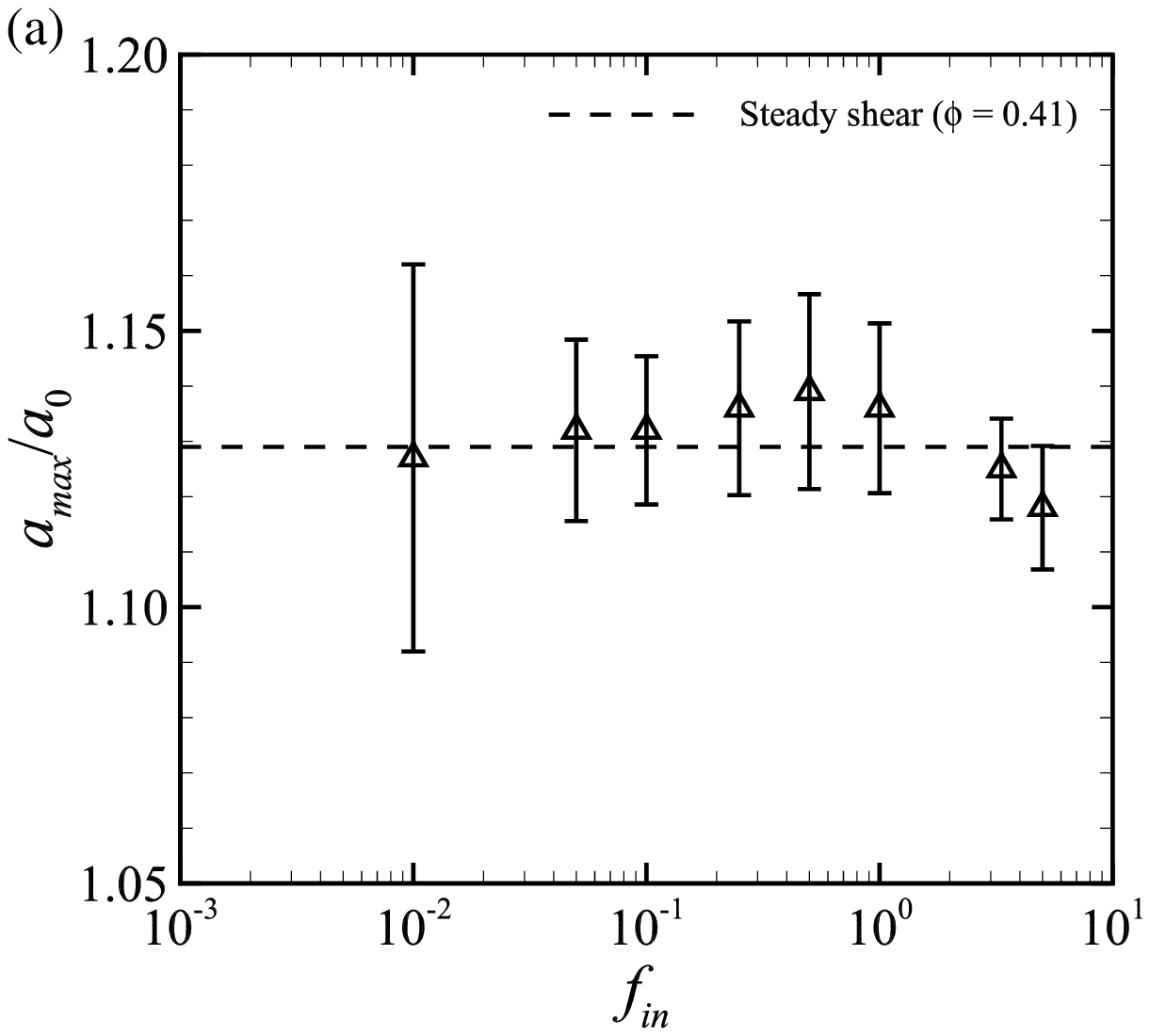}
  \includegraphics[height=5.5cm]{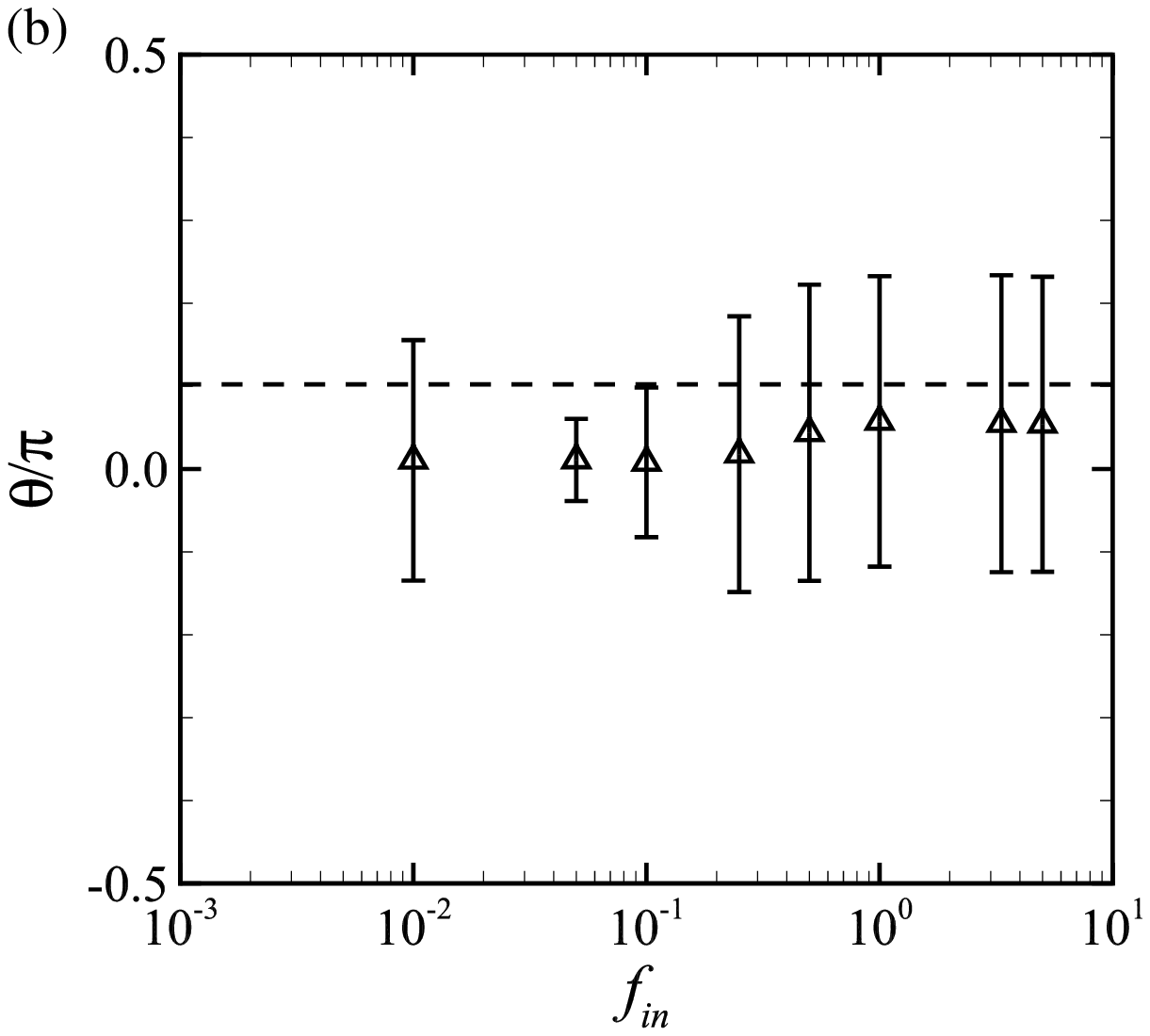}
  \includegraphics[height=5.5cm]{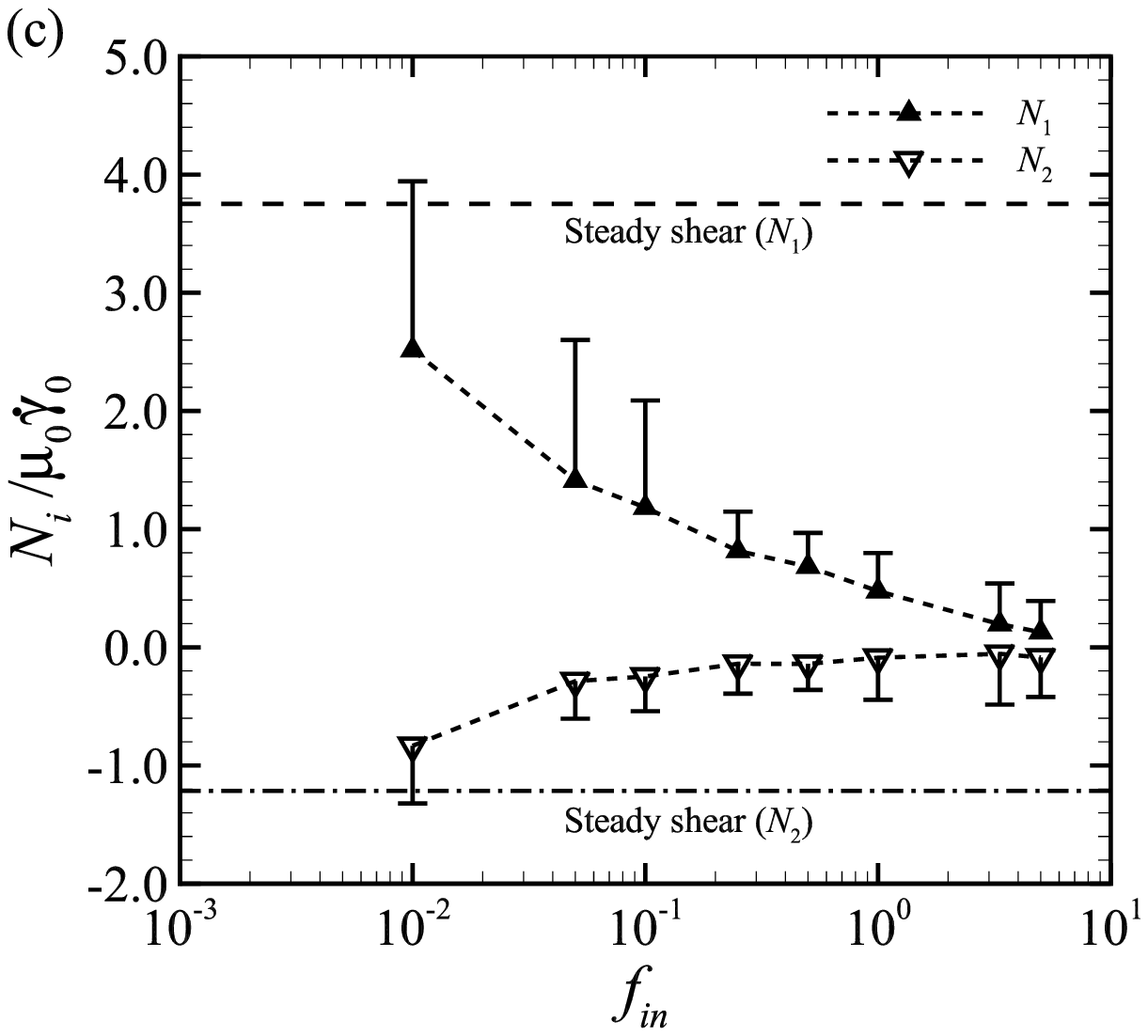}
  \includegraphics[height=5.5cm]{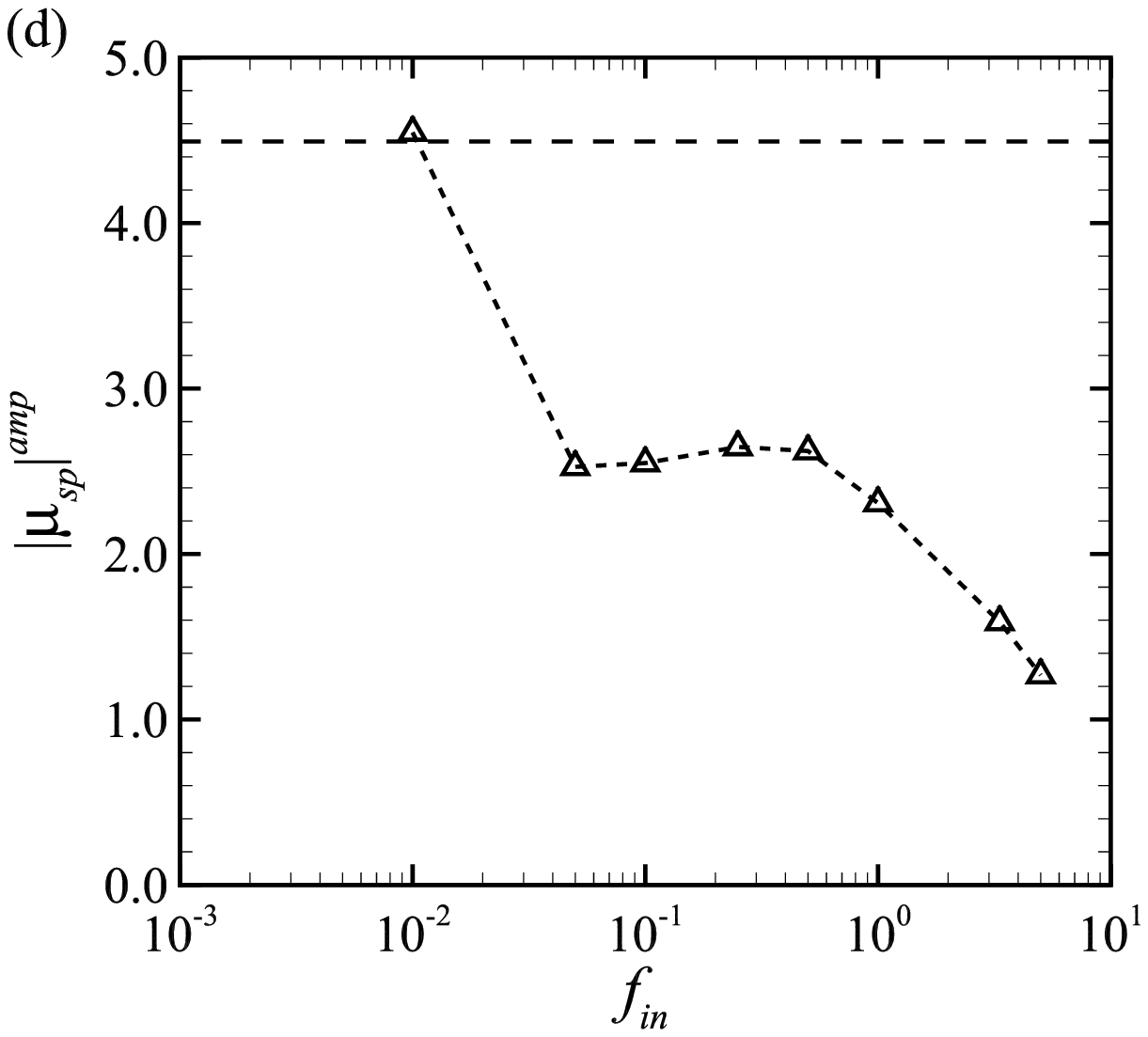}
  \includegraphics[height=5.5cm]{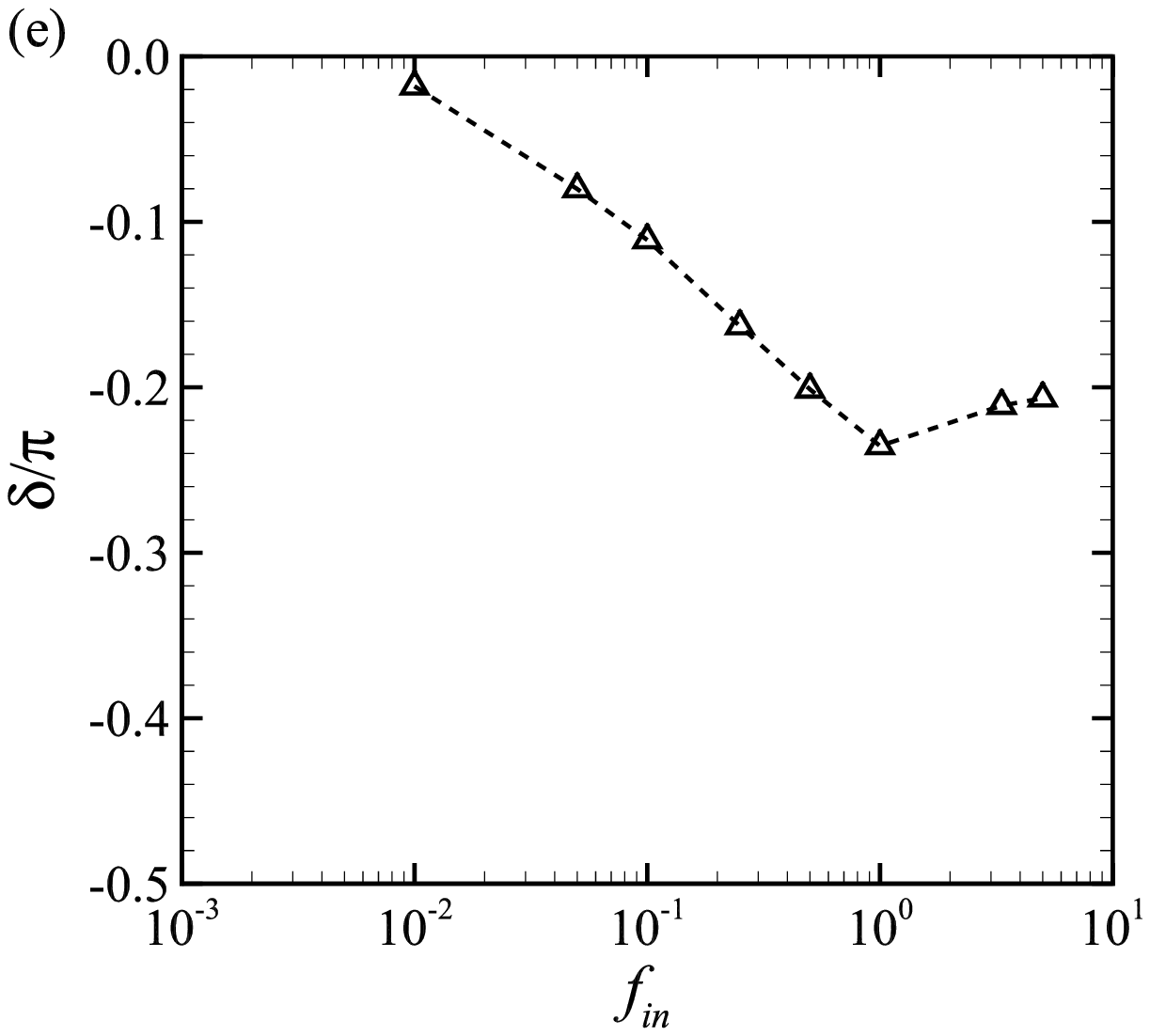}
  \includegraphics[height=5.5cm]{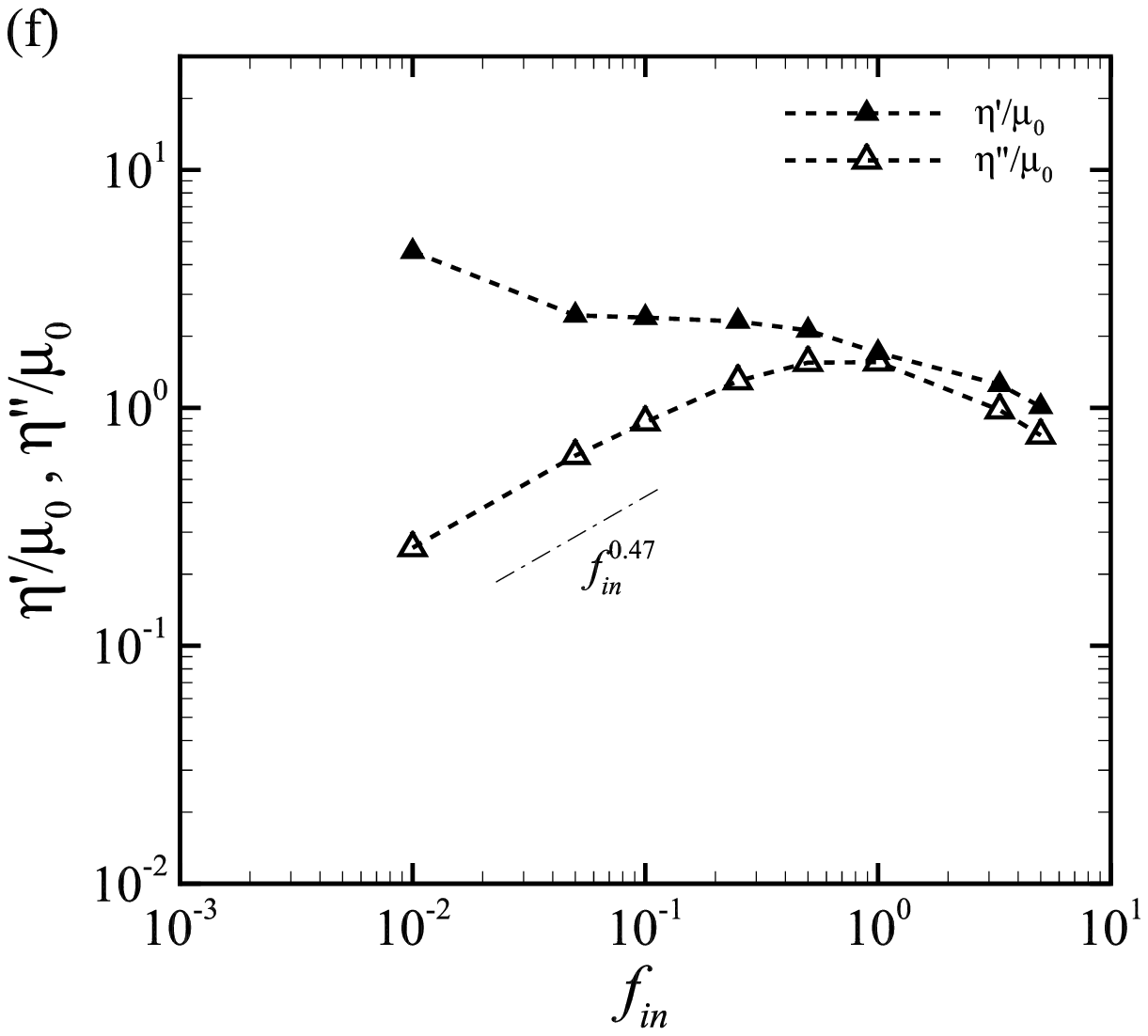}
  \caption{
	  (a) Spatial-temporal average of the deformation index $\langle a_{max} \rangle/a_0$,
	  (b) orientation angle $\langle \theta \rangle/\pi$ between the major axis of the deformed RBC and the shear direction,
	  (c) the first and second normal stress differences $\langle N_i \rangle/\mu_0 \dot{\gamma}_0$ ($i$ = 1 and 2),
	  (d) amplitude of the specific viscosity $\mu_{sp}$,
	  (e) phase difference $\delta$, and
	  (f) complex viscosity $\eta^{\prime}/\mu_0$ and $\eta^{\prime\prime}/\mu_0$ as a function of the input frequency $f_{in}$.
	  The results obtained under steady shear flow are also displayed in (a)--(d) as dashed (or dash-dot) lines.
	  The error bars in (a)--(c) represent standard deviations.
	  The error bars in panel (c) are displayed only on one side of the mean value for major clarity.
	  The results are obtained with $\phi$ = 0.41, $\lambda$ = 5 and $Ca_0$ = 0.05. 
  }
  \label{fig:hct04ca005lam5}
\end{figure*}

We display the time history of the orientation angle $\theta$, of the specific viscosity $\mu_{sp}$ and of the normal stress difference $N_i$ (with $i$ = 1, 2) in Fig.~\ref{fig:phase_hct04ca005lam5}; the instantaneous values are averaged over the different RBCs, normalized by their amplitude $\chi_{amp}$, and shifted by their mean value $\chi_m$. The period is also normalized by the inverse frequency $1/f_{in}$ of each case. The plots show that $\theta$ is slightly delayed from the imposed shear (Fig.~\ref{fig:phase_hct04ca005lam5}a), and this delay becomes greater for larger frequencies. Thus, the time signal of the oscillatory rheology $\mu_{sp}$ adapts to the change in the imposed shear faster than the RBCs orientation $\theta$. Despite this delay, $\theta$ has the same period as $\mu_{sp}$, independently of the applied frequency. The normal stress differences $N_i$, on the other hand, exhibit a period which is almost half that of the imposed shear ($\mu_{sp}$ and $\theta$), and it increases with the oscillation frequency. The first normal stress difference $N_1$ tends to synchronise with the imposed shear ($\mu_{sp}$ and $\theta$), while the second normal stress difference $N_2$ is basically in opposite phase. These results remain the same in dilute conditions ($\phi$ = 0.11).

\begin{figure}
  \centering
  \includegraphics[height=5.5cm]{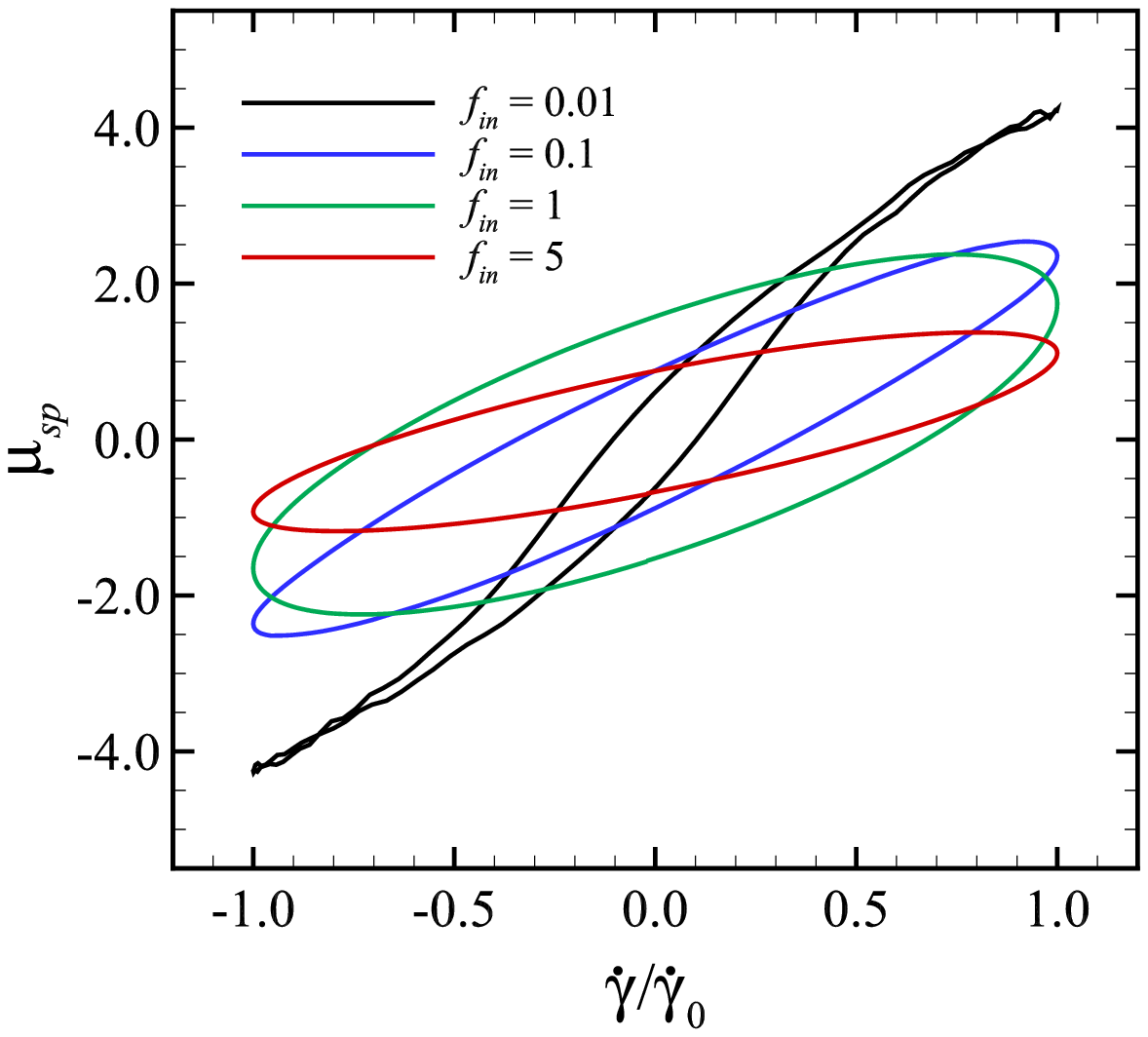}
  \caption{
	  Lissajous-Bowditch plots of $\mu_{sp}$ versus $\dot{\gamma}/\dot{\gamma}_0$ for different $f_{in}$.
	  The results are obtained with $\phi$ = 0.41, $\lambda$ = 5 and $Ca_0$ = 0.05.
  }
  \label{fig:Lissajous_hct04}
\end{figure}

The frequency-dependent deformation and orientation of individual RBCs are investigated to understand their link to the flow rheology. The deformation is characterised by the maximal radius $a_{max}$, with the ratio $a_{max}/a_0$ chosen as a deformation index. The spatial-temporal averages of these two observables are shown in Figs.~\ref{fig:hct04ca005lam5}(a) and \ref{fig:hct04ca005lam5}(b) versus the oscillation frequency $f_{in}$.  The value of $\langle a_{max} \rangle/a_0$ is slightly non-monotonic, first growing till $f_{in}\approx 1$ and then decreasing. Especially for the middle range of $f_{in}$ (0.05 $\leq f_{in} \leq$ 1),  $\langle a_{max} \rangle/a_0$ is greater than in steady shear flow. However, the frequency-induced mean deformation differs by only 0.8\% from that in steady shear flow even for the maximum $\langle a_{max} \rangle/a_0$ at $f_{in}$ = 0.5 (Fig.~\ref{fig:hct04ca005lam5}a). The amplitude of the fluctuations around the mean keeps instead reducing as $f_{in}$ grows.  The orientation angles $\theta/\pi$ fluctuate around zero for all $f_{in}$, and thus, the values are lower than those obtained under steady shear flow, with no significant frequency dependence, as shown in Fig.~\ref{fig:hct04ca005lam5}(b).

The first and second normal stress differences $\langle N_i \rangle/\mu_0\dot{\gamma_0}$, the resultant amplitude $|\mu_{sp}|^{amp}$ and phase difference $\delta$ are shown in Figs.~\ref{fig:hct04ca005lam5}(c), \ref{fig:hct04ca005lam5}(d) and \ref{fig:hct04ca005lam5}(e) for the different frequencies under consideration. The magnitude of $|\langle N_i \rangle|/\mu_0\dot{\gamma_0}$ decreases monotonically from that in the steady shear flow as $f_{in}$ increases, reaching nearly zero for the highest value of $f_{in}$. The first normal stress difference $\langle N_1 \rangle$ is positive, while $\langle N_2 \rangle$ is negative (Fig.~\ref{fig:hct04ca005lam5}c). It is known that the ratio $N_2/N_1 = -0.5$ can be typically observed in liquid crystals consisting of rod-like polymer solutions~\citep{Maklad2021}. Although our numerical results do no follow this, the order of magnitude of the ratio is $O(-\langle N_2 \rangle/\langle N_1 \rangle) \approx 10^{-1}$, and almost independent of $\phi$. The error bars for $\langle N_i \rangle/\mu_0\dot{\gamma_0}$ are displayed only on one side of the mean value for major clarity. The specific viscosity $|\mu_{sp}|^{amp}$ also decreases from the value in steady shear flow as $f_{in}$ increases above the lowest value (Fig.~\ref{fig:hct04ca005lam5}d). The specific viscosity $|\mu_{sp}|^{amp}$ only weakly changes in the intermediate range of $f_{in}$ (0.05 $\leq f_{in} \leq$ 0.5), before decreasing again at higher $f_{in}$. The phase delay $\delta$ is also found to be a function of the shear frequency, as documented in Fig.~\ref{fig:hct04ca005lam5}(e), where we observe that $\delta$ decreases from almost zero at the lowest $f_{in}$ (= 0.01) until $f_{in}$ = 1, and then increases for $f_{in} >$ 1.

The frequency-dependent $|\mu_{sp}|^{amp}$ and $\delta$ define the complex viscosity, with real and imaginary part $\eta^\prime$ and $\eta^{\prime\prime}$, which are reported in Fig.~\ref{fig:hct04ca005lam5}(f). The variations of $\eta^\prime$ with $f_{in}$ well follow $|\mu_{sp}|^{amp}$, whereas $\eta^{\prime\prime}$ first increases,  reaches a maximum at $f_{in} \approx $  1, and then decreases again for larger frequencies. While $\eta^{\prime\prime}$ almost coincides with $\eta^\prime$ at large $f_{in}$,  the increase of $\eta^{\prime\prime}$ at low $f_{in}$ can be well approximated by a power law: $\eta^{\prime\prime} \propto f_{in}^{0.47}$. This result indicate that dense suspensions of RBCs cannot be fully modelled as Maxwell fluids, where the complex shear moduli $\eta^{\prime\prime} (= G^\prime/(2\pi \mathrm{f})) \approx \mathrm{f}$ and $\eta^\prime (= G^{\prime\prime}/(2\pi\mathrm{f})) \approx const$ for low $\mathrm{f}$; on the other hand, the results of $\eta^{\prime\prime}$ are qualitatively similar to those obtained with the Oldroyd-B fluids~\citep{Bird1987, Maklad2021}. However, since an Oldroyd-B fluid is characterised by zero $N_2$~\citep{Bird1987, Maklad2021}, RBCs (or SK capsule) suspension differ from the aforementioned non-Newtonian models.

\begin{figure*}
  \centering
  \includegraphics[height=5.5cm]{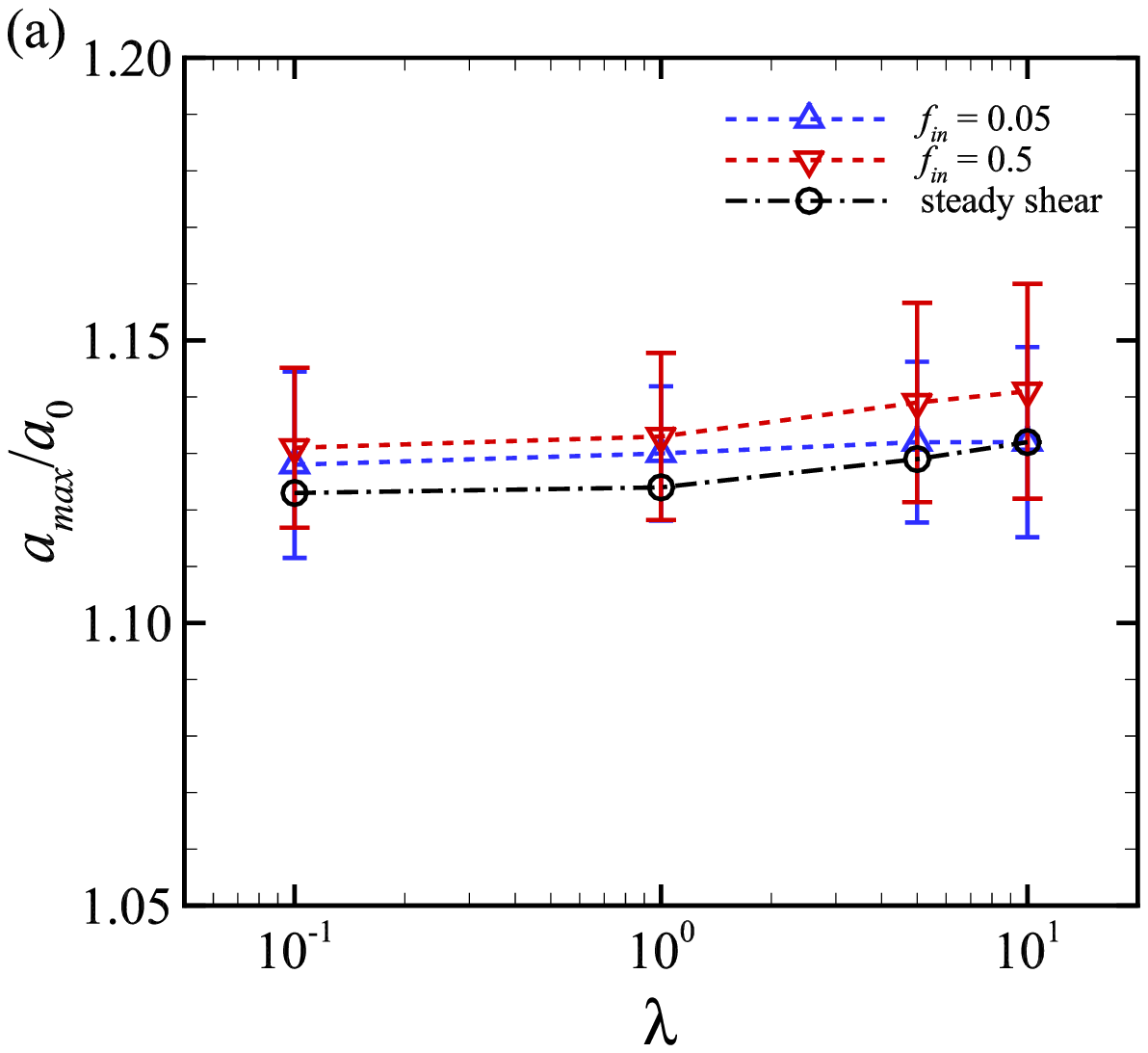}
  \includegraphics[height=5.5cm]{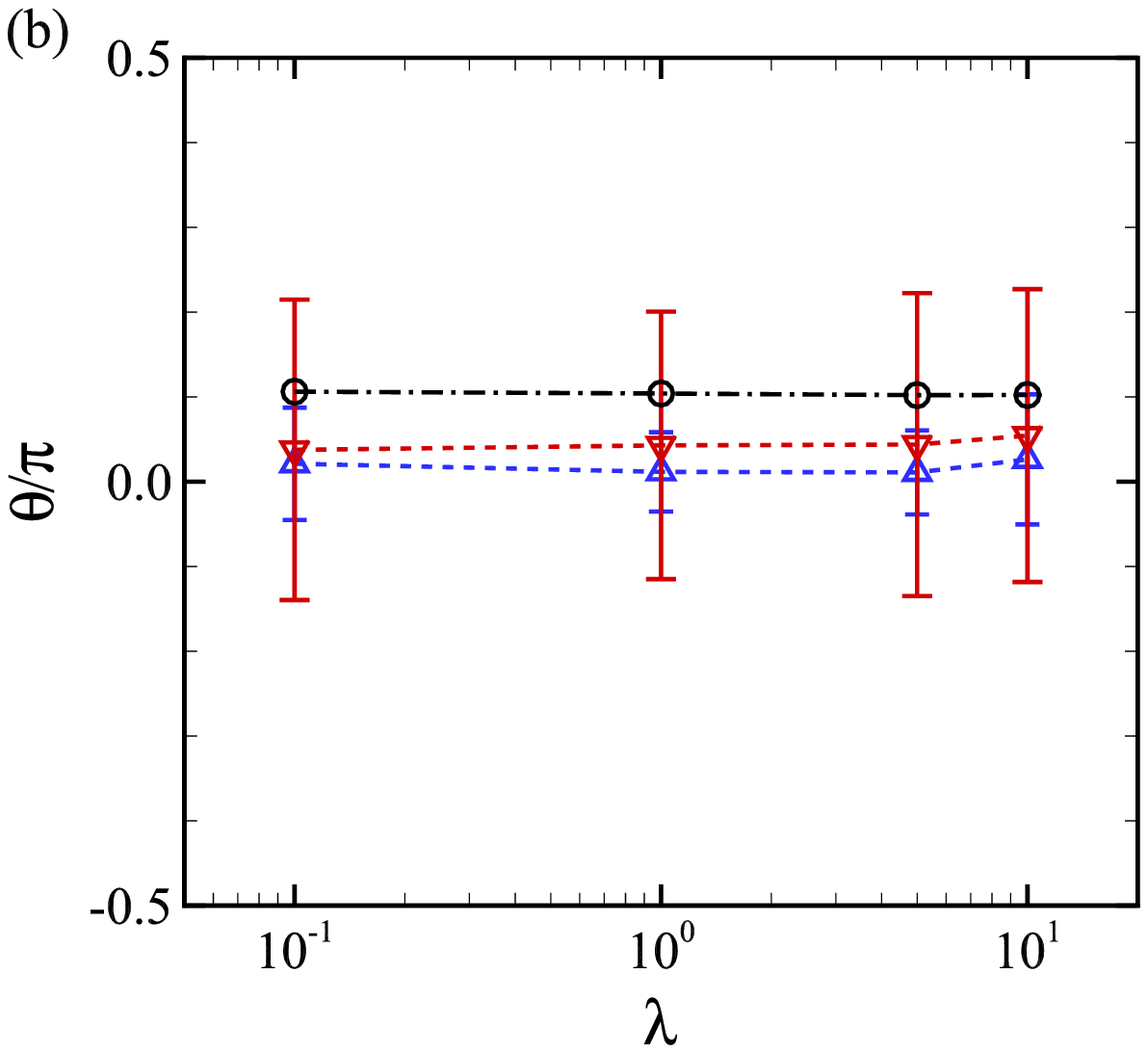}
  \includegraphics[height=5.5cm]{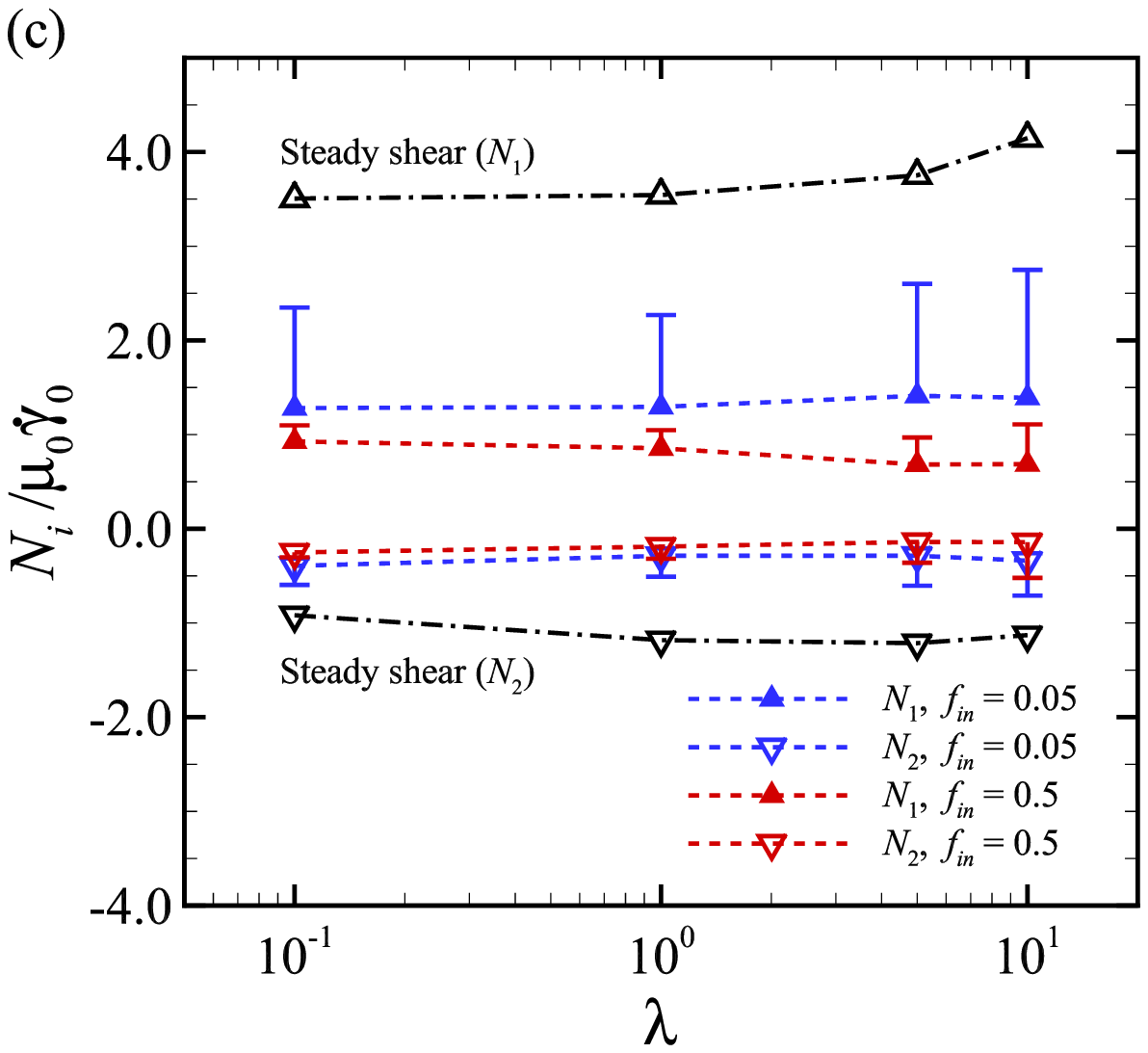}
  \includegraphics[height=5.5cm]{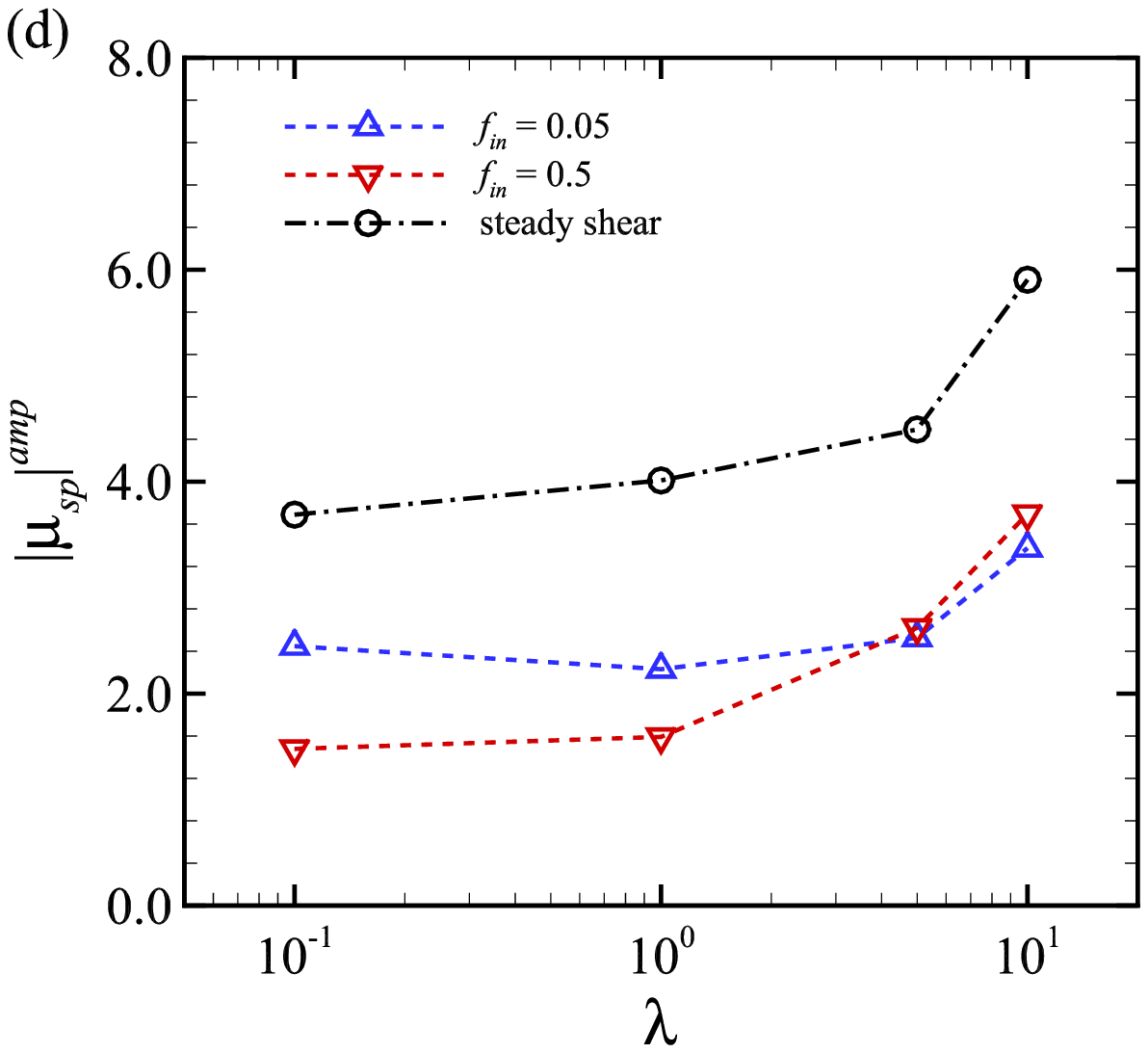}
  \includegraphics[height=5.5cm]{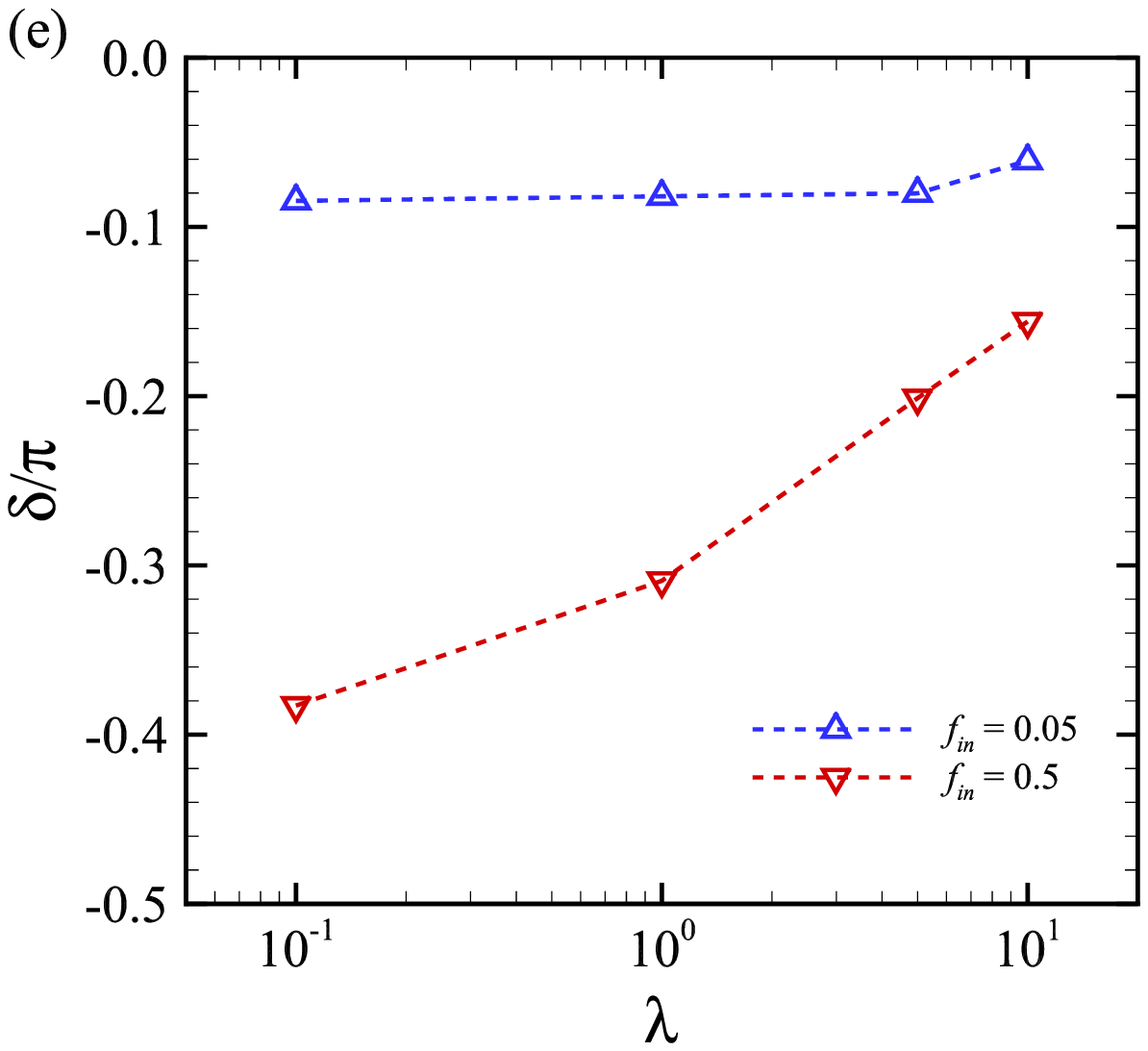}
  \includegraphics[height=5.5cm]{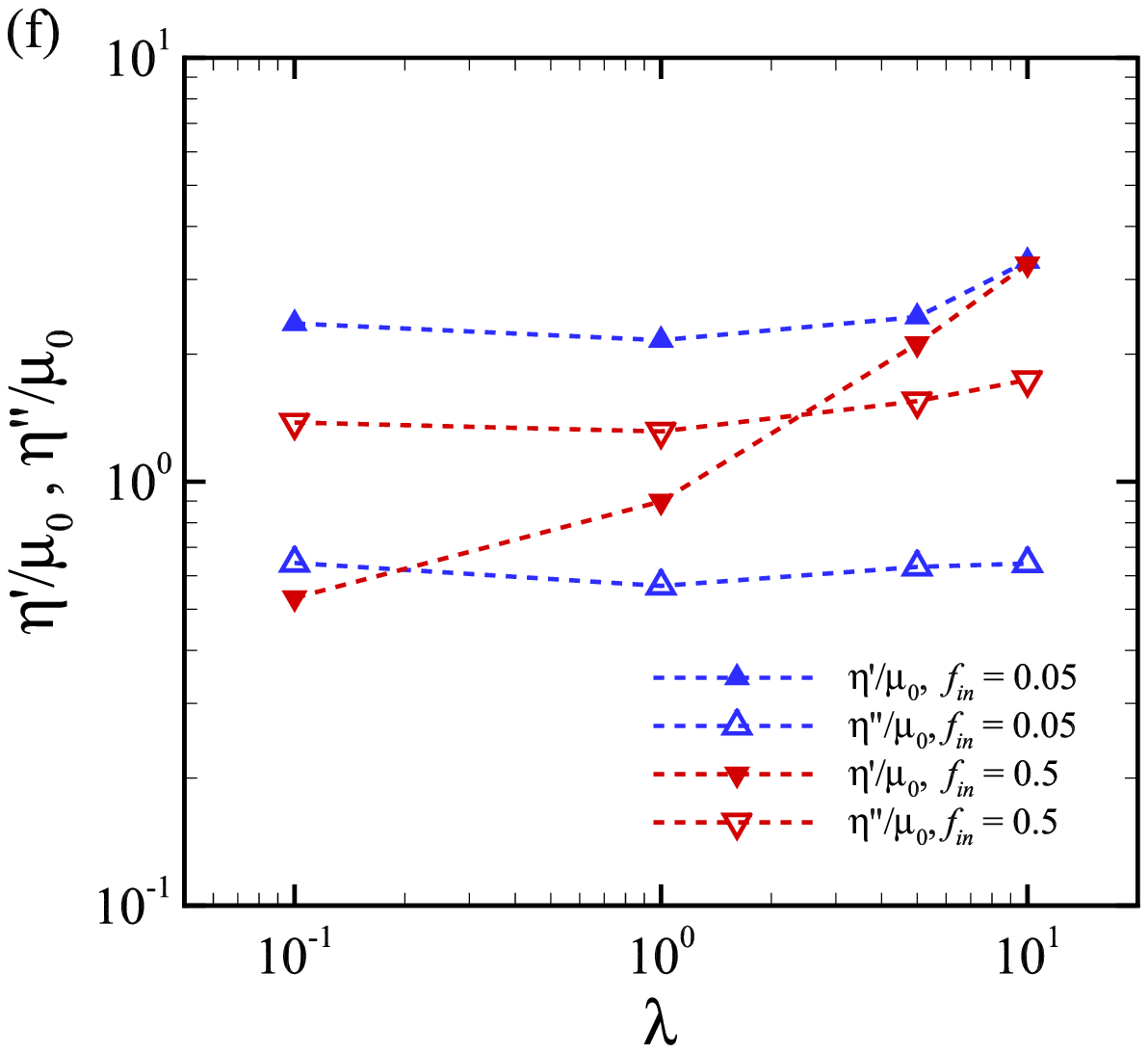}
  \caption{
	  (a) Spatial-temporal average of the deformation index $\langle a_{max} \rangle/a_0$,
	  (b) orientation angle $\langle \theta \rangle/\pi$,
         (c) the first and second normal stress differences $\langle N_i \rangle/\mu_0 \dot{\gamma}_0$ ($i$ = 1 and 2),
	  (d) amplitude $|\mu_{sp}|^{amp}$,
	  (e) phase difference $\delta/\pi$, and
	  (f) complex viscosity $\eta^{\prime}/\mu_0$ and $\eta^{\prime\prime}/\mu_0$ as a function of viscosity ratio $\lambda$ (= 0.1, 1, 5 and 10).
	  The results obtained under steady shear flow at each $\lambda$ are also displayed in (a)--(d) as black dash-dot lines.
	  The results are obtained with $\phi$ = 0.41 and $Ca_0$ = 0.05 for relatively low and high shear frequencies, $f_{in}$ = 0.05 and $f_{in}$ = 0.5. 
  }
  \label{fig:effect_lam}
\end{figure*}

At the microstructure level, the RBCs fully orient towards the shear directions at the low frequency ($f_{in}$ = 0.05) except for the lowest one ($f_{in}$ = 0.01), while their orientations vary for high frequencies ($f_{in} \geq$ 0.1) because individual RBCs are forced to reorient before reaching a stable condition. Consequently, the standard deviations of the orientation angle $\theta$ are greater at low frequency and smaller at high frequencies (Fig.~\ref{fig:hct04ca005lam5}b). The results are consistent with those described later in Figs.~\ref{fig:effect_lam}(b), \ref{fig:effect_ca_shape}(d) and \ref{fig:effect_ca_shape}(e). Although it is known that the RBC deformation alone is sufficient to give rise to shear-thinning~\citep{Omori2014}, the resulting complex viscosity shown in the Fig.~\ref{fig:hct04ca005lam5}(f) weakly depends on the frequency-modulated deformations or orientations of individual RBCs (Figs.~\ref{fig:hct04ca005lam5}a and \ref{fig:hct04ca005lam5}b, respectively), but rather depends on combinations of the frequency-dependent amplitude $|\mu_{sp}|^{amp}$ and phase difference $\delta$ (Figs.~\ref{fig:hct04ca005lam5}d and \ref{fig:hct04ca005lam5}e, respectively).

To conclude, we present the Lissajous-Bowditch plots~\citep{Mujumdar2002, Lee2003} of $\mu_{sp}$ versus $\dot{\gamma}/\dot{\gamma}_0$, following the classical way to analyse nonlinear oscillations. We recall that an elastic solid would correspond to a straight line, a Newtonian fluid to a perfect circle, and viscoelastic materials under SAOS to an ellipsoid. Nonlinear responses can generally lead to a more complex shape, which itself is a characterization of the nonlinear viscoelasticity of the suspension~\citep{Mujumdar2002},  as usually observed under LAOS. The diagrams for our RBC suspensions are shown in Fig.~\ref{fig:Lissajous_hct04}, where the Lissajous plots reveal deviation from the ellipsoidal to the sigmoidal shape as $f_{in}$ decreases.

\subsection{Effect of the viscosity ratio}
Next, we investigate the effect of the viscosity ratio $\lambda$ (= 0.1, 1, 5 and 10) on the system viscoelasticity for relatively low and high frequencies $f_{in}$ (= 0.05 and 0.5). The behavior of $\langle a_{max} \rangle/a_0$ and $\langle \theta \rangle/\pi$ as a function of $\lambda$ is reported in Figs.~\ref{fig:effect_lam}(a) and \ref{fig:effect_lam}(b). The mean deformation of individual RBCs progressively increases with $\lambda$, with only a slight increase at low $f_{in}$ (= 0.05), and a more significant one for $f_{in}$= 0.5,  when comparing with the case of steady shear flow (Fig.~\ref{fig:effect_lam}a). 

The average cell orientation $\langle \theta \rangle/\pi$ decreases under oscillatory shear flow, with no significant differences in the mean values for the different $\lambda$ considered (Fig.~\ref{fig:effect_lam}b).  When the shear frequency is small ($f_{in}$ = 0.05), the RBCs are able to attain a similar orientation (i.e., small standard deviations of $\langle \theta \rangle/\pi$), while when the shear frequency is large ($f_{in}$ = 0.5),  the ordered orientation is disrupted, as shown by the large value of standard deviations in Fig.~\ref{fig:effect_lam}(b). This is consistent with the observations from the data for $\lambda$ = 5 discussed above, cf.\ Fig.~\ref{fig:hct04ca005lam5}(b).

As also noted before from the data in Figs.~\ref{fig:hct04ca005lam5}(c) and \ref{fig:hct04ca005lam5}(d), $|\langle N_i \rangle|/\mu_0\dot{\gamma_0}$ and $|\mu_{sp}|^{amp}$ decrease under oscillatory shear flow, and this behavior is consistently observed also for the different values of $\lambda$ under consideration (Figs.~\ref{fig:effect_lam}c and \ref{fig:effect_lam}d). $\langle N_1 \rangle$ only slightly increases with $\lambda$ at low $f_{in}$, while it exhibits an opposite tendency at high $f_{in}$; furthermore, there are no significant variations in $\langle N_2 \rangle$ between high and low $f_{in}$. In particular, $|\mu_{sp}|^{amp}$ increases with $\lambda$, similarly to what observed for steady shear flow. For low viscosity ratios, $|\mu_{sp}|^{amp}$ progressively reduces with $f_{in}$, while for large viscosity ratios, it does not vary significantly with the frequency of the imposed shear $f_{in}$. The phase difference $\langle \delta \rangle/\pi$ instead does not vary significantly with the viscosity ratio for small values of $f_{in}$, while it does change significantly (i.e., the magnitude of $\delta$ decreases as $\lambda$ increases) for the largest value of $f_{in}$ (see Fig.~\ref{fig:effect_lam}e). In other words, large values of the viscosity ratios help to reduce the time-lag between the input and output signal  at high shear frequency as compared to steady shear.

The complex viscosity is displayed in Fig.~\ref{fig:effect_lam}(f) as a function of $\lambda$ for the two values of $f_{in}$ under investigation. For all the values of  the viscosity ratio, $\eta^\prime$ well follows $|\mu_{sp}|^{amp}$, while $\eta^{\prime\prime}$ is only weakly dependent on $\lambda$. At the large $f_{in}$ (= 0.5), $\eta^\prime$ overcomes $\eta^{\prime\prime}$ between $\lambda$ = 1 and 5, while such crossover is not found at low $f_{in}$ (= 0.05). The viscous component $\eta^\prime$ is higher at the low $f_{in}$ than at the large one, while $\eta^{\prime\prime}$ follows an opposite trend, i.e., its value is higher at the large $f_{in}$ than at the low frequencies.

Overall, the rheological behavior of the RBC suspension in oscillating shear flows appears to be only weakly dependent of the viscosity ratio, except for the lower phase lag observed at high $\lambda$.

\begin{figure*}
  \centering
  \includegraphics[height=5.5cm]{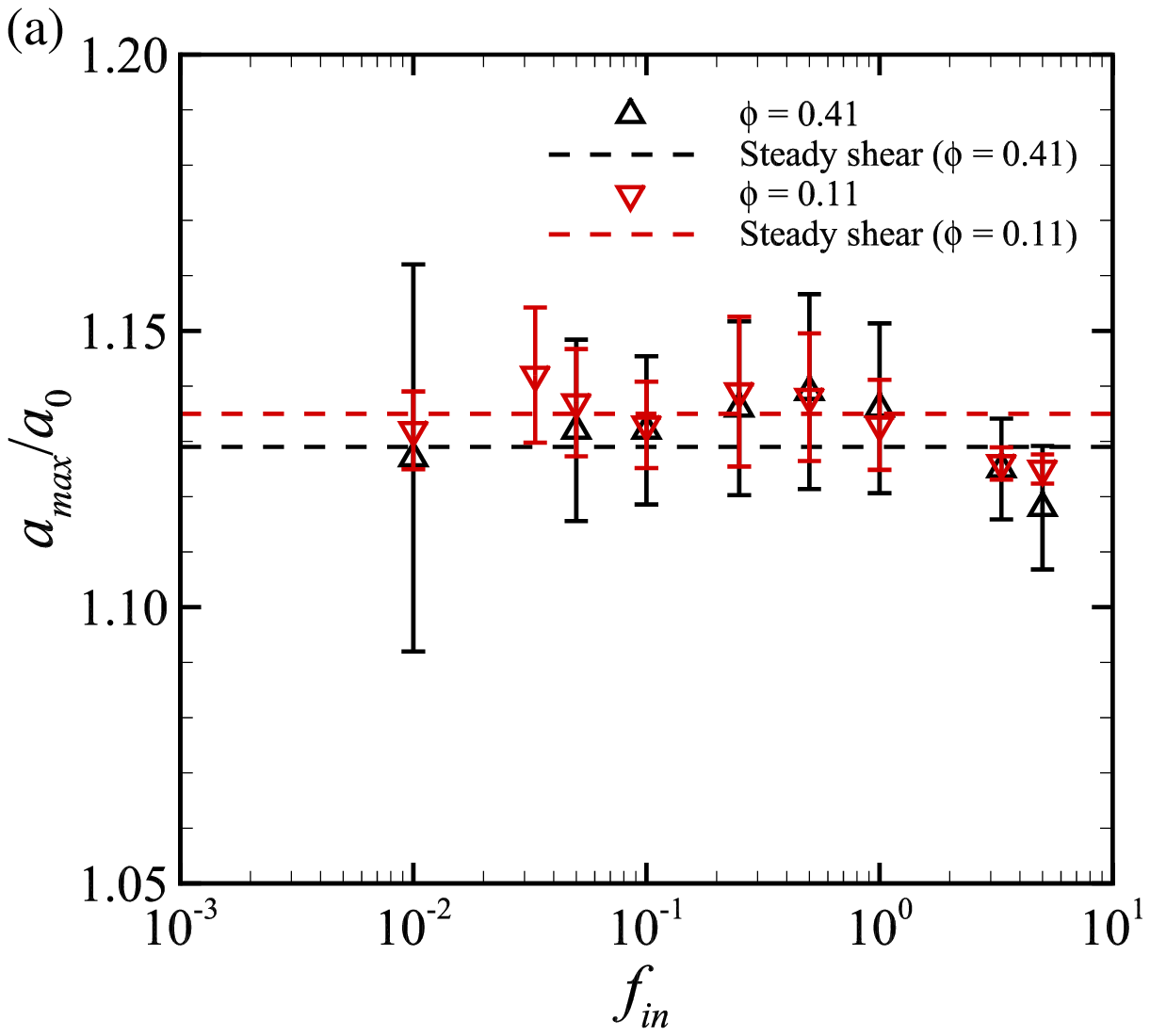}
  \includegraphics[height=5.5cm]{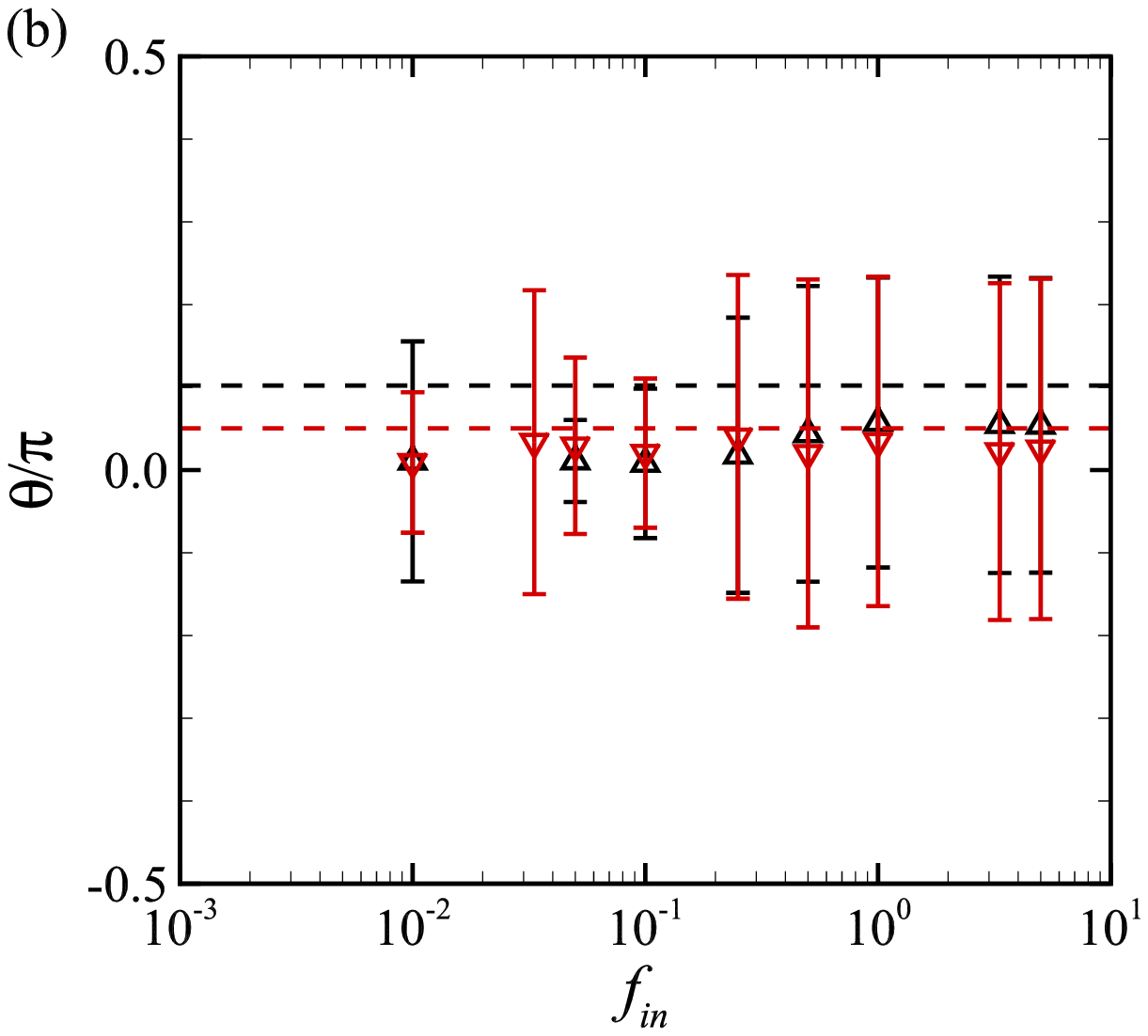}
  \includegraphics[height=5.5cm]{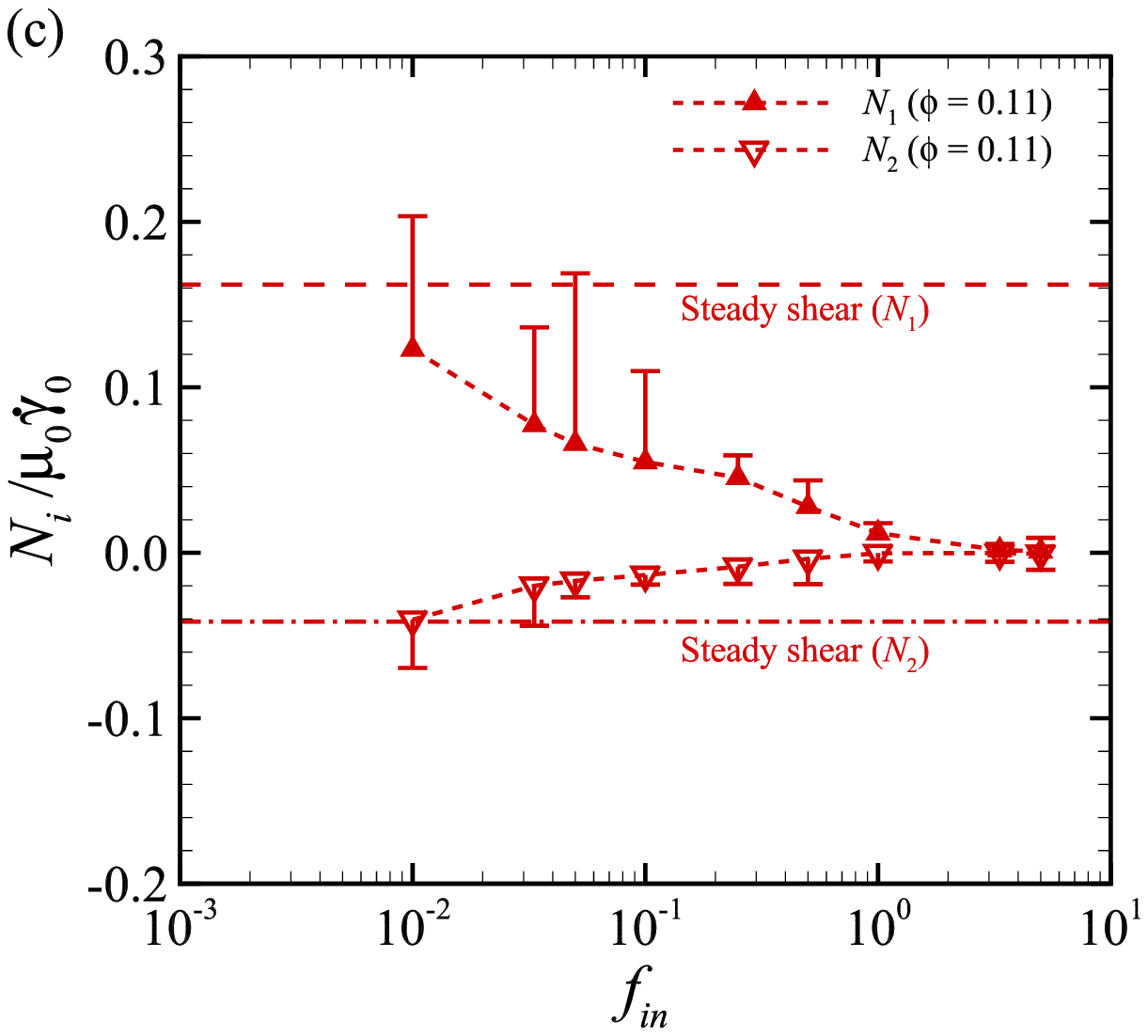}
  \includegraphics[height=5.5cm]{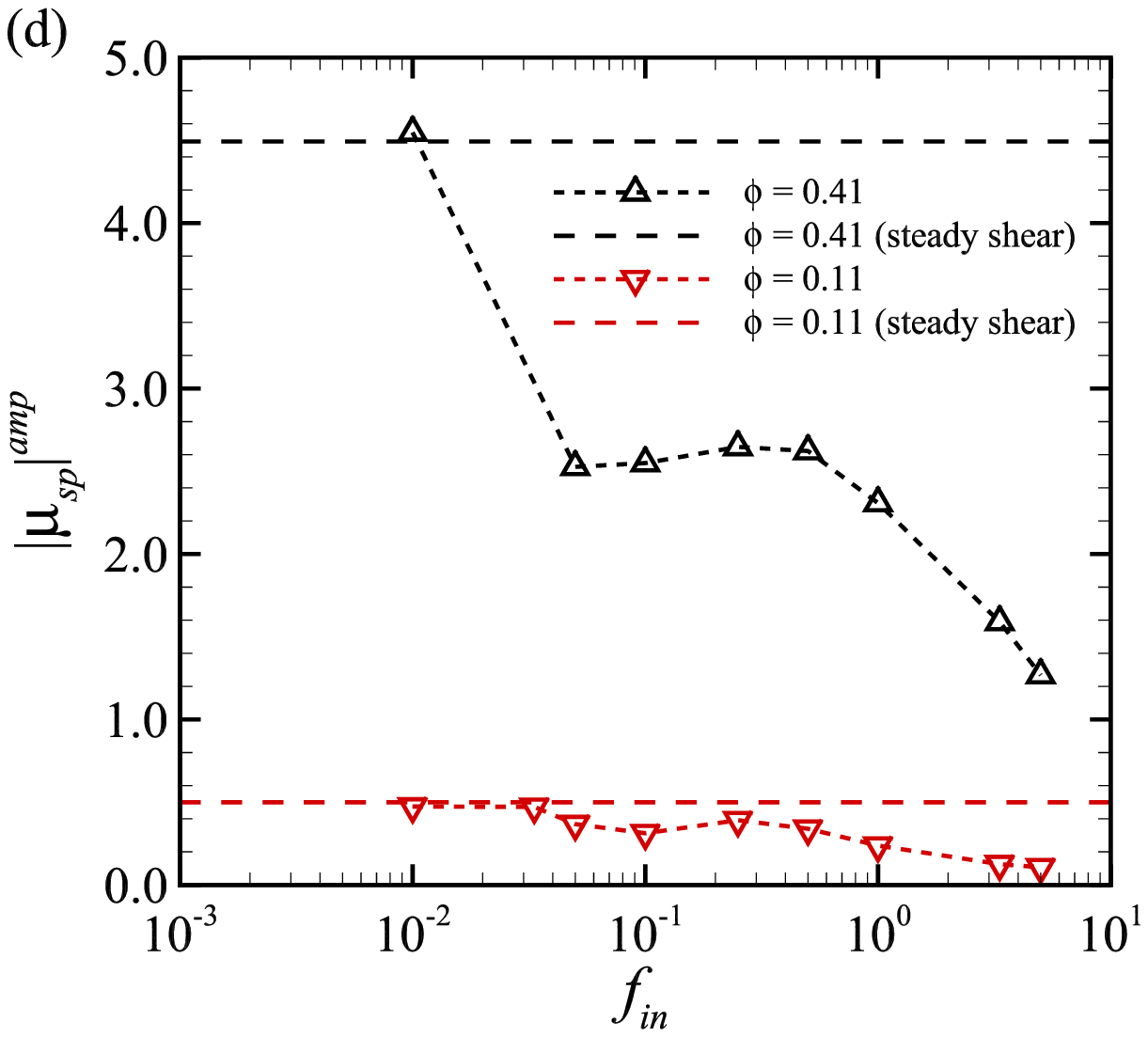}
  \includegraphics[height=5.5cm]{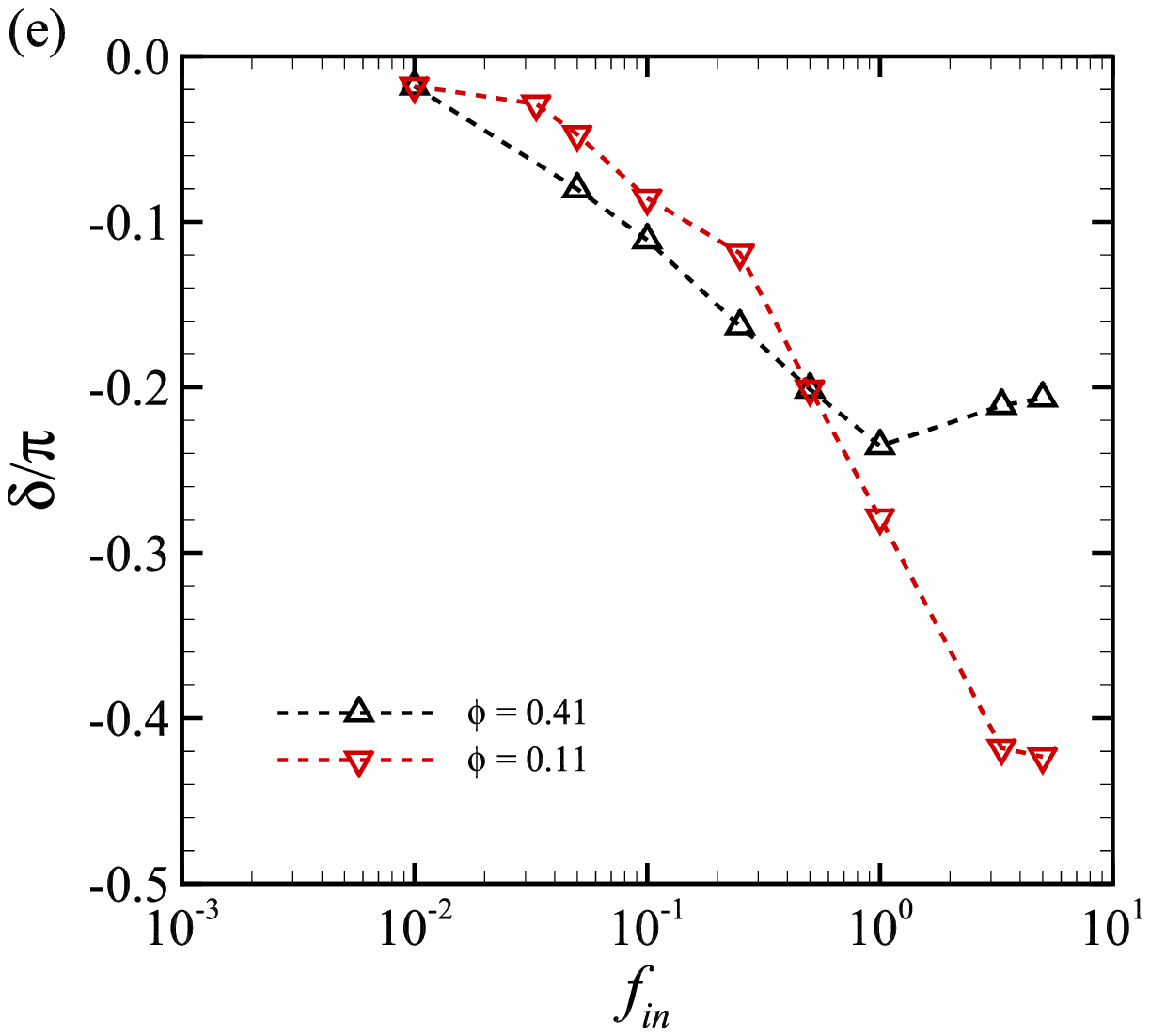}
  \includegraphics[height=5.5cm]{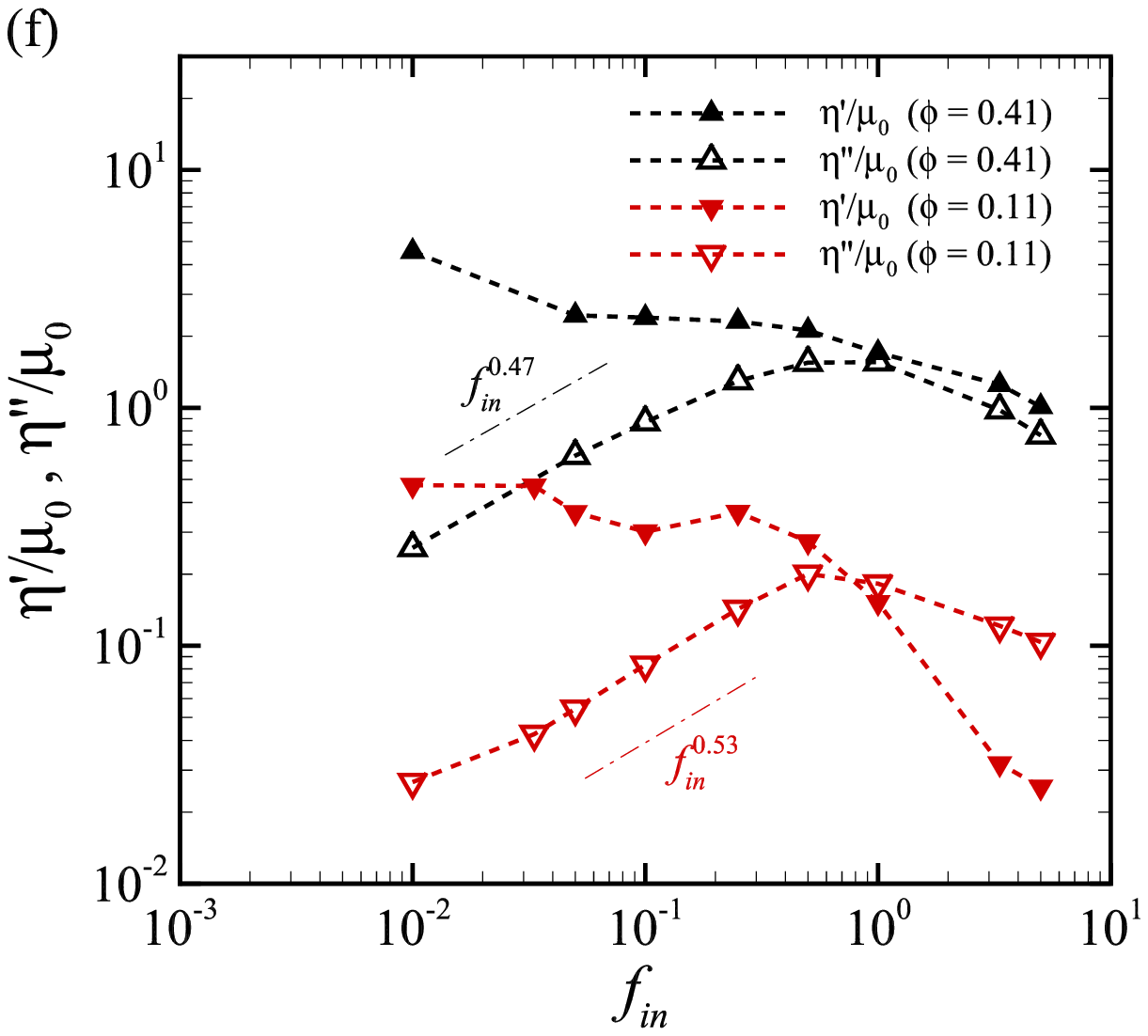}
  \caption{
	  (a) Spatial-temporal average of the deformation index $\langle a_{max} \rangle/a_0$,
	  (b) orientation angle $\langle \theta \rangle/\pi$,
	  (c) the first and second normal stress differences $\langle N_i \rangle/\mu_0 \dot{\gamma}_0$ ($i$ = 1 and 2),
	  (d) amplitude $|\mu_{sp}|^{amp}$,
	  (e) phase difference $\delta/\pi$, and
	  (f) complex viscosity $\eta^{\prime}/\mu_0$ and $\eta^{\prime\prime}/\mu_0$ as a function of $f_{in}$ for dilute ($\phi$ = 0.11) and dense ($\phi$ = 0.41) suspensions.
	  The results obtained under steady shear flow are displayed in panels (a)--(d) with dashed lines.
	  The results are obtained with $Ca_0$ = 0.05, and $\lambda$ = 5.
  }
  \label{fig:hct01ca005lam5}
\end{figure*}

\subsection{Effect of the volume fraction}
We performed simulations in a dilute condition ($\phi$= 0.11) to investigate the behavior at low volume fractions, characterised by fewer cell-cell interactions. The results for different oscillation frequencies at $\phi$ = 0.11 are compared to those at $\phi$= 0.41 in Fig.~\ref{fig:hct01ca005lam5}, following the same order as in Figs.~\ref{fig:hct04ca005lam5} and \ref{fig:effect_lam}.

In dilute conditions, the deformation index, $\langle a_{max} \rangle/a_0$, remains almost the same as in dense conditions, but with smaller fluctuations than at larger volume fractions, as indicated by the error bar (Fig.~\ref{fig:hct01ca005lam5}a). Although frequency-induced deformations are found in the intermediate range of $f_{in}$ (0.0333 $\leq f_{in} \leq$ 1), all the variations with respect to the case of steady shear flow are always less than 1\%. The mean orientation $\langle \theta \rangle/\pi$ and its fluctuations also remain similar to those in the dense suspension, as  shown in Fig.~\ref{fig:hct01ca005lam5}(b), with $\langle \theta \rangle/\pi$ fluctuating around zero for all $f_{in}$, resulting in a lower average $\langle \theta \rangle/\pi$ than  under steady shear flow.

The spatio-temporal average of the first and second normal stress differences $\langle N_i \rangle$, the specific viscosity $|\mu_{sp}|^{amp}$ and of the phase shift $\delta$ are shown as a function of $f_{in}$ in Figs.~\ref{fig:hct01ca005lam5}(c), \ref{fig:hct01ca005lam5}(d) and \ref{fig:hct01ca005lam5}(e). As for the flow of dense suspensions, $\langle N_i \rangle/\mu_0\dot{\gamma_0}$ and $|\mu_{sp}|^{amp}$ decrease from the steady-shear flow value as $f_{in}$ increases. The values of $\langle N_i \rangle$ decrease by one order of magnitude from $\phi$ = 0.41 to $\phi$ = 0.11. The specific viscosity $|\mu_{sp}|^{amp}$ basically decreases with $f_{in}$. The trend is similar to that observed in dense conditions, although $|\mu_{sp}|^{amp}$ is clearly lower at low volume fraction. For both volume fractions under investigation, a first plateau with lower $|\mu_{sp}|^{amp}$ amplitude occurs at $f_{in} \approx$ 0.05 before the final decay for high frequencies. 

The behavior of $\delta/\pi$ in dilute conditions, on the other hand, is different from that in the dense suspension: in this case, $\delta/\pi$ monotonically decreases as $f_{in}$ increases, see panel (e) of Fig.~\ref{fig:hct01ca005lam5}. Because of this, at relatively high forcing frequencies, $f_{in} \geq$ 1, the magnitude of $\delta/\pi$ becomes greater in the dilute condition than in the dense case (cf.\ Fig.~\ref{fig:hct01ca005lam5}e).

Figure~\ref{fig:hct01ca005lam5}(f) reports the complex viscosity. The real part, $\eta^{\prime}$, again well follows the trend discussed for $|\mu_{sp}|^{amp}$, whereas $\eta^{\prime\prime}$ attains its maximum value around $f_{in}$ = 0.5 and then crosses over $\eta^{\prime}$ for larger frequencies. At low frequencies, we estimate $\eta^{\prime\prime}$ to increase with $f_{in}$ following a power law with exponent 0.53, i.e., $\eta^{\prime\prime} \propto f_{in}^{0.53}$.

Overall, the volume fraction clearly affects the mean specific viscosity $\mu_{sp}$ in SAOS flows; however the differences due to volume fraction reduce at higher oscillations frequencies. The other observables appear instead to be almost independent of the volume fraction. High frequency or dilute conditions decrease the rate of hydrodynamic interaction between RBCs, resulting in a reduced contribution of the RBCs to the suspension bulk properties, as observed for the normal stress differences $\langle N_i \rangle$.

A frequency sweep data typically reveals the characteristic microstructural relaxation time scales in the system, often identified as the intersection of the two complex moduli. In our case, $\eta^\prime$ and $\eta^{\prime\prime}$ show change in behavior for $f_{in} \geq$ 1 for both $\phi$ = 0.11 and 0.41,  (Fig.~\ref{fig:hct01ca005lam5}f). Also, the normal stress differences $\langle N_i \rangle$ vanish above this value of $f_{in}$, and $\delta$ changes behavior (Fig.~\ref{fig:hct01ca005lam5}e). Intriguingly, this frequency corresponds to the rotation time scale $\dot{\gamma}_0^{-1}$, which is equal to the tumbling time scale of rigid particles suspended in a simple shear flow.

\begin{figure}
  \centering
  \includegraphics[height=5.5cm]{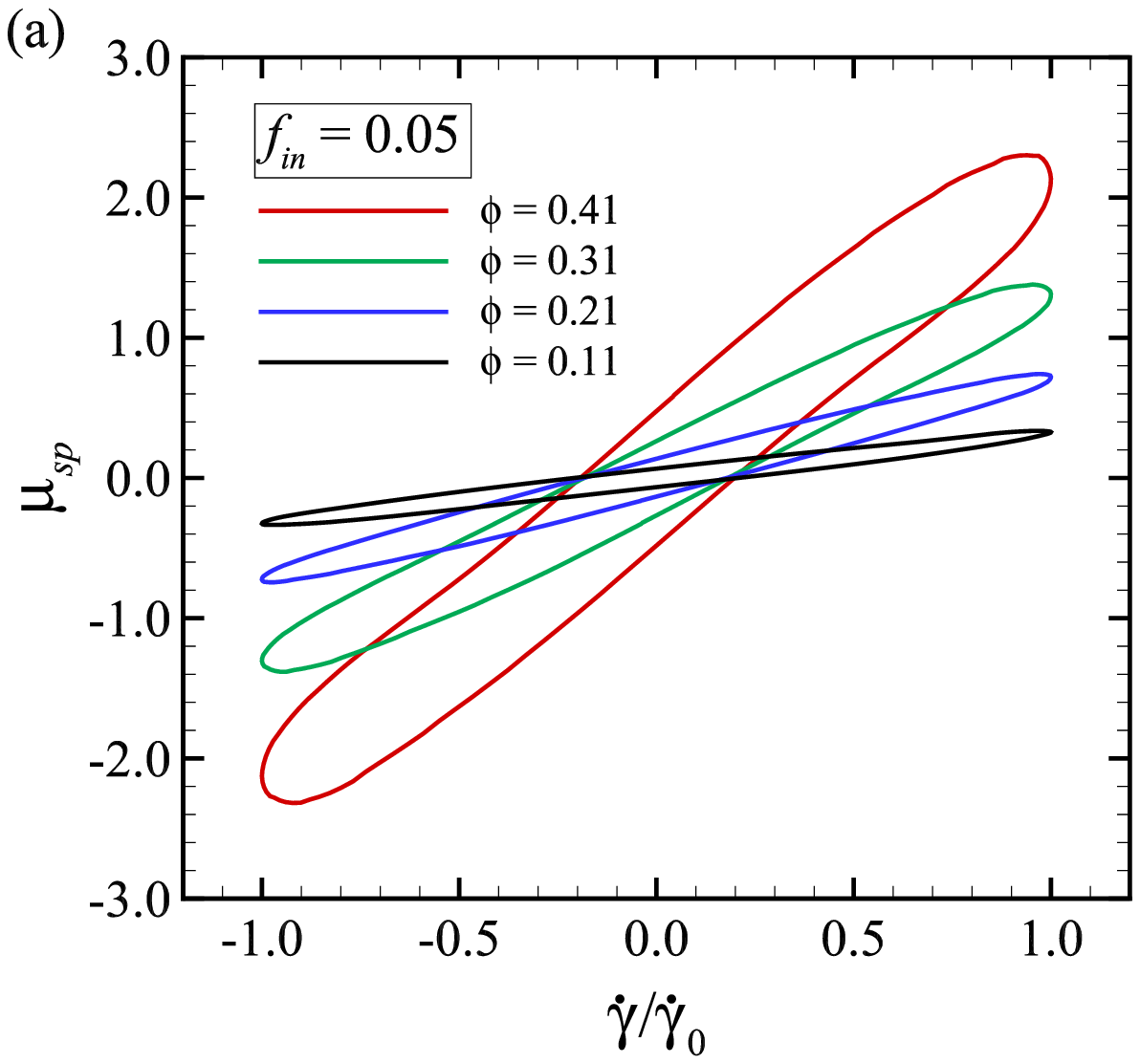}
  \includegraphics[height=5.5cm]{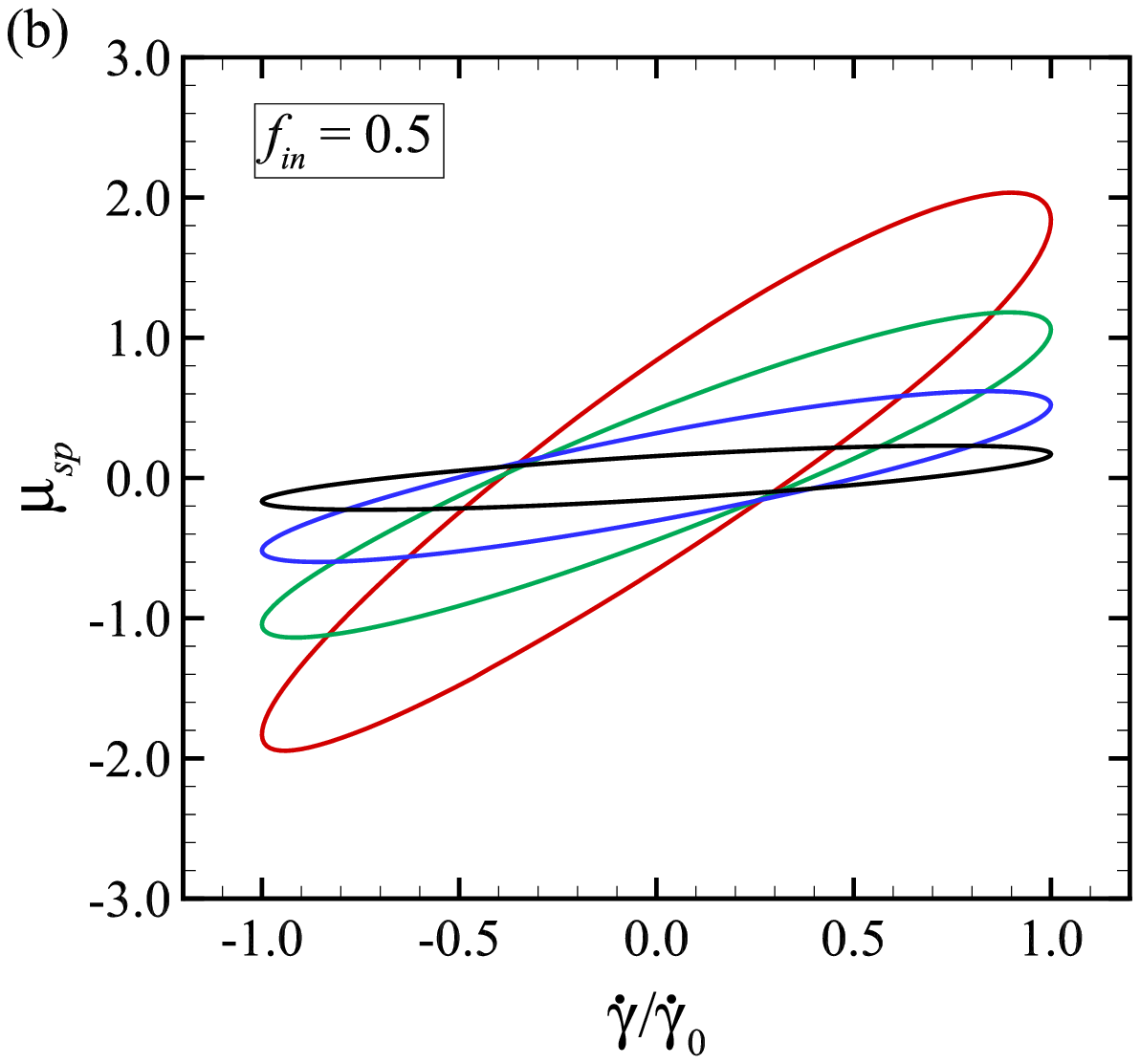}
  \caption{
       Lissajous-Bowditch plots of $\mu_{sp}$ versus $\dot{\gamma}/{\dot{\gamma}}_0$ at $Ca_0$ = 0.8 for different $\phi$ at ($a$) $f_{in}$=0.05 and ($b$) $f_{in}$ = 0.5. The results are obtained with viscosity ratio $\lambda$ = 5.
  }
  \label{fig:Lissajous_hct}
\end{figure}

\begin{figure*}
  \centering
  \includegraphics[height=6.0cm]{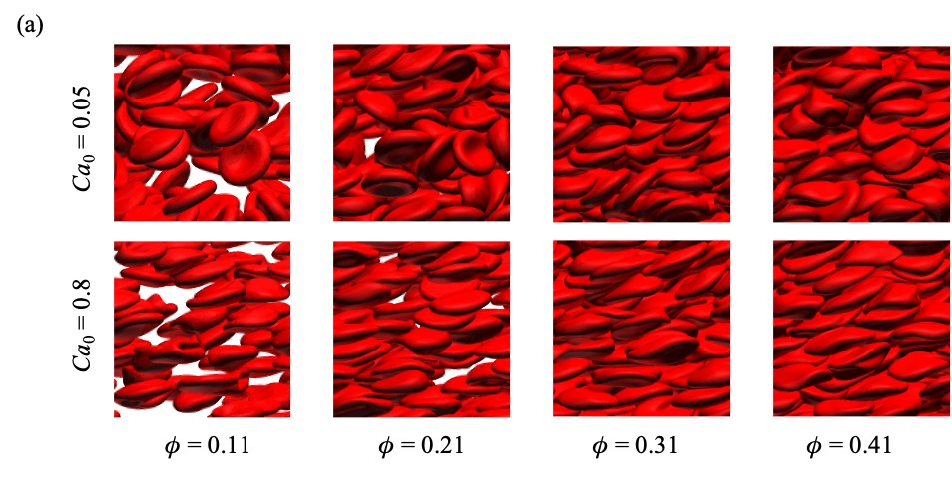}\\
  \includegraphics[height=5.5cm]{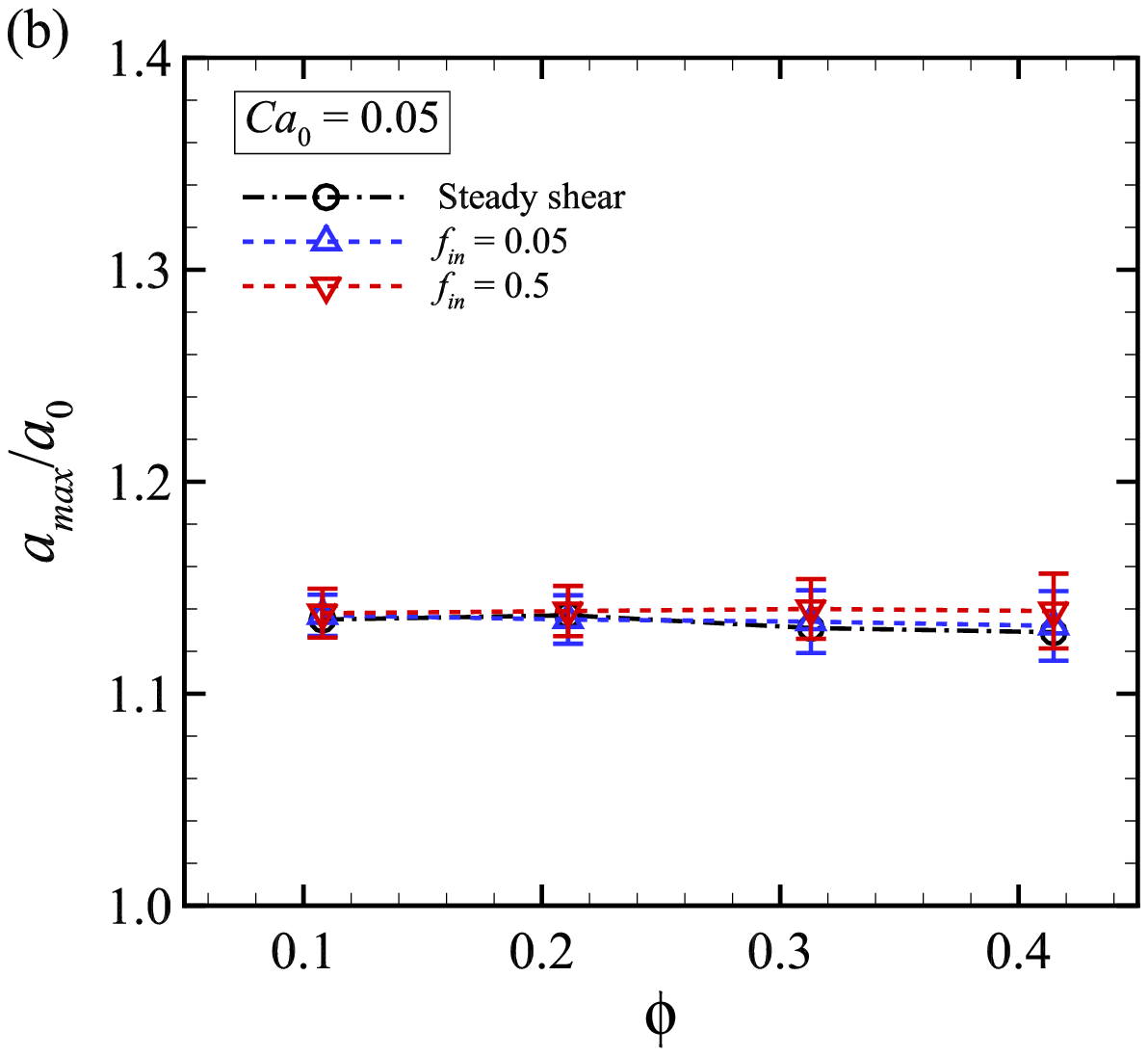}
  \includegraphics[height=5.5cm]{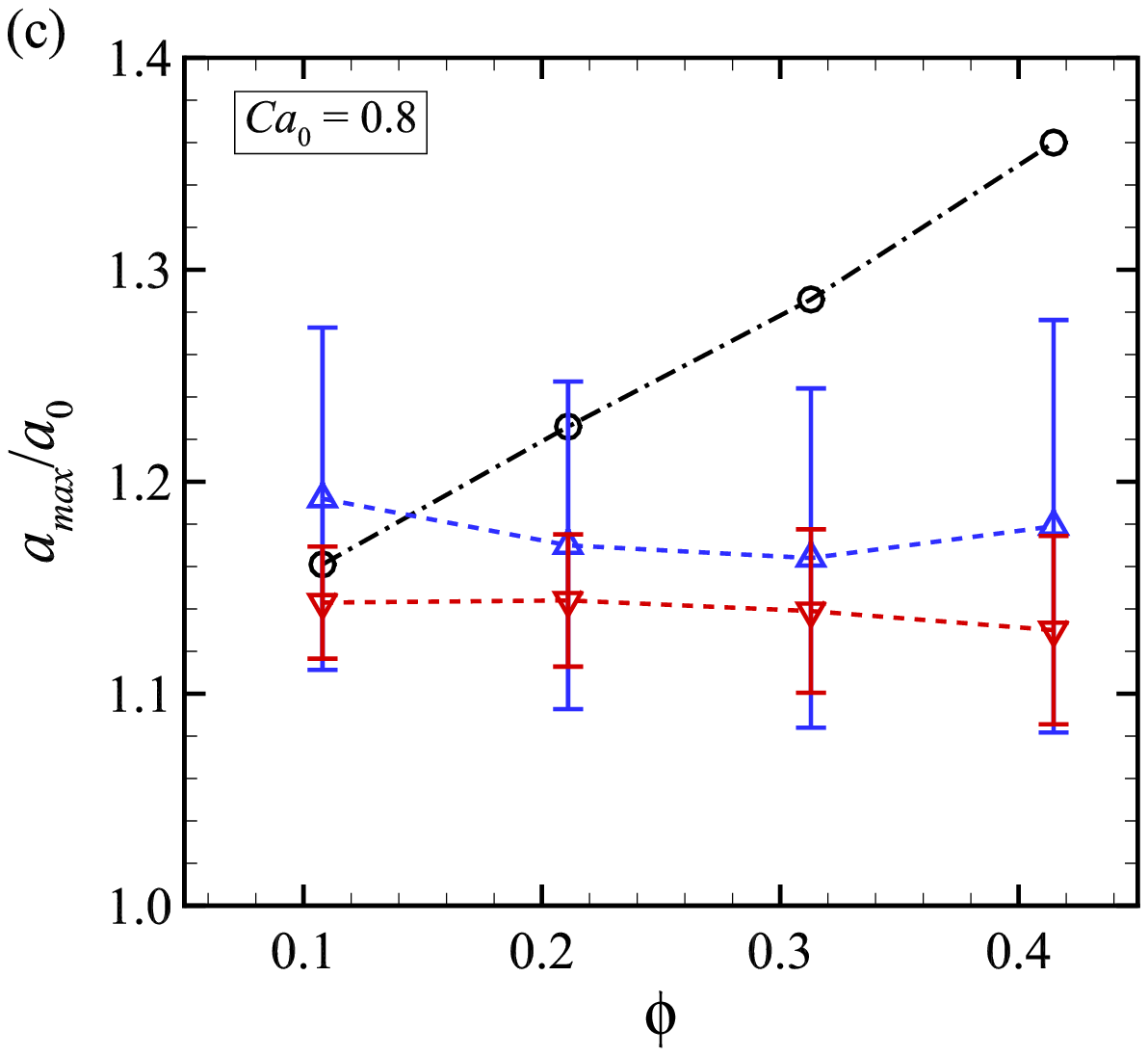}
  \includegraphics[height=5.5cm]{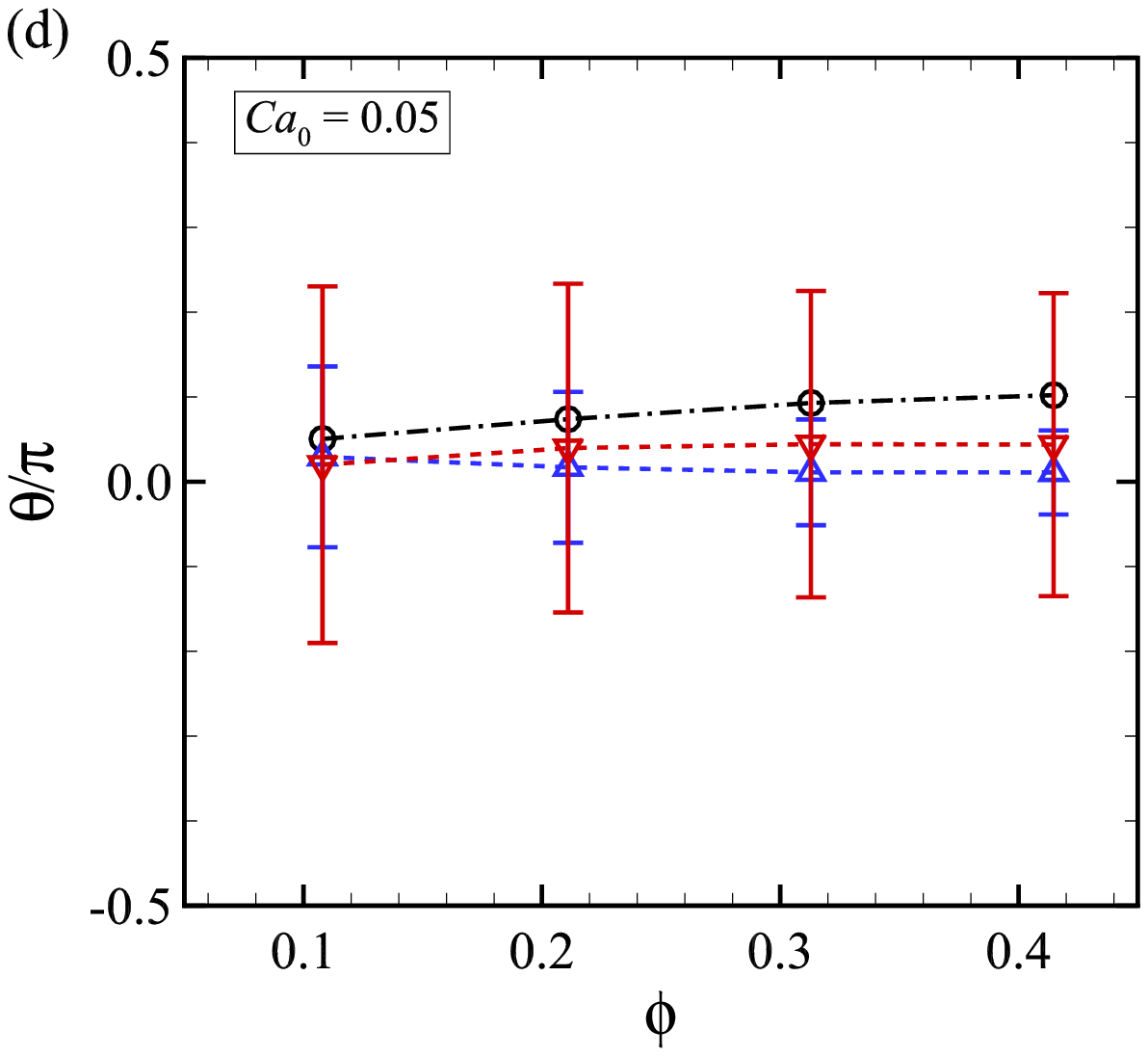}
  \includegraphics[height=5.5cm]{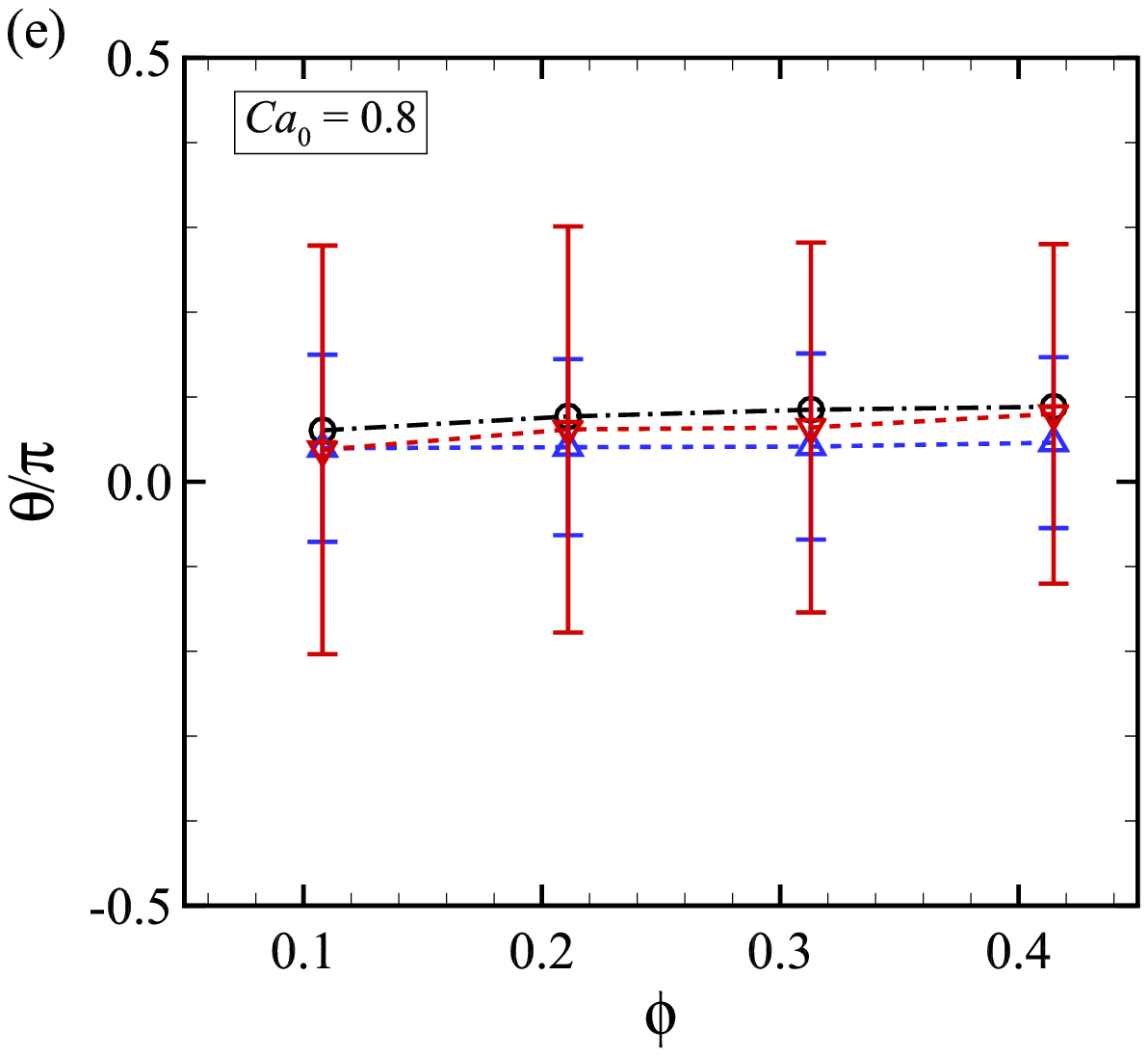}
  \caption{
	  (a) Snapshots from our numerical results of suspensions with different $\phi$=0.11 and 0.41 at imposed shear frequency $f_{in}$ = 0.5. The top raw shows the data at the lowest $Ca_0$ = 0.05 and the bottom raw those at the highest $Ca_0$= 0.8.
	  (b and c) Spatial-temporal average of the deformation index $\langle a_{max} \rangle/a_0$, and
	  (d and e) orientation angle $\langle \theta \rangle/\pi$ as function of $\phi$ for two values of $f_{in}$ (0.05 and 0.5).
	  The left column (panels b, and d) displays the results for $Ca_0$ = 0.05, and the right column (c, and e) those for $Ca_0$ = 0.8. The results are obtained for viscosity ratio $\lambda$ = 5.
  }
  \label{fig:effect_ca_shape}
\end{figure*}

\begin{figure*}
  \centering
  \includegraphics[height=5.5cm]{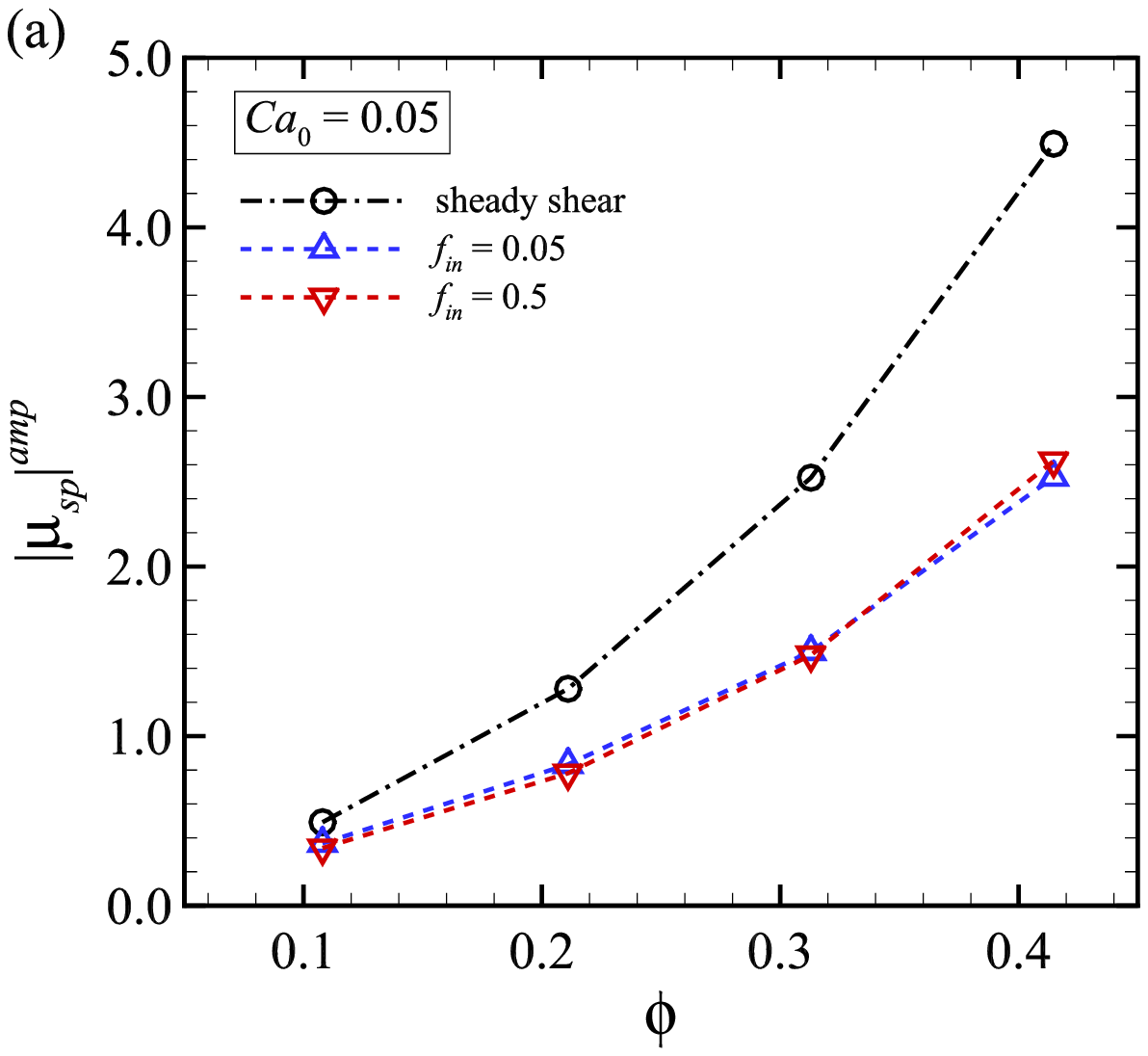}
  \includegraphics[height=5.5cm]{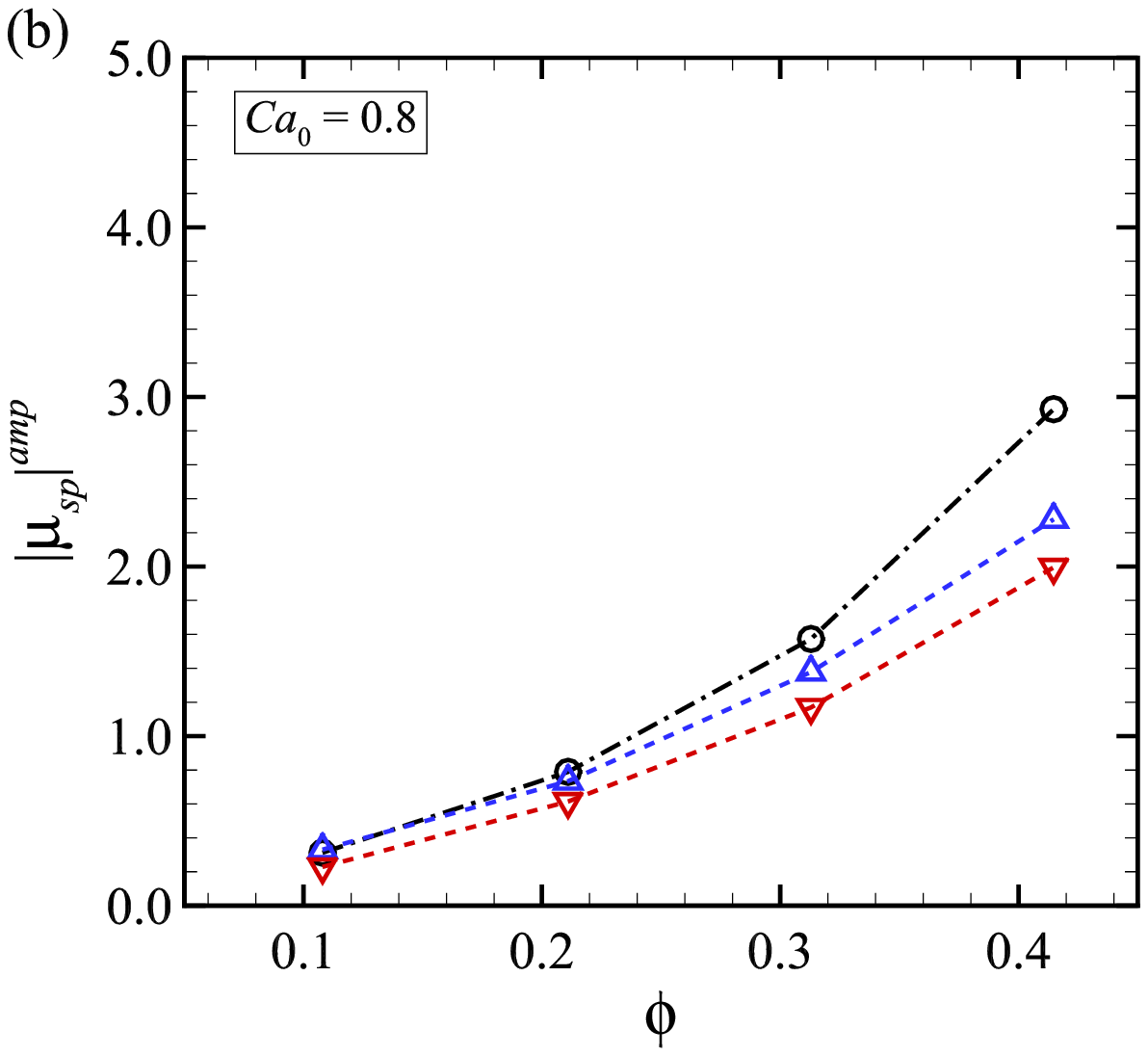}
  \includegraphics[height=5.5cm]{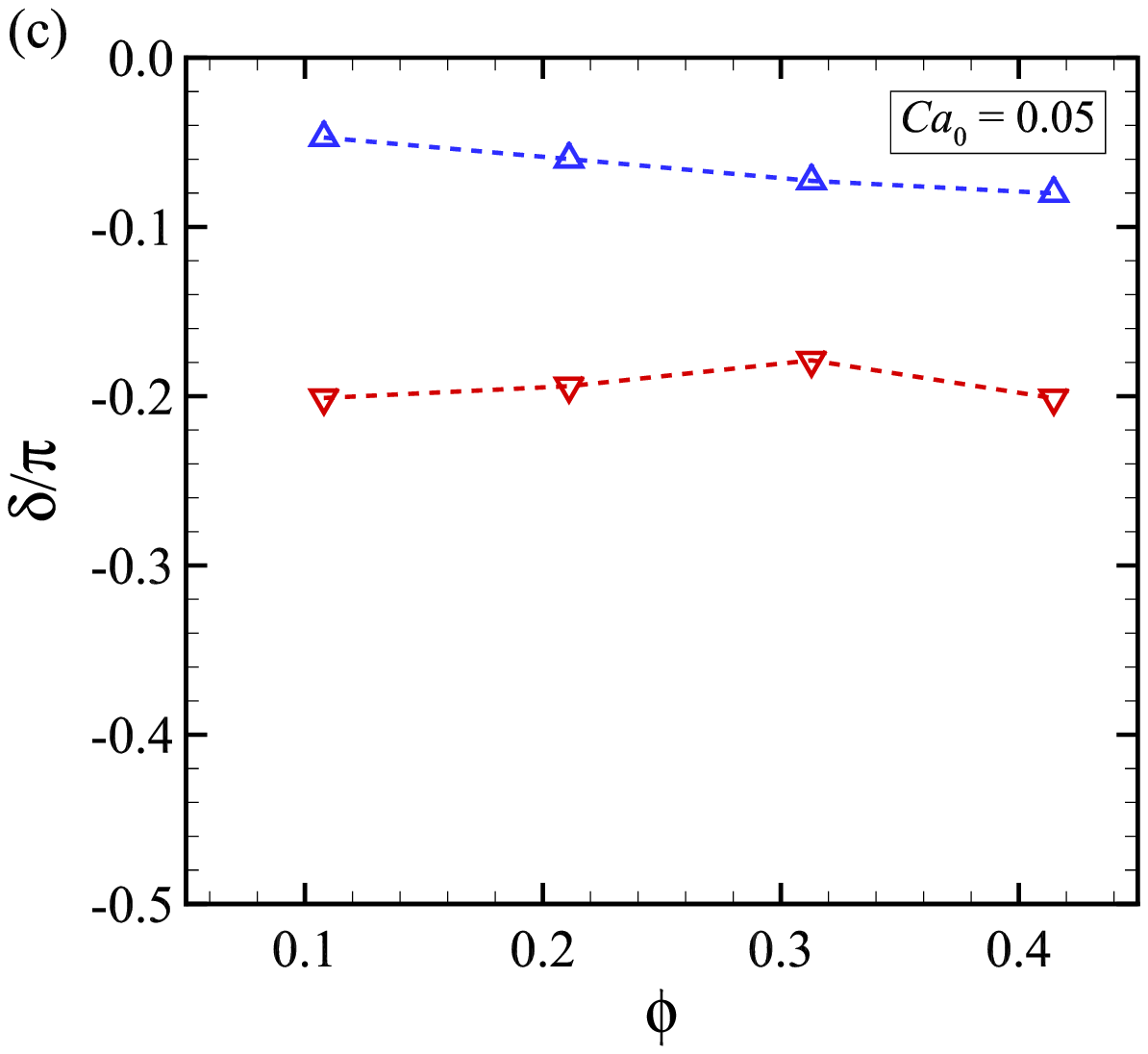}
  \includegraphics[height=5.5cm]{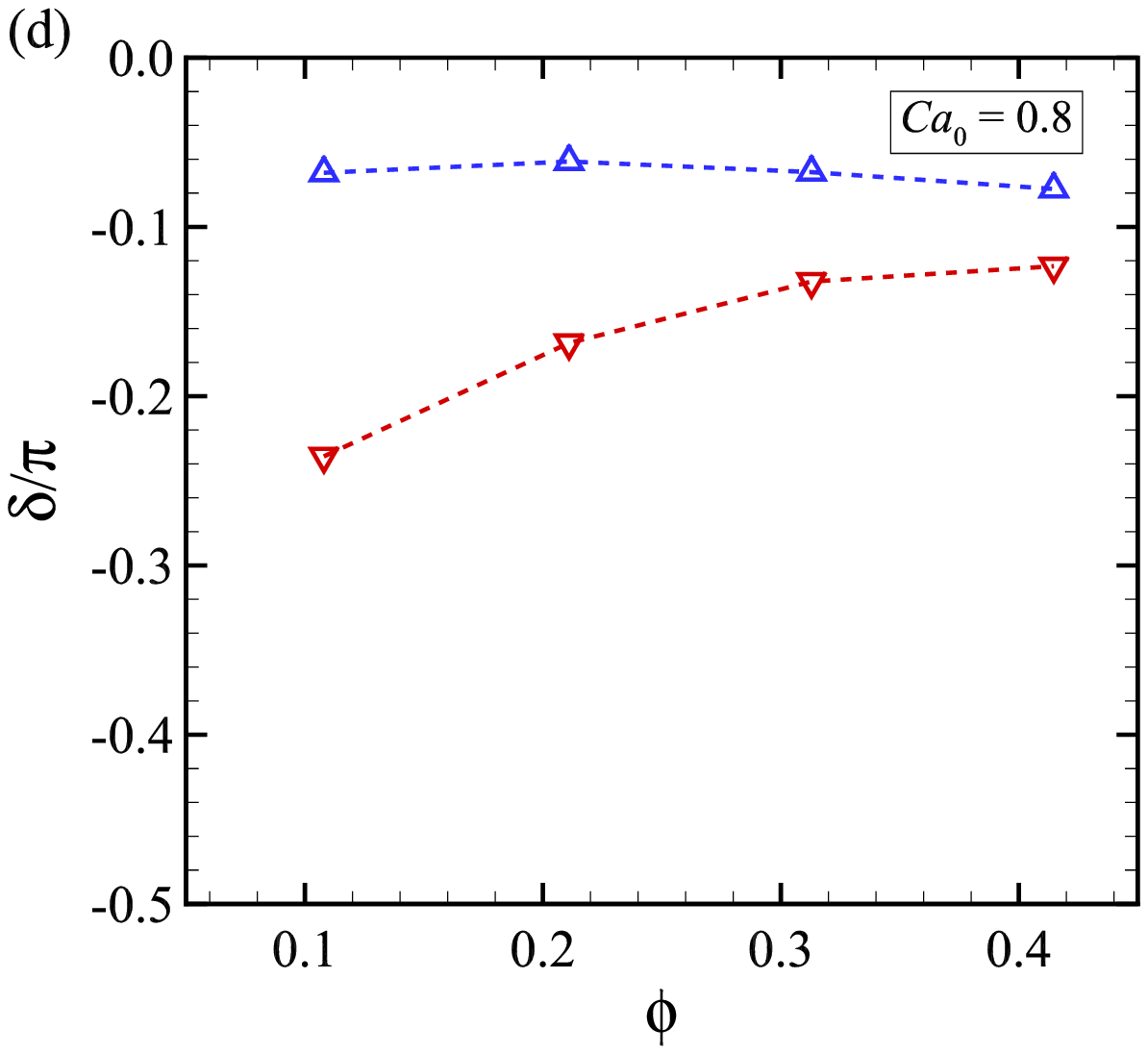}
  \includegraphics[height=5.5cm]{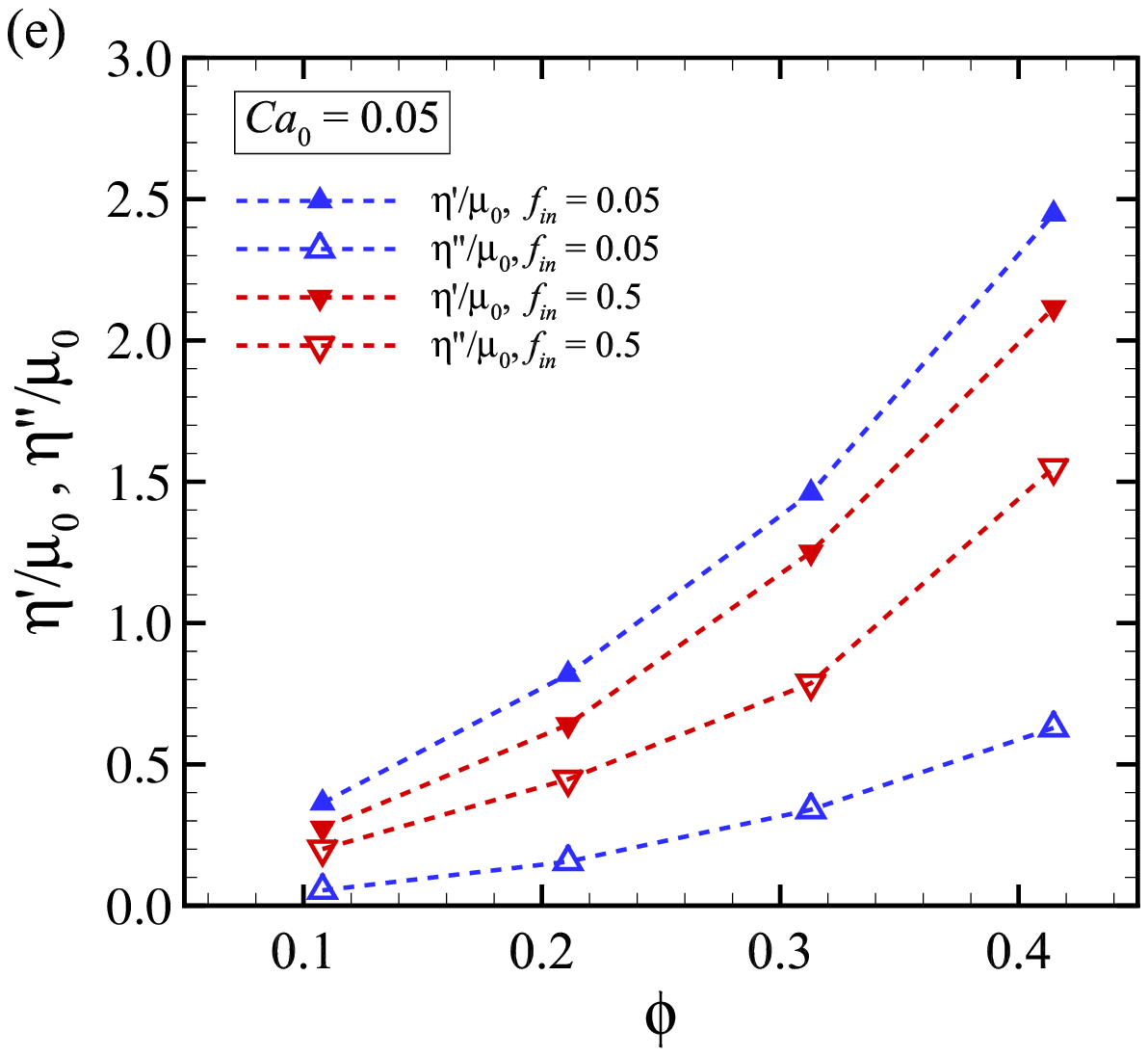}
  \includegraphics[height=5.5cm]{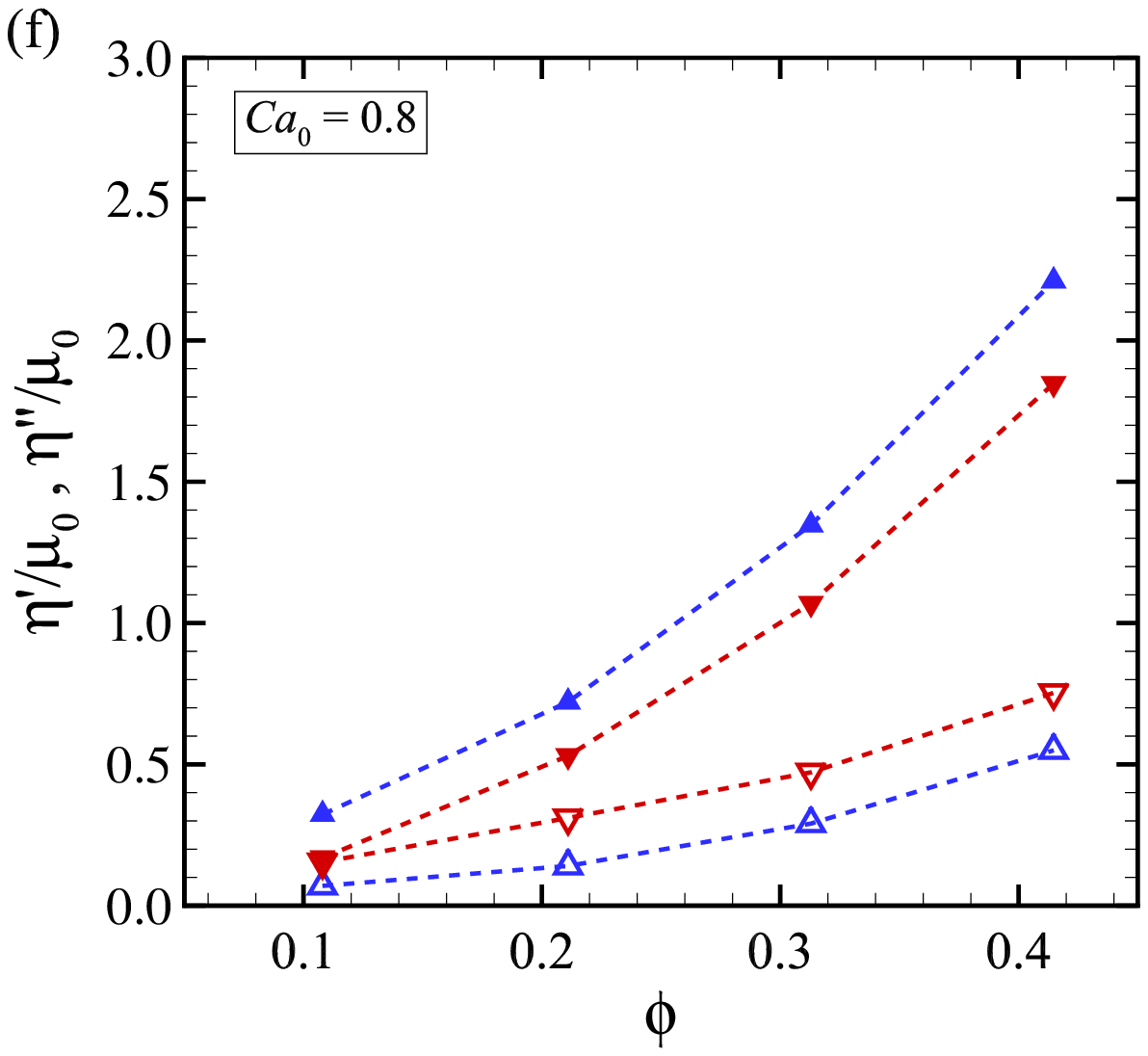}  
  \caption{
	  (a and b) Spatial-temporal average of the amplitude of the specific viscosity $|\mu_{sp}|^{amp}$,
	  (c and d) phase difference between applied shear and corresponding stress $\delta/\pi$, and 
	  (e and f) complex viscosity $\eta^{\prime}/\mu_0$ and $\eta^{\prime\prime}/\mu_0$ as function of the volume fraction $\phi$ and the oscillatory frequency $f_{in}$ (= 0.05 and 0.5).
	  The left column (panels a, c, and e) displays the results for $Ca_0$ = 0.05, and the right column (b, d, and f) those for $Ca_0$ = 0.8. The results are obtained for viscosity ratio $\lambda$ = 5.
  }
  \label{fig:effect_ca}
\end{figure*}

\subsection{Effect of the capillary number}
In this section, we consider the relatively large capillary number $Ca_0$ = 0.8, and quantify its effect on the complex viscosity for different values of the volume fraction $\phi$ (= 0.11--0.41). Even in such large $Ca_0$ condition, the Lissajous-Bowditch plots shown in Fig.~\ref{fig:Lissajous_hct} exhibit an elliptical shape, which is the characteristic response in SAOS, independently of the value of $f_{in}$. Snapshots of the suspension of RBCs subject to low and large capillary number, $Ca_0$ = 0.05 and 0.8, at $f_{in}$ = 0.5 and for different volume fractions $\phi$ are shown in Fig.~\ref{fig:effect_ca_shape}(a). Also, the spatio-temporal average of the deformation $\langle a_{max} \rangle/a_0$ and of the orientation angle $\langle \theta \rangle/\pi$ are shown in Figs.~\ref{fig:effect_ca_shape}(b)--\ref{fig:effect_ca_shape}(e) for both low $Ca_0$ (= 0.05) and high $Ca_0$ (= 0.8).

Starting from Fig.~\ref{fig:effect_ca_shape}(b), we note that the deformation $\langle a_{max} \rangle/a_0$ remains almost the same for low $Ca_0$ (= 0.05), independently of $f_{in}$ (= 0.05 and 0.5) and $\phi$; the values are close to those obtained for steady shear flow (see the black dashed line in the figure). 
 On the other hand, for the large $Ca_0$ (= 0.8) condition, we observe a difference between the low and high $f_{in}$; in particular, $\langle a_{max} \rangle/a_0$ is larger at $f_{in}$ = 0.05 than at $f_{in}$ = 0.5, also with larger variations around the mean (see error bars in Fig.~\ref{fig:effect_ca_shape}c). We therefore observe that the deformation is impeded by oscillatory flows. For the lowest value considered, $\phi$ = 0.11, we do not observe relatively large variations of the average deformation at $f_{in} = 0.5$, suggesting that the decrease of deformation with frequency is due to hydrodynamic interactions among the different cells.

In the results presented so far, $\langle \theta \rangle/\pi$ is found to fluctuate around zero, resulting in a lower value of $\langle \theta \rangle/\pi$ than that for steady shear flow. This is observed also when $f_{in}$, $\phi$, and $Ca_0$ change as shown in Figs.~\ref{fig:effect_ca_shape}(d) and \ref{fig:effect_ca_shape}(e). The fluctuations of $\langle \theta \rangle/\pi$ are larger for high-frequency oscillatory shear, which is similar to the results for different viscosity ratios $\lambda$ presented in Fig.~\ref{fig:effect_lam}(b). As mentioned above or in appendix~\S\ref{appA_distance}, high frequency or dilute conditions decrease the rate of hydrodynamic interaction between RBCs, resulting in a reduced contribution of the RBCs to the suspension bulk properties. Thus, large fluctuations of $\langle \theta \rangle/\pi$ at high frequency come from various orientations of the individual RBCs due to weak hydrodynamic interactions, independently of $\phi$ and $Ca_0$.

The spatio-temporal average of the specific viscosity $|\mu_{sp}|^{amp}$ and of the phase delay $\delta/\pi$ are compared for low $Ca_0$ (= 0.05) and high $Ca_0$ (= 0.8)  in panels (a)--(d) of Fig.~\ref{fig:effect_ca}. As expected, the specific viscosity increases with the RBC volume fraction: $|\mu_{sp}|^{amp}$ is maximum for steady shear flow and decreases under oscillatory shear flow. For the smallest $Ca_0$ (= 0.05), $|\mu_{sp}|^{amp}$ is almost independent of the frequency of the imposed shear $f_{in}$, at least for the range considered in this study, deemed relevant for physiological flows. For the large $Ca_0$ (= 0.8) flow, $|\mu_{sp}|^{amp}$ appears to decrease with $f_{in}$. For both low and high $Ca_0$, the frequency-dependent decrease of $|\mu_{sp}|^{amp}$ becomes more evident as $\phi$ increases. We can also note in the figure that $|\mu_{sp}|^{amp}$ is lower for the largest value of the capillary number investigated at steady shear rate. As high $Ca_0$ corresponds to large deformability in the limit $f_{in}\to 0$, this result confirms the shear-thinning behavior with deformability in the case of suspensions of deformable objects~\citep{Rosti_Brandt_Mitra2018, Rosti_Brandt2018, Chiara2020}.

As discussed when examining the results in Fig.~\ref{fig:effect_lam}(e), the magnitude of $\delta/\pi$ increases at high $f_{in}$ for every volume fraction $\phi$, independently of $Ca_0$ as shown in Figs.~\ref{fig:effect_ca}(c) and \ref{fig:effect_ca}(d). Furthermore, the magnitude of $\delta/\pi$ also depends on $\phi$: at low $f_{in}$ (= 0.05), $|\delta|/\pi$ slightly increases as $\phi$ increases for low $Ca_0$ (Fig.~\ref{fig:effect_ca}c). On the other hand, at high $f_{in}$ (= 0.5), $|\delta/\pi|$ decrease as $\phi$ increases, especially for large $Ca_0$ conditions. Overall, the phase difference $|\delta/\pi|$ is largest for the lowest $\phi$ (= 0.11), high $f_{in}$ (= 0.5), and high $Ca_0$ (= 0.8).

The complex viscosity is reported in Figs.~\ref{fig:effect_ca}(e) and \ref{fig:effect_ca}(f) versus the volume fraction $\phi$ for the two values of $Ca_0$ under consideration. Consistently with the results described so far, $\eta^\prime$ follows the trend displayed by $|\mu_{sp}|^{amp}$ for both high and low $Ca_0$; in particular, $\eta^\prime$ is higher for low $Ca_0$ than for high $Ca_0$, confirming once more the shear thinning with deformability of elastic objects. Furthermore, $\eta^\prime$ at low $f_{in}$ (= 0.05) is always higher than at high $f_{in}$ (= 0.5) for each $\phi$, while this trend is opposite in $\eta^{\prime\prime}$, which always higher at high $f_{in}$ than those at low $f_{in}$. Overall, conditions of small $Ca_0$ and high $f_{in}$ are those to lead to the largest $\eta^{\prime\prime}$.

\citet{Thurston1972} adopted the metrics of viscoelasticity discussed in this work, i.e., $\eta^\ast = \eta^\prime - i\eta^{\prime\prime}$~\eqref{eq:eta}, and measured the value of $\eta^\ast$ of normal human blood (i.e., suspension of RBCs in the plasma) for varying amplitude of the shear rates (or velocity gradient) at a temperature of 25 $^\circ$C and with concentrations ranging from 0 to 100\%. The measurements were carried out at a frequency of 10 Hz, corresponding to $f_{in} \approx$ 0.25 in this study. The experimental results showed that for volume fractions above 20\% ($\phi \geq$ 0.2) and shear rates less than 2 s$^{-1}$, corresponding to $Ca_0 \approx$ 2.4 $\times$ 10$^{-3}$ in our study ($a_0 = 4$ $\mu$m and $G_s = 4$ $\mu$N/m are assumed here), the elastic component $\eta^{\prime\prime}$ increased with $\phi$ approximately cubically (i.e., $\eta^{\prime\prime} \propto \phi^3$) while the viscous component $\eta^\prime$  increases exponentially. Such dramatical increases in both metrics ($\eta^\prime$ and $\eta^{\prime\prime}$) are seen for low shear rates, which potentially causes the shear-thinning character of the blood~\citep{Chien1970}. In our study, the minimum shear rate condition is $\approx$ 42 s$^{-1}$ ($Ca_0$ = 0.05), which is still over 20 times greater than the experimental condition of~\citet{Thurston1972}, and the increase of $\eta^\prime$ and $\eta^{\prime\prime}$ as a function of $\phi$ (see Figs.~\ref{fig:effect_ca}e and \ref{fig:effect_ca}f) are more moderate than in the experiment. For precise comparison with experimental data, modeling of aggregation and disaggregation of RBCs would also be important.

\begin{figure*}
  \centering
  \includegraphics[height=5.5cm]{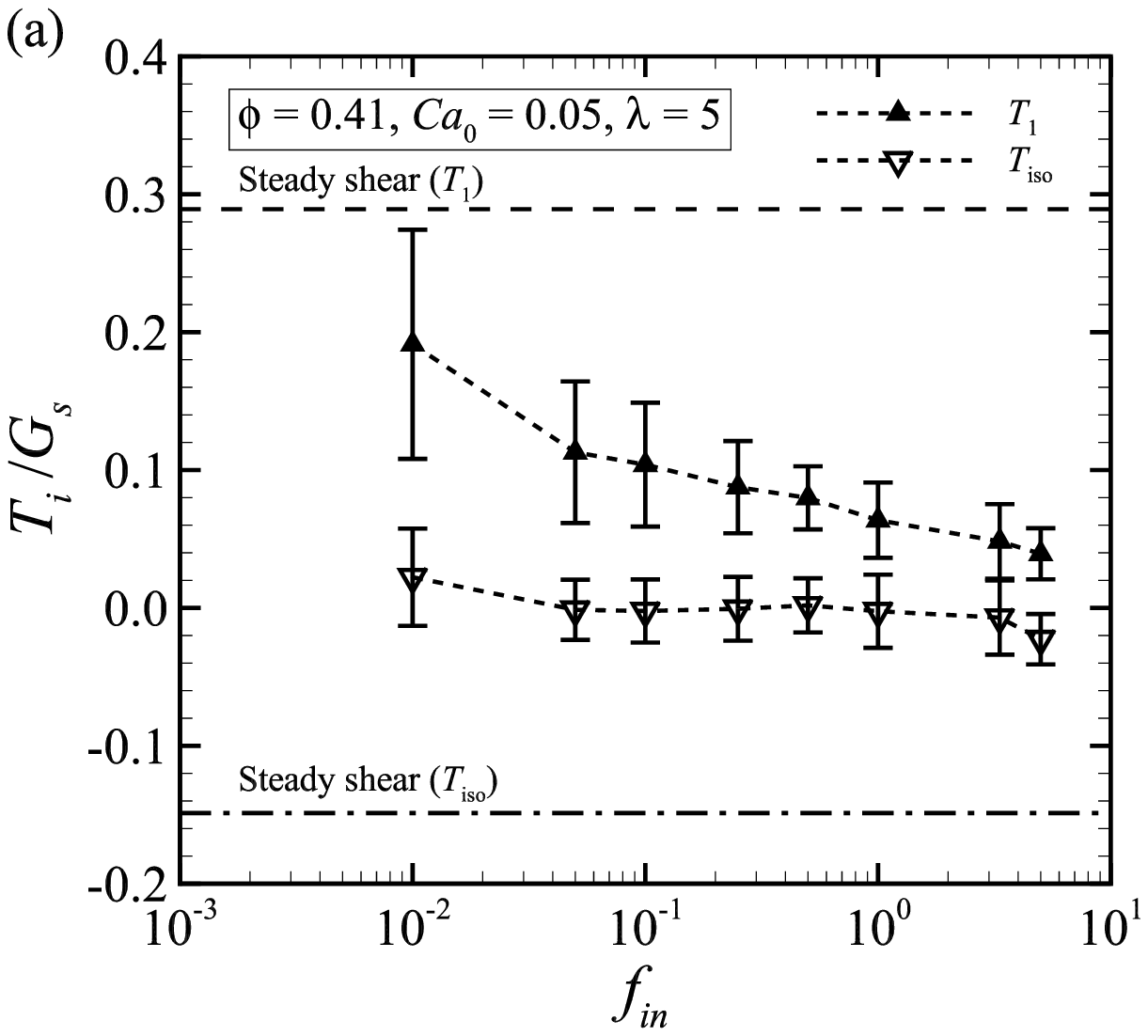}
  \includegraphics[height=5.5cm]{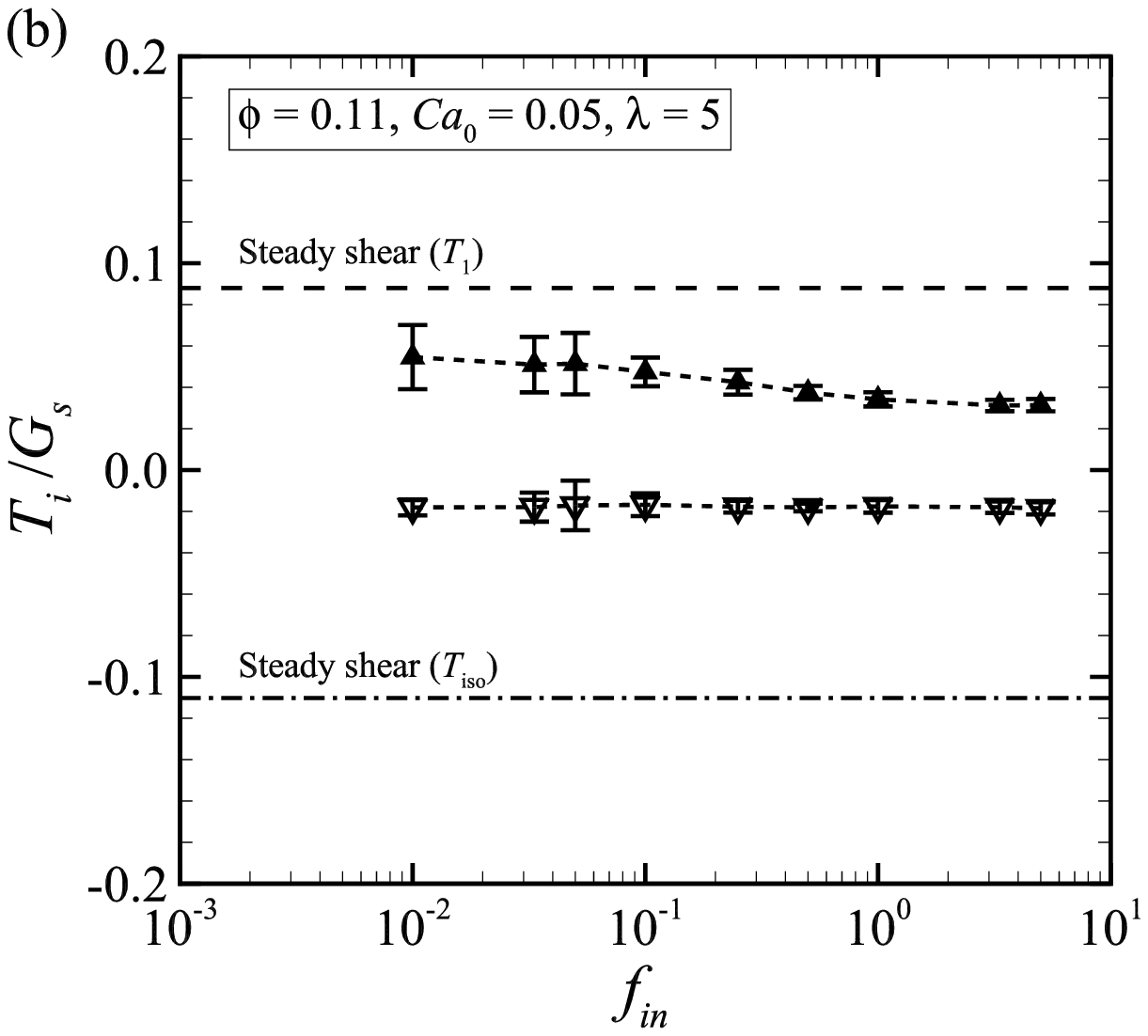}
  \includegraphics[height=5.5cm]{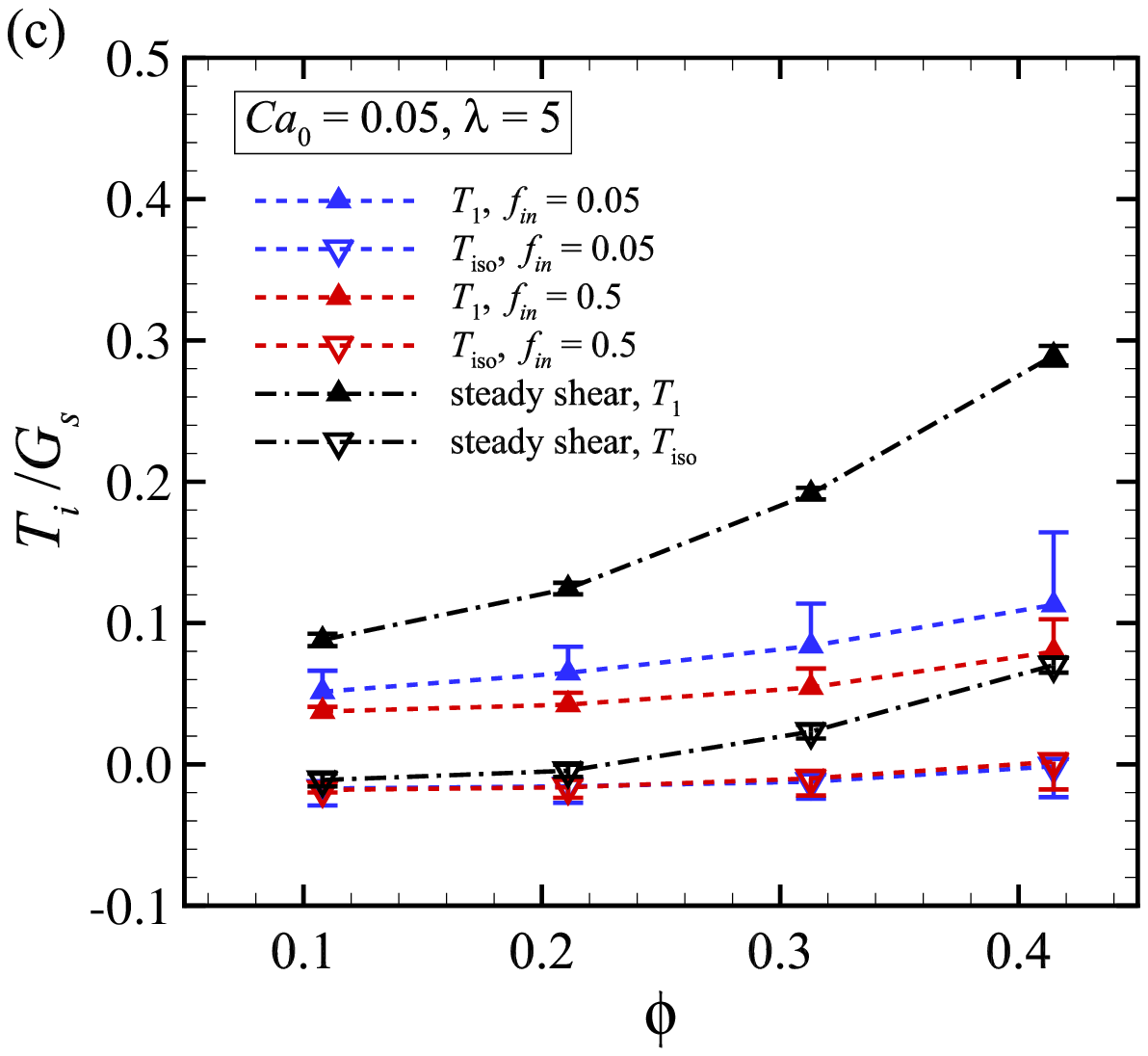}
  \includegraphics[height=5.5cm]{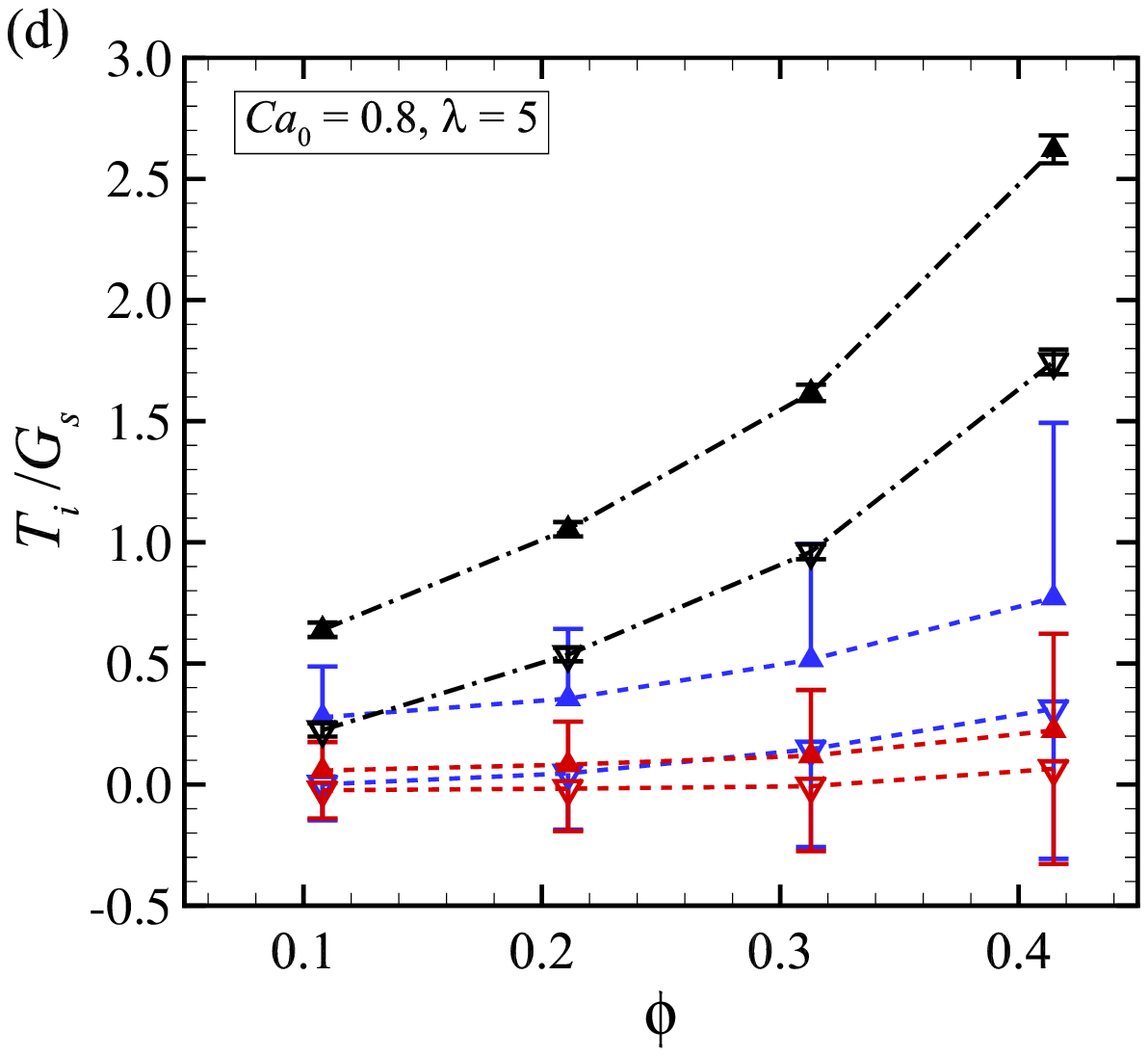}
  \includegraphics[height=5.5cm]{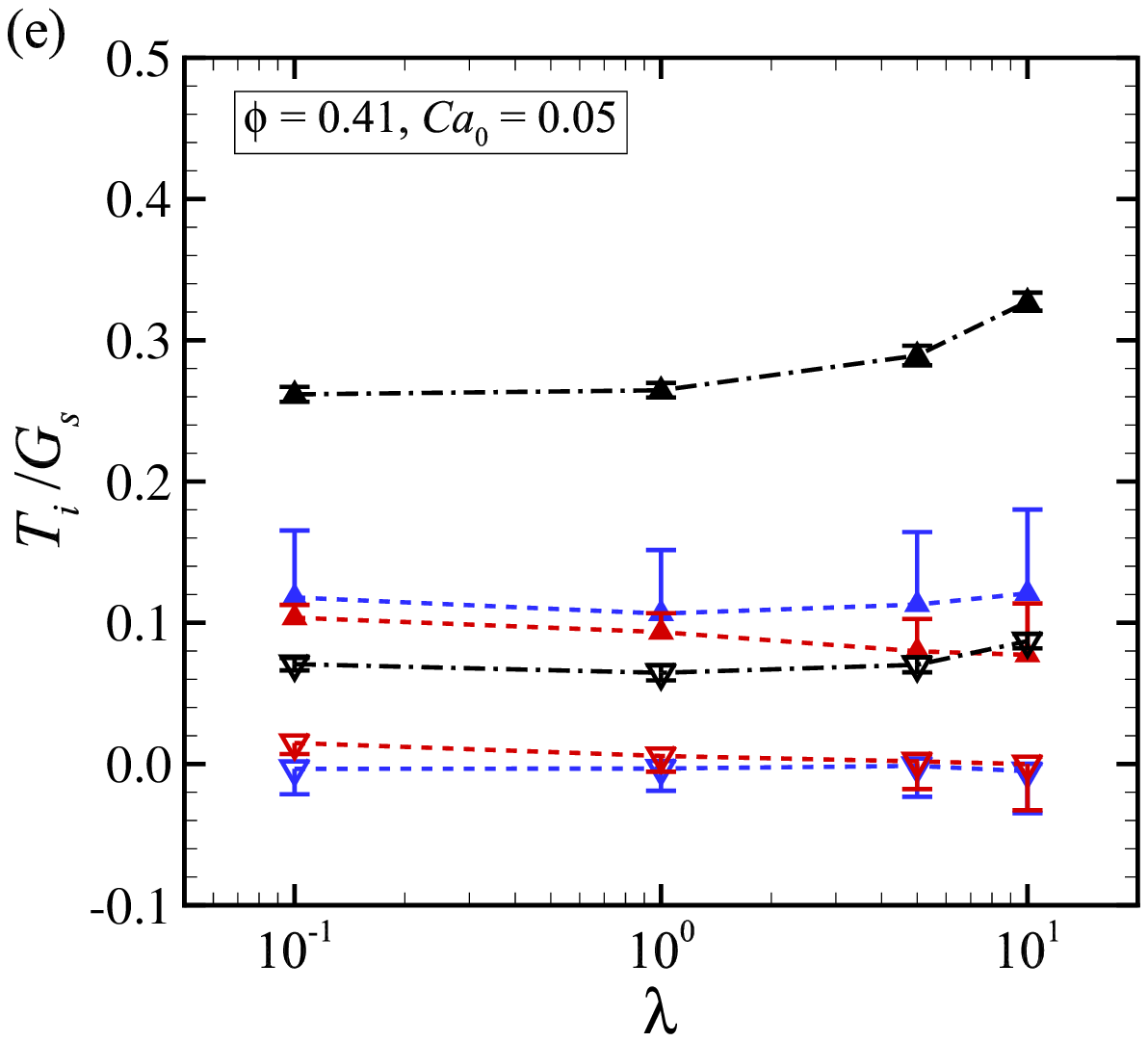}
  \caption{
	  (a and b) Spatial-temporal average of the first and isotropic tensions ($T_1$ and $T_{iso}$, respectively) as a function of $f_{in}$ for (a) dense ($\phi$ = 0.41) and (b) dilute ($\phi$ = 0.11) conditions. The results are obtained for SAOS ($Ca_0$ = 0.05).
	 In  (c) and (d), the membrane tensions are shown as function of $\phi$ and $f_{in}$ (= 0.05 and 0.5) for (c) $Ca_0$ = 0.05 and (d) $Ca_0$ = 0.8.
	  The results shown in (a)--(d) are obtained for $\lambda$ = 5.
  	  (e) The membrane tensions as function of the viscosity ratio $\lambda$ and $f_{in}$ (= 0.05 and 0.5) for dense condition ($\phi$ = 0.41).
	  The result in (e) is obtained for SAOS ($Ca_0$ = 0.05).
  }
  \label{fig:tension}
\end{figure*}

\subsection{Membrane tension under oscillatory shear flow}
Finally, we also investigate the maximal in-plane principal tension $T_1$ ($\geq T_2$) and the isotropic tension $T_{iso} (= T_1 + T_2)/2$ in the deformed RBC, and display the results in Fig.~\ref{fig:tension}. We indicate the spatial-temporal average of those tensions as $\langle T_1 \rangle$ and $\langle T_{iso} \rangle$, which are computed using equation~\eqref{eq:Ti}. It is known that these tensions monotonically increase with $Ca$, as observed in narrow rectangular microchannels~\citep{Takeishi2019MDPI}.
Similarly to the normal stress differences $\langle N_i \rangle$, the values of $\langle T_i \rangle$ also decrease with $f_{in}$, independently of $\phi$ (Figs.~\ref{fig:tension}a and \ref{fig:tension}b), clearly indicating the decrease of the RBC deformation and a reduction in their hydrodynamic interactions. Indeed, the magnitude of $|\langle T_i \rangle|$ decreases for all $f_{in}$, compared to those in a steady shear flow, with the maximum principal tension $\langle T_1 \rangle$ being much lower especially in the dilute condition ($\phi$ = 0.11). A frequency-dependent decrease of the membrane tensions is also found for all $\phi$ both for $Ca_0$ = 0.05 and $Ca_0$ = 0.8 (Figs.~\ref{fig:tension}c and \ref{fig:tension}d, respectively). 

The state of membrane tension is further investigated for different viscosity ratios $\lambda$, the results shown in Fig.~\ref{fig:tension}(e). Compared with the previous results of single RBC in Refs.~\onlinecite{Omori2012, Takeishi2019MDPI}, the similarities or discrepancies in the values of $\langle T_i \rangle$ for different $\lambda$ arise here from the different input frequencies $f_{in}$. The observations above about the reduction in membrane tension with the oscillations are valid also for the different values of $\lambda$ examined. Considering these results shown in~Fig.\ref{fig:tension}, SAOS flow basically impedes the deformation of individual RBCs as well as the magnitude of fluid-membrane interactions, resulting in a lower specific viscosity and first and second normal stress differences than in steady shear flow (Figs.~\ref{fig:hct01ca005lam5}c and \ref{fig:hct01ca005lam5}d).

Overall, the input frequency results to be the main parameter affecting the tension of the RBC membrane. Although we could not observe any particular frequency-induced variation of the deformation of the RBCs in terms of the index $\langle a_{max} \rangle/a_0$, the membrane tension is strongly affected by the input frequency. Therefore, the full viscoelastic behavior of RBC suspensions cannot be simply estimated by the geometrical properties of individual deformed RBCs, but it should include the state of the membrane tension, which appears to be strongly related to the normal-stress differences (see Figs.~\ref{fig:tension}a and \ref{fig:tension}b).

\section{Conclusions}
We have performed numerical simulations of dense suspension of RBCs under SAOS for a wide rage of shear frequencies $f_{in}$, and quantified the viscoelastic character of the bulk suspension by its complex viscosity; this is defined in terms of the amplitude of the particle stress and the phase difference between the output particle stress and the sinusoidal applied shear~(see Eq.~\ref{eq:eta_def2}). The flow is assumed to be inertialess, with the fluids inside and outside the RBCs modeled as Newtonian. The role of the viscosity ratio $\lambda$, the volume fraction of the RBCs $\phi$, and the $Ca_0$ on the complex viscosity have been evaluated and discussed.

The first important question we focused on is whether the cell deformation might be enhanced by varying the oscillatory frequency. Although frequency-induced deformation can be found both in dense and dilute conditions, especially for the intermediate range of frequencies, 0.05 $\leq f_{in} \leq$ 1 corresponding to 2.1 Hz $\leq \mathrm{f} \leq$ 42 Hz for $Ca_0$ = 0.05 ($\approx$ 42 s$^{-1}$ assuming the reference radius of $a_0$ = 4 $\mu$m and the surface shear elastic modulus of $G_s$ = 4 $\mu$N/m), the differences with the deformations in steady shear flow are always less than 1\%. Thus, enhancement of cell deformation under oscillatory shear is unlikely to occur for RBCs. Previous numerical analysis of a single spherical capsule, whose membrane follows the neo-Hookean (NH) constitutive law (i.e., strain-softening property), demonstrated that frequency-dependent deformations become evident for high $Ca$ and large values of  $\lambda$~\citep{Matsunaga2015}, specifically $Ca$ = 2.0 and $\lambda$ = 10, with a 26\% deformation increase observed at $f_{in}$ = 0.01. Our results, however, show that even for relatively large $Ca_0$ (= 0.8), the cell deformation by oscillatory shear flow is very limited for biconcave capsules, which is possibly due to the SK law (i.e., strain-hardening membrane property). Focusing on biological systems, the oxygen-dependent regulation of RBC properties should be taken into considerations~\citep{Parthasarathi1999}. For instance, it is known that the elongation of RBCs in response to shear stresses increases as oxygen tension is decreased \citep{Wei2016}. Thus, it will be interesting to study whether deformation of RBCs under pulsatile flows results in passive regulation for oxygen transport. Differently from what observed for the cell deformation, the tension of the membrane decreases significantly with the input frequency, also resulting in a decrease of the normal stress differences. This result suggests that the viscoelastic character of RBC suspensions can be fully understood only by accounting for the state of the membrane tension.

Our next question is how the bulk suspension rheology is altered in oscillatory flows when compared to the case of steady shear flow, where most experimental measures are taken. As $f_{in}$ increases, $\eta^\prime$ gradually decreases, while $\eta^{\prime\prime}$ attains its  maximum value at frequency $f_{in}$ = 0.5 corresponding to $\mathrm{f}$ = 21 Hz (Fig.~\ref{fig:hct01ca005lam5}f). A similar trend was reported for a single spherical capsule with NH law in~\citet{Matsunaga2020}, where $\eta^{\prime\prime}$ was shown to reach its maximum at $f_{in}$ = 0.2. Interestingly, the local maximum $f_{in}^{max}$ (= 0.5) remains the same in dilute condition ($\phi$ = 0.11) for the RBC suspensions considered here (Fig.~\ref{fig:hct01ca005lam5}f); therefore, the discrepancy in the frequency of the maximum $\eta^{\prime\prime}$ with the previous numerical study by \citet{Matsunaga2020} might be due to the different membrane constitutive law or the capsule shape rather than an effect of the volume fraction. Furthermore, the curves obtained here for $\eta^{\prime\prime}$ are qualitatively similar to those in Oldroyd-B fluids~\citep{Bird1987, Maklad2021}, whose $N_2$ is however null. While the ratio $-N_2/N_1$ is not exactly equal to 0.5 as typically observed in liquid crystals consisting of rod-like polymer solution~\citep{Maklad2021}, it has the same order of magnitude $O(-\langle N_2 \rangle/\langle N_1 \rangle) = 10^{-1}$, and it is almost independent of the volume fraction $\phi$. Overall, SAOS flow basically impedes the deformation of individual RBCs as well as the magnitude of fluid-membrane interactions (Fig.\ref{fig:tension}), resulting in a lower specific viscosity and first and second normal stress differences than in steady shear flow (Figs.~\ref{fig:hct01ca005lam5}c and \ref{fig:hct01ca005lam5}d).

This study provides the first conclusive evidence of oscillatory rheology of suspension of RBCs in SAOS. Although it is known that the RBC deformation alone is sufficient to give rise to shear-thinning~\citep{Omori2014}, the resulting complex viscosity shown in the Fig.~\ref{fig:hct01ca005lam5}(f) weakly depends on the frequency-modulated deformations or orientations of individual RBCs (Figs.~\ref{fig:hct01ca005lam5}a and \ref{fig:hct01ca005lam5}b, respectively), but rather depends on combinations of the frequency-dependent amplitude $|\mu_{sp}|^{amp}$ and phase difference $\delta$ (Figs.~\ref{fig:hct01ca005lam5}d and \ref{fig:hct01ca005lam5}e, respectively). Effects of particle deformability, inertia, and LAOS on the viscoelastic character in suspensions of RBCs will be reported in a future study. A physiological interpretation for the local maximum of the $\eta^{\prime\prime}$ viscosity at a frequency of $\approx$ 21 Hz is still missing. Since this is 20 times higher than the physiological heart beat ($\approx$ 1 Hz), such high frequency hydrodynamic interactions may only occur in local vascular areas, e.g., aneurisms, or in an artificial blood pumps. Thus, in the future, it will be interesting to also study whether such high-frequency oscillations can induce or delay blood clots.

The values of the dimensional frequency studied in this work $\mathrm{f} (= f_{in} \dot{\gamma}_0 = f_{in} Ca_0 \left\{ G_s/(\mu_0 a_0) \right\})$ are estimated using $G_s$ = 4 $\mu$N/m (see also Figure~12 in~\citet{Takeishi2019}), which is determined by fitting the lengths of stretched RBCs by optical tweezers at steady state~\citep{Suresh2005}. Thus, the exact values of the frequency depend on the estimation of $G_s$, which varies with the membrane constitutive laws and is also sensitive to different experimental methodologies, e.g., atomic force microscopy, micropipette aspiration and etc~\citep{Bao2003}. Direct comparison with experiments is therefore not trivial and left as a future study.

Existing works on oscillatory rheology focus on building macroscale models for dilute suspensions: soft particle glasses~\citep{Khabaz2018}, viscoelastic particles~\citep{Kammer2020}, bubbles~\citep{Lin2019}, vesicles~\citep{Farutin2012}, and NH capsule~\citep{Matsunaga2020}. Although a suspension of particles can be described as a continuum at length scales large compared to the size of the constituent particles, none of these works successfully derived constitutive equations for the oscillatory rheology of dense suspensions. Misbah~\cite{Misbah2006} derived an equation to describe the suspension viscosity $\mu_{eff}$ in the tank-treading regime under the assumption of small deformations of vesicles in steady shear flow: 
$(\mu_{eff} - \mu_0)/(\phi \mu_0) = 2.5 - \mathscr{A} (23 \lambda + 32)/(16 \pi)$, where $\mathscr{A}$ is the excess area.
The next challenge will therefore be extending the applicability of such models to cover a wide range of volume fractions and large deformations of RBCs. Our numerical results and quantitative model analysis of the viscoelastic property of dense suspensions of RBCs will be also helpful to build more sophisticated non-Newtonian constitutive laws that consider multi-scale dynamics, and to gain insights not only into the passive cellular flow in physiological systems~\citep{Chien1967, Chien1970, Cokelet1968, Goldsmith1975}, but also into the design of novel artificial blood pumps~\citep{Deutsch2006}. We hope that our numerical results will stimulate further numerical and experimental studies, able to shed light on the complex interplay between cellular dynamics and oxygen transport in the microcirculation. 

\begin{acknowledgments}
This research was supported by JSPS KAKENHI Grant Number JP20H04504, and by the Keihanshin Consortium for Fostering the Next Generation of Global Leaders in Research (K-CONNEX), established by Human Resource Development Program for Science and Technology. M.E.R. acknowledges the support of the Okinawa Institute of Science and Technology Graduate University (OIST) with subsidy funding from the Cabinet Office, Government of Japan. L.B. acknowledges the financial support by the Swedish Research Council (VR) via the multidisciplinary research environment INTERFACE, Hybrid multiscale modelling of transport phenomena for energy efficient processes, Grant no. 2016-06119. N.T. is grateful for the financial support of UCL-Osaka Partner Funds. The presented study was partially funded by Daicel Corporation. Last but not least, N.T. thanks Dr. Hiroshi Yamashita, Dr. Toshihiro Omori and also Prof. Masako Sugihara-Seki for helpful discussions.
\end{acknowledgments}

\section*{AUTHOR DECLARATIONS}
\subsection*{Conflict of Interest}

The authors report no conflict of interest.

\subsection*{Author Contributions}
\textbf{Naoki Takeishi:} Conceptualization (equal); Funding acquisition (lead); Data curation (lead); Formal analysis (lead); Investigation (lead); Methodology (lead); Resources (lead); Software (lead); Validation (lead); Visualization (lead); Writing - original draft (lead); Writing - review \& editing (equal).
\textbf{Marco Edoardo Rosti:} Conceptualization (equal); Funding acquisition (supporting); Data curation (supporting); Formal analysis (supporting); Investigation (supporting); Methodology (supporting); Validation (supporting); Writing - review \& editing (equal).
\textbf{Naoto Yokoyama:} Conceptualization (equal); Data curation (supporting); Formal analysis (supporting); Investigation (supporting); Methodology (supporting); Validation (supporting); Writing - review \& editing (equal).
\textbf{Luca  Brandt:} Conceptualization (lead); Funding acquisition (lead); Data curation (supporting); Formal analysis (supporting); Investigation (supporting); Methodology (supporting); Validation (supporting); Writing - review \& editing (lead).

\appendix


\section{\label{appA_domain_size}Effect of the domain size}
\begin{figure}
  \centering
  \includegraphics[height=5.5cm]{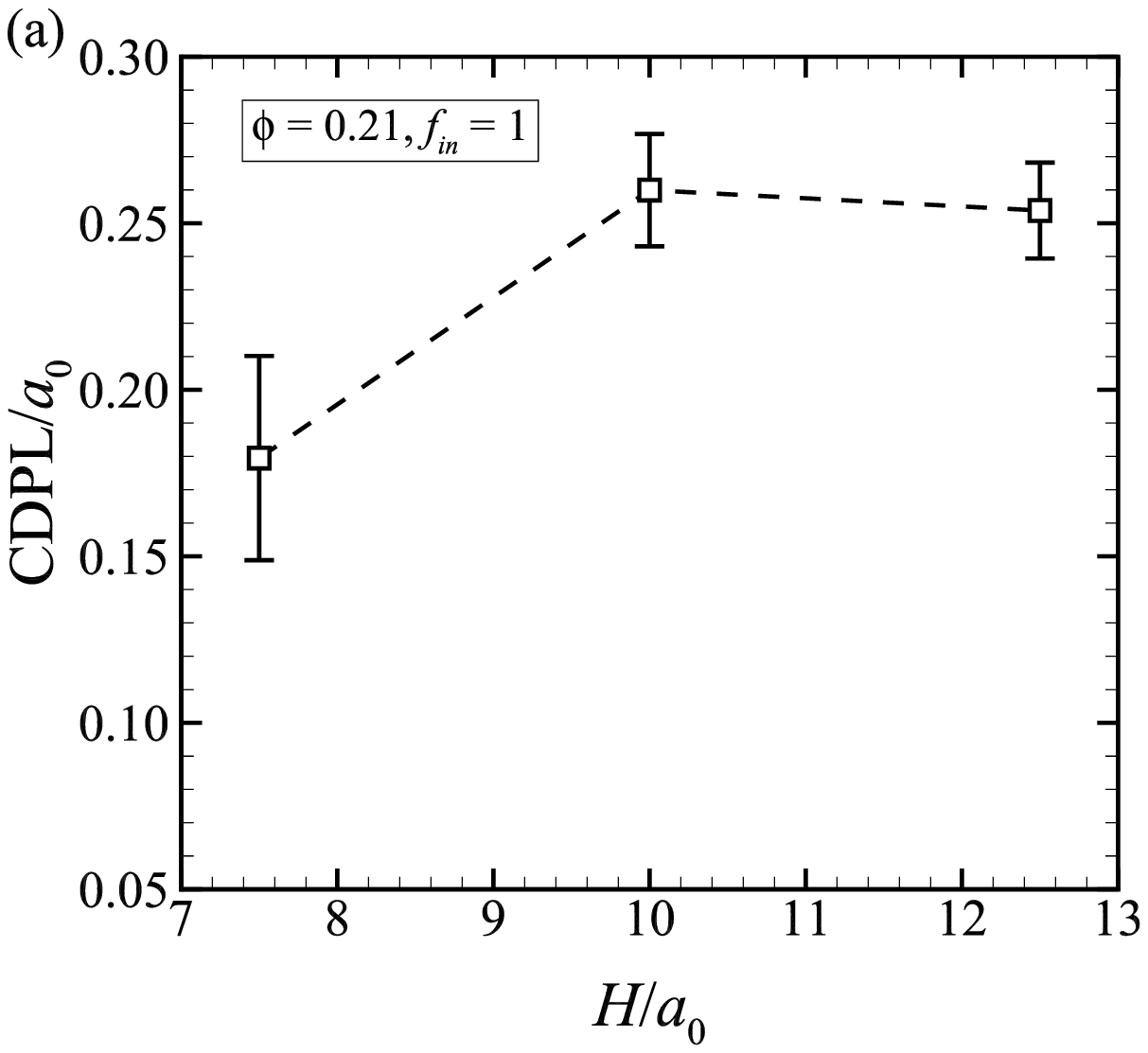}
  \includegraphics[height=5.5cm]{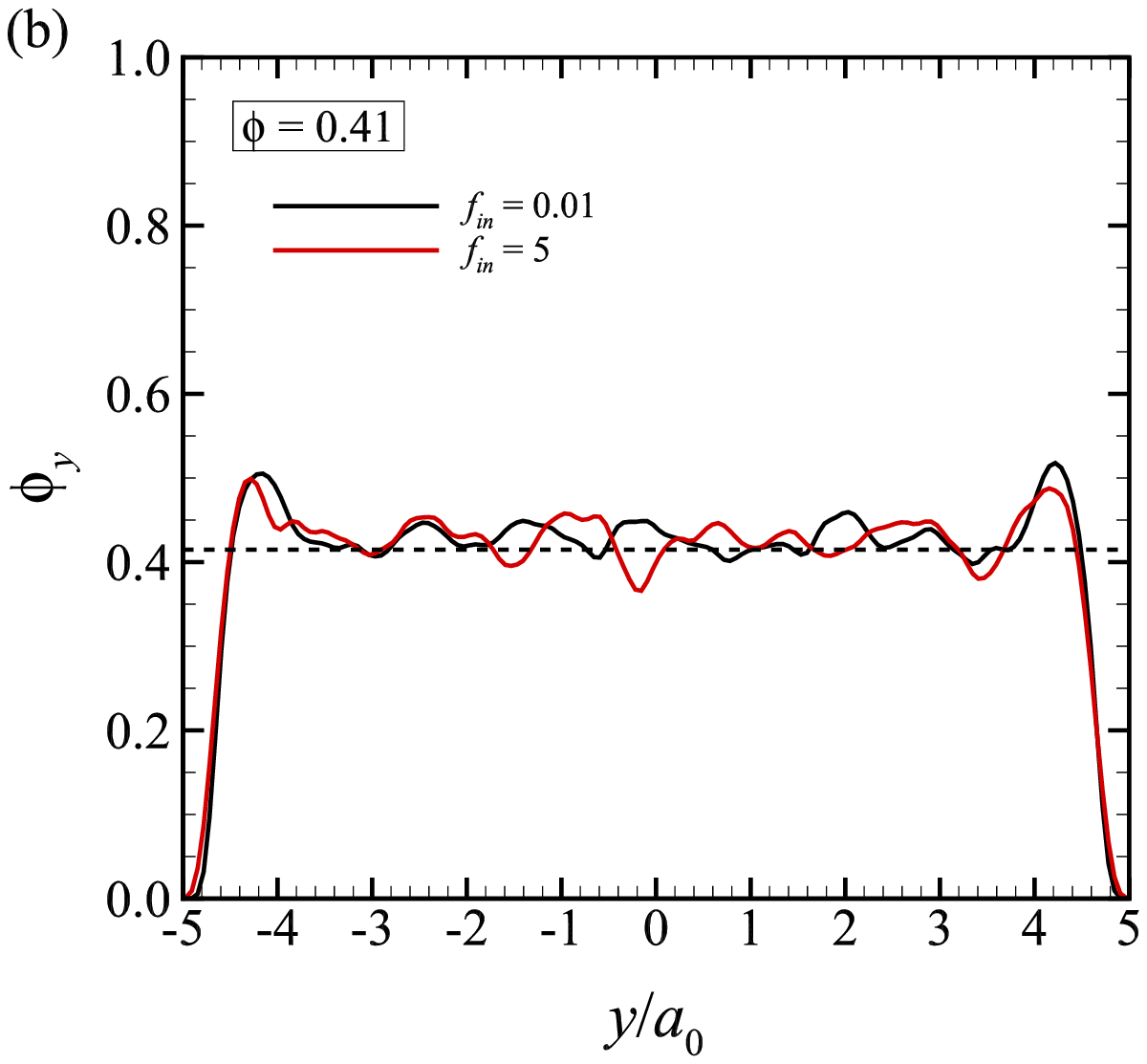}
  \caption{(a) Thickness of the CDPL for different domain height $H/a_0$ (= 7.5, 10, and 12.5) at $\phi$ = 0.21, $Ca_0$ = 0.4, and $f_{in}$ = 1. The error bars represent the standard deviation over time. (b) The effective volume fraction profile $\phi_y$ along the domain height $y$ at the minimum and maximum frequency $f_{in}$ (= 0.01 and 5, respectively) for a channel height $H$ = $10 a_0$, where the dashed line denotes the volume fraction (or $Hct_D$) $\phi$ = 0.41 (dashed line). The results are obtained on the fully developed flow and pertain the simulations with $\lambda$ = 5.}
  \label{fig:effect_height}
\end{figure}

The presence of walls bounding the suspension causes a finite CDPL. We quantify the thickness of the CDPL as done in a previous work~\citep{Takeishi2014}. The result is shown in Fig.~\ref{fig:effect_height}(a), where the thicknesses of the CDPL at $\phi$ = 0.21 and $f_{in}$ = 1 are investigated for different domain heights $H$ (= 7.5$a_0$, 10$a_0$, and 12.5$a_0$). We observe that the change in CDPL between the present height ($H$ = 10$a_0$) and a larger one ($H$ = 12.5$a_0$) $|\mathrm{CDPL}_{H = 10 a_0}/\mathrm{CDPL}_{H = 12.5 a_0} - 1|$ is less than 3\%, so we can conclude that the thickness of CDPL remains almost the same for $H \geq 10a_0$ .

Effective volume fraction profiles along the domain height $\phi_y$ are also shown at the minimum and maximum $f_{in}$ (= 0.01 and 5) for $\phi$ = 0.41 in Fig.~\ref{fig:effect_height}(b). The data show that $\phi_y$ is slightly greater than $\phi$ (or $Hct_D$), i.e., $\phi_y > \phi$, near the wall regions, with no big differences observed for the two $f_{in}$. We also evaluate the value of several observables (the phase difference $\delta$, the amplitude of the specific viscosity $|\mu_{sp}|^{amp}$, and the moduli of the complex viscosity $\eta^\prime/\mu_0$ and $\eta^{\prime\prime}/\mu_0$) for different values of the wall-to-wall distance $H$, whose results with the corresponding relative errors are listed in table~\ref{tab:effect_height}. Since the difference between the case with the largest height ($H$ = 12.5$a_0$) and our reference case are less than 2\% for all the quantities, the results presented in this study are all obtained with the domain height $H = 10a_0$, as also done in our previous numerical analysis~\citep{Takeishi2019}.

\begin{table}
  \caption{\label{tab:effect_height}Effect of the domain height $H$ on the phase difference $\delta$, amplitude of the specific viscosity $|\mu_{sp}|^{amp}$, and moduli of the complex viscosity ($\eta^\prime/\mu_0$ and $\eta^{\prime\prime}/\mu_0$). The error parameter $\varepsilon (\chi)$ is calculated based on $H = 10a_0$. The simulations are performed at $\phi$ = 0.21, $Ca_0$ = 0.4, $\lambda$ = 5, and $f_{in}$ = 1.}
  \begin{center}
  \def~{\hphantom{0}}
  \begin{ruledtabular}
  \begin{tabular}{cccc}  \\
  Height $H$ & 7.5$a_0$ & 10$a_0$ (reference) & 12.5$a_0$ \\ \hline
  Number of RBCs & 258 & 344 & 442 \\
  $\delta/\pi$ & -0.19032 & -0.19069 & -0.18874 \\
  $\varepsilon (\delta/\pi)$ & 0.00189 & - & 0.01021 \\
  $|\mu_{sp}|^{amp}$ & 0.59599 & 0.53463 & 0.54107 \\
  $\varepsilon (|\mu_{sp}|^{amp})$ & 0.11477 & - & 0.01205 \\
  $\eta^\prime/\mu_0$ & 0.49259	& 0.44153 & 0.44871 \\
  $\varepsilon (\eta^\prime/\mu_0)$ & 0.11563 & - & 0.01626 \\
  $\eta^{\prime\prime}/\mu_0$ & 0.33550 & 0.30146 & 0.30235 \\
  $\varepsilon (\eta^{\prime\prime}/\mu_0)$ & 0.11292 & - & 0.00297 \\
  \end{tabular}
  \end{ruledtabular}
  \end{center}
\end{table}

\section{\label{appA_stokes_problem}Stokes boundary layer}
\begin{figure}
  \centering
  \includegraphics[height=5.5cm]{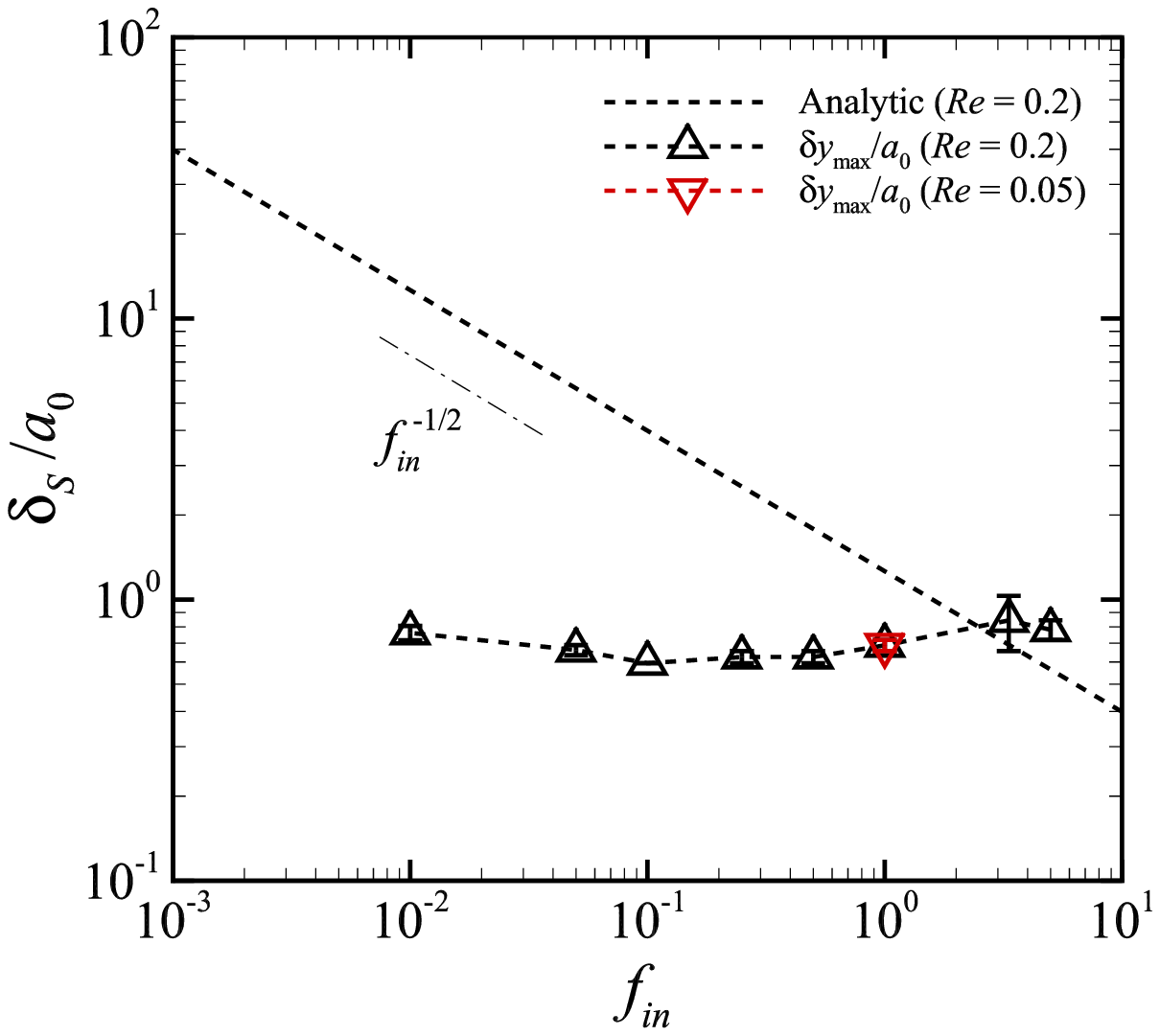}
  \caption{Time average of the distance $\delta y_{max}$ at $Re = 0.2$ along the domain height $y$ when the effective volume fraction profile $\phi_y$ is maximized, where the dashed line denotes the SBL layer thickness $\delta_S$ at $Re = 0.2$. The result of $\delta y_{max}$ at $Re = 0.05$ and $f_{in} = 1$ is also superposed. The results are obtained on the fully developed flow and pertain the simulations with $\lambda$ = 5.}
  \label{fig:stokes_layer}
\end{figure}

In the laminar regime, a boundary layer generated by the harmonic oscillations is denoted by a viscous length $\delta_S$ or Stokes boundary layer (SBL)~\citep{Stokes1851}. A conventional viscous length $\delta_S$ is defined using the kinematic viscosity $\nu$ of the (external) fluid; $\delta_S = \sqrt{2\nu/\omega}$. Therefore, a non-dimensional length can be defined as $\delta_S^\ast = \delta_S/a_0 = 1/\sqrt{\pi Re f_{in}}$. We quantify the distance $\delta y_{max}$ at $Re = 0.2$ along the wall-normal corrdinate $y$ where the effective volume fraction profile $\phi_y$ is maximum, and compare with $\delta$. The result is shown in Fig.~\ref{fig:stokes_layer}. All numerical results of $\delta y _{max}$ are smaller than or comparable to the estimated SBL. Furthermore, the result of $\delta y _{max}$ at $Re = 0.2$ is consistent with that obtained with smaller $Re$ (= 0.05). The distance $\delta y_{max}$ is independent of $f_{in}$, indicating that the CDPL is irrelevant to the SBL.

\section{\label{appA_distance}Distance between neighboring membrane nodes}
\begin{figure}
  \centering
  \includegraphics[height=5.5cm]{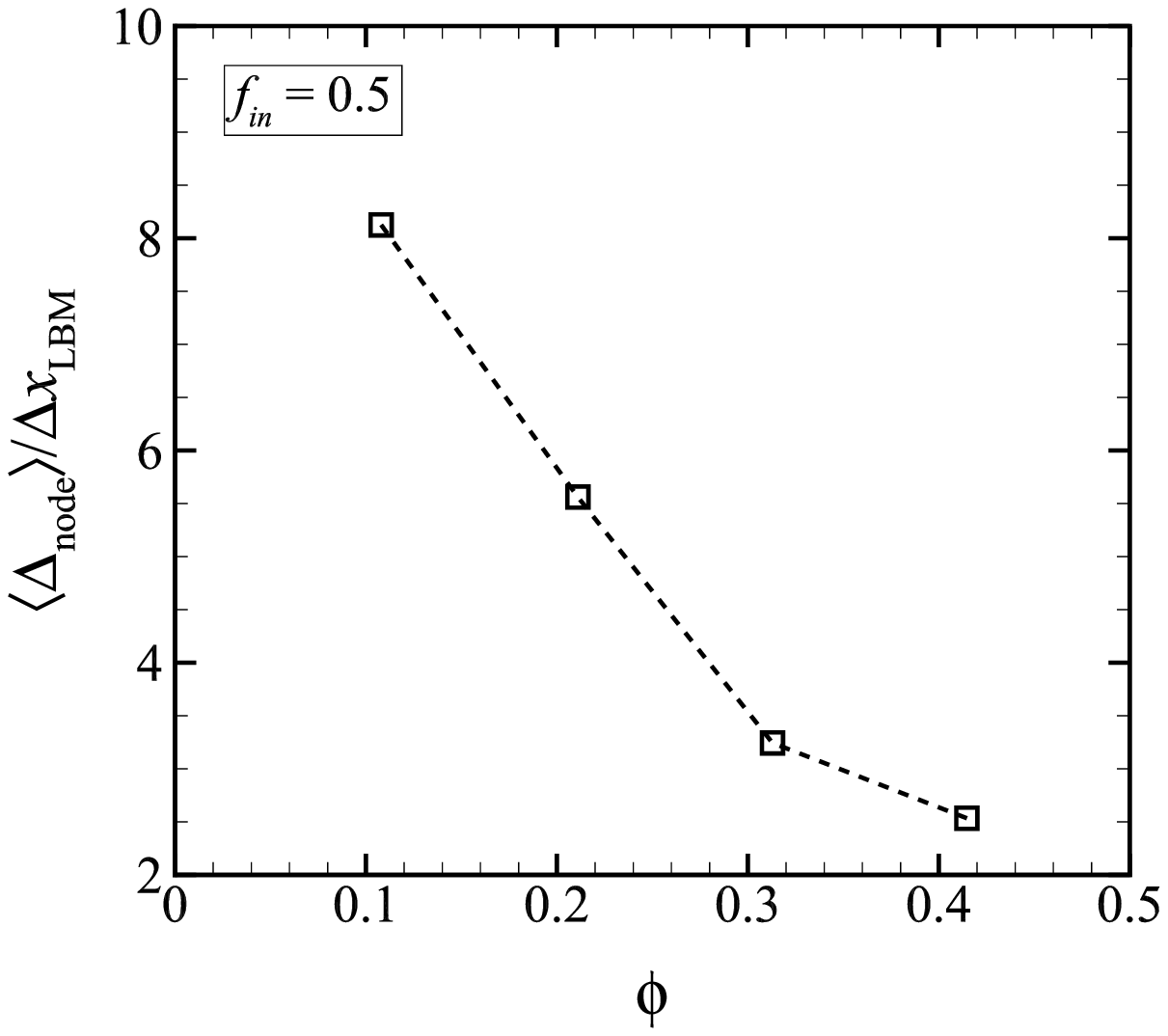}
  \caption{The time-averaged distance of neighboring membrane nodes $\langle \Delta_\mathrm{node} \rangle/\Delta x_\mathrm{LBM}$ as a function of $\phi$, where $\Delta x_\mathrm{LBM}$ is the mesh size of the LBM. The results are obtained with $Ca = 0.05$, viscosity ratio $\lambda = 5$, and frequency $f_{in} = 0.5$.}
  \label{fig:ave_distance}
\end{figure}

We further investigated the distance between neighboring membrane nodes. The result of time-averaged distance of neighboring membrane nodes $\langle \Delta_\mathrm{node} \rangle$ for different volume fractions $\phi$ is shown in Fig.~\ref{fig:ave_distance}. The result is obtained with $Ca = 0.05$, viscosity ratio $\lambda$ = 5, and frequency $f_{in} = 0.5$. Even at the highest volume fraction $\phi = 0.41$, the average distance still maintains over two fluid lattices. Thus, at least for $\phi \leq 0.41$, our coupling of LBM and IBM successfully resolve the distance between neighboring membrane nodes.

\section{\label{appA_DFT}Discrete Fourier transform}
\begin{figure}
  \centering
  \includegraphics[height=5.5cm]{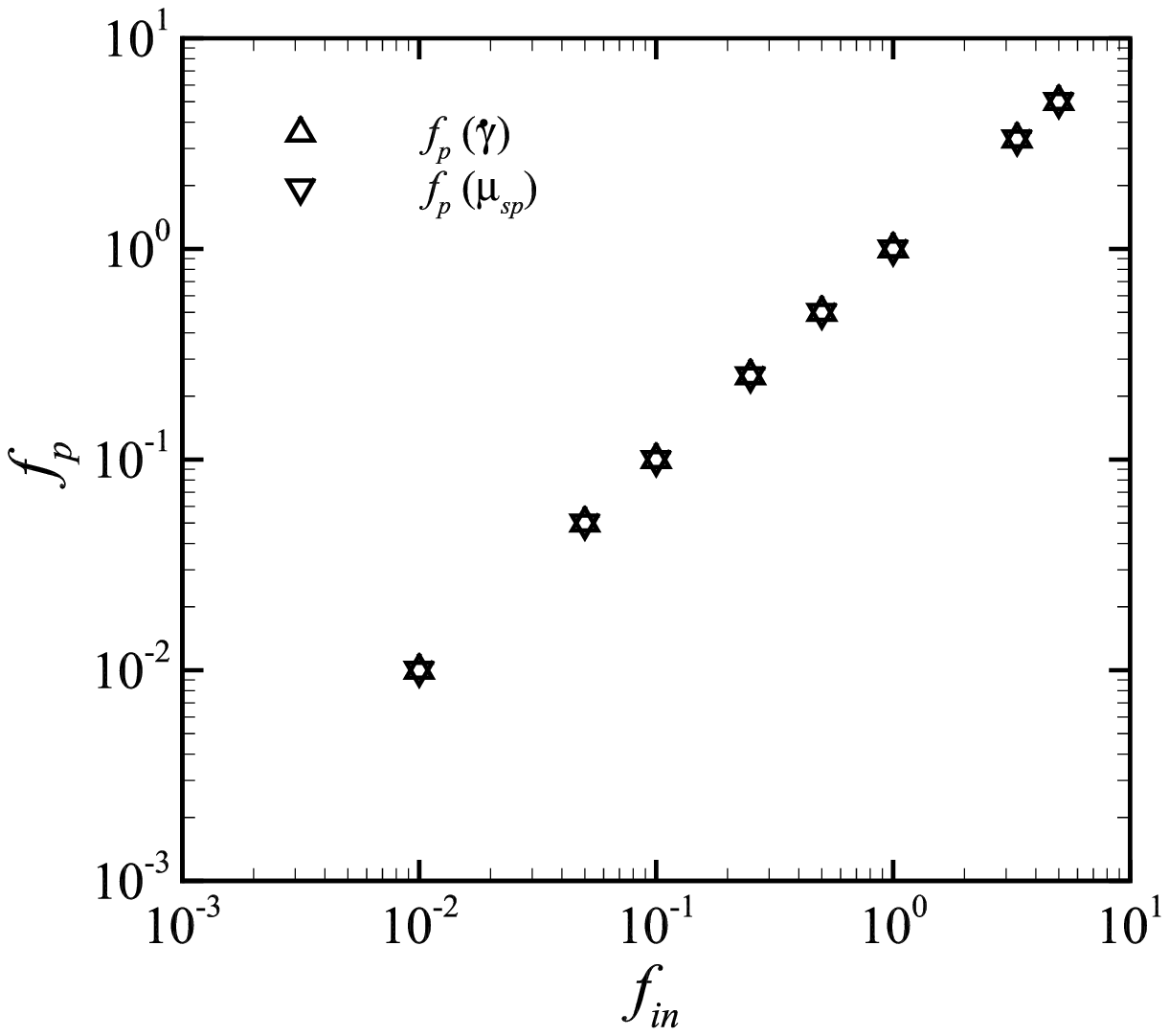}
  \caption{The peak frequency $f_p$ of the input shear rate $\dot{\gamma} (t)$ versus the peak frequency of the output specific viscosity $\mu_{sp} (t)$ calculated by the DFT analysis as a function of shear rate frequency $f_{in}$ ($= \mathrm{f}/\dot{\gamma}_0$). The results pertain the simulations with $\phi$ = 0.41, $\lambda$ = 5, and $Ca_0$ = 0.05.}
  \label{fig:fpeak_hct04ca005lam5}
\end{figure}

The peak frequency and amplitude of the output signal, the specific viscosity $\mu_{sp}$, and of the imposed shear rate $\dot{\gamma}$ are calculated by the discrete Fourier transform (DFT), defined as:
\begin{align}
	X(f_k) &= \sum^{n-1}_{j = 0} x(t_j) \exp{\left( -i 2 \pi j k/n \right)},
\end{align}
where $X(f_k)$ is the Fourier coefficient of the variable $x$ for the $k$-th mode,
and $n$ is the number of data points. 
Considering only the signal dominant component of frequency $f_p$, the one of the imposed shear as shown below, we can write the input and output signals as $\dot{\gamma} (t) \approx \dot{\gamma}_0 \cos{(2 \pi f_p t + \Theta_{\dot{\gamma}})}$ and $\mu_{sp} (t) \approx |\mu_{sp}|^{amp} \cos{(2 \pi f_p t + \Theta_{\mu_{sp}})}$, with the phase difference $\delta$ ($-\pi$/2 $\leq \delta \leq$ 0) expressed as $\delta = \Theta_{\mu_{sp}} - \Theta_{\dot{\gamma}}$. The phases $\Theta$ (= $\Theta_{\dot{\gamma}}$ or $\Theta_{\mu_{sp}}$) can be defined as the angle between the real and imaginary parts of  the Fourier coefficient $X(f_p)$ at the peak frequency $f_p$ of $\dot{\gamma}(t)$ or $\mu_{sp}(t)$:
\begin{align}
 \Theta = \mathrm{atan2}(\mathrm{Im}(X(f_p)), \mathrm{Re}(X(f_p))).
\end{align}
Indeed, the calculated peak frequencies of $\dot{\gamma}$ and $\mu_{sp}$ coincide with the input strain frequency $f_{in} (= \mathrm{f}/\dot{\gamma}_0)$ as shown in Fig.~\ref{fig:fpeak_hct04ca005lam5}.

\begin{table*}
  \caption{\label{tab:error}Effect of the number of wave periods used in the DFT analysis on the phase difference $\delta$, amplitude of the specific viscosity $|\mu_{sp}|^{amp}$, and moduli of the complex viscosity ($\eta^\prime/\mu_0$ and $\eta^{\prime\prime}/\mu_0$). The error parameter $\varepsilon (\chi)$ is defined as $\varepsilon (\chi) = |\chi/\chi_{ref} - 1|$, where the subscript $ref$ indicates the reference values. The simulations are performed at $\phi$ = 0.41, $Ca$ = 0.05, $\lambda$ = 5, and $f_{in}$ = 0.05.}
  \begin{center}
  \def~{\hphantom{0}}
  \begin{ruledtabular}
  \begin{tabular}{ccccccc}  \\
  Number of wave periods & 5 & 10 & 15 & 20 & 25 & 30 ({\it ref}) \\ \hline
  $\delta/\pi$ & -0.07956 & -0.08127 & -0.08117 & -0.08080 & -0.08045 & -0.08014 \\
  $\varepsilon (\delta/\pi)$ & 0.00730 & 0.01400 & 0.01283 & 0.00818 & 0.00382 & - \\
  $|\mu_{sp}|^{amp}$ & 2.76387 & 2.43931 & 2.46789 & 2.49496 & 2.51260 & 2.52682 \\
  $\varepsilon (|\mu_{sp}|^{amp})$ & 0.09381 & 0.03463 & 0.02332 & 0.01261 & 0.00563 & - \\
  $\eta^\prime/\mu_0$ & 2.67798 & 2.36025 & 2.38808 & 2.41501 & 2.43278 & 2.44715 \\
  $\varepsilon (\eta^\prime/\mu_0)$ & 0.09433 & 0.03551 & 0.02414 & 0.01313 & 0.00587 & - \\
  $\eta^{\prime\prime}/\mu_0$ & 0.68364 & 0.61603 & 0.62254 & 0.62654 & 0.62830 & 0.62950 \\
  $\varepsilon (\eta^{\prime\prime}/\mu_0)$ & 0.08600 & 0.02141 & 0.01107 & 0.00470 & 0.00191 & -\\
  \end{tabular}
  \end{ruledtabular}
  \end{center}
\end{table*}

\section{\label{appA_wave_num}Effect of wave periods on the solutions by DFT analysis}
The accuracy of the analysis  is important especially for small values of $\delta$ because the complex modulus is proportional to $\sin{\delta} \approx \delta$ for $\delta \ll$ 1. We have therefore checked the effect of the number of wave periods used for the input and output signals in the  DFT analysis: $\delta/\pi$, $|\mu_{sp}|^{amp}$, $\eta^\prime$, $\eta^{\prime\prime}$. Specifically, we have computed these values for a number of periods ranging from 5 to 30, where each output wave is resolved by a sufficiently large number of discrete points ($\geq$ 100). The values and errors obtained with respect to the data from the longest signal (30 periods) are summarised in table~\ref{tab:error}. All errors decrease as the number of wave periods increases.
For a reasonable  analysis, we deem necessary to use at least 10 periods, which gives a difference lower than 4\% with respect to using the full 30 wave periods.

\bibliography{reference}

\end{document}